\documentclass[11pt]{article}

\usepackage[margin=1in]{geometry}
\usepackage{amsmath}
\usepackage{amssymb}
\usepackage{amsthm}
\usepackage{booktabs}
\usepackage{graphicx}
\usepackage{url}
\usepackage{hyperref}
\usepackage[numbers,sort&compress]{natbib}
\usepackage{algorithm}
\usepackage{algpseudocode}
\usepackage{verbatim}
\usepackage{placeins}

\newtheorem{proposition}{Proposition}
\newtheorem{assumption}{Assumption}

\newcommand{\X}{\mathcal{X}}
\newcommand{\A}{\mathcal{A}}
\newcommand{\E}{\mathbb{E}}
\newcommand{\R}{\mathbb{R}}

\newcommand{\qcphast}{QC-PHAST}
\newcommand{\qcphastexpanded}{Quantum-Classical Phase-space and Stability-Threshold Search}
\newcommand{\bbhtfull}{Boyer--Brassard--H{\o}yer--Tapp}
\newcommand{\bbht}{\bbhtfull{} (BBHT)}

\newcommand{\cientry}[3]{\shortstack[r]{#1\\[-1pt]{\scriptsize[#2,}\\[-1pt]{\scriptsize #3]}}}
\newcommand{\readableci}[3]{\shortstack[r]{#1\\[-1pt][#2,\\[-1pt]#3]}}

\title{QC-PHAST Search: Classical--Quantum Query Benchmarks for Finite-Pool Rare-Regime Discovery}
\author{
\begin{minipage}[t]{0.46\textwidth}
\centering
Harsh Milind Tirhekar\\
\small Department of Computer Science\\
\small College of Natural Sciences\\
\small The University of Texas at Austin\\
\small Austin, TX, USA\\
\small \href{mailto:harsh.tirhekar@utexas.edu}{harsh.tirhekar@utexas.edu}
\end{minipage}
\and
\begin{minipage}[t]{0.46\textwidth}
\centering
Chandrajit Bajaj\\
\small Department of Computer Science\\
\small Oden Institute for Computational Engineering and Sciences\\
\small The University of Texas at Austin\\
\small Austin, TX, USA\\
\small \href{mailto:bajaj@cs.utexas.edu}{bajaj@cs.utexas.edu}
\end{minipage}
}
\date{}

\begin{document}
\maketitle

\begin{abstract}
Rare-regime discovery in parameterized dynamical systems is an active-search problem: find one verified parameter at which a scientifically defined qualitative threshold is crossed, even when acceptable candidates are rare, nonconvex, or fragmented. We introduce \emph{\qcphastexpanded{}} (\qcphast), an evidence-gated decision protocol and query-accounting framework for finite candidate libraries. A candidate induces a dynamical object, simulator-derived criticality score, and verified first-hit predicate. Scientific metadata and charged pilot evidence are used to assess whether equation-aware search, scalar-score active search, predicate-only search, or only a query-model comparison is admissible. The quantum row is the inherited Grover/\bbhtfull{} (BBHT) unknown-$M$ marked-set query reference; it is not a new quantum-search theorem, materialized circuit, or hardware-speedup claim.

The result is a regime map. In an offline controlled sweep of 875 configurations over seven canonical systems, five pool sizes, five target fractions, and five resampling seeds, the exact finite-pool replay gives a point estimate of 2.71 for the included non-quantum / BBHT ratio at $M/N=0.001$. Paired hierarchical resampling by system, seed, and size gives 2.71 [1.89, 3.68] for the mean ratio and 2.39 [1.76, 3.31] for the geometric mean. Under stronger scalar-score GP access, the configuration-level ratio is 2.24 [2.02, 2.47], with BBHT favorable in 0.71 of configurations. Four confirmation sweeps span 7,175 base configurations covering fixed thresholds, fully charged pilot calibration, continuous structure-aware routing, and predicate noise; the noise study further expands each of its 875 base configurations across three noise models and eight rates. A 5\% noisy-predicate ablation gives 0.29 [0.27, 0.32], a predicate-only replication gives 0.17 [0.15, 0.20], and coherent-oracle costs above roughly 2--3 classical score checks remove total-cost headroom. Direct boundary constructions, geometry controls, online simulator loops, and learned-label accounting further identify when classical structure, false positives, calibration cost, or state preparation erases the query-model margin. \qcphast{} is therefore an auditable protocol for deciding when a finite-pool marked-set reference is informative and when classical or resource-aware search should dominate.
\end{abstract}

\section{Introduction}

Many scientific-ML workflows are first-discovery problems. A modeler may need one parameter vector that crosses a verified qualitative threshold, such as a stability boundary, oscillatory onset, conservative limit, saddle-node condition, or Hopf-type boundary, rather than the global minimum of a simulator score. When the candidate library is finite and the acceptable set is rare, exhaustive sweeps and unguided random search spend most evaluations away from the scientific target. Classical active-search and Bayesian-optimization methods can exploit smooth score geometry, but this help is uneven: smooth coherent targets can be easy, while thin, fragmented, or boundary-like rare sets remain difficult.

The paper asks one question: \emph{after classical policies are allowed to exploit finite-pool score geometry and equation-level structure, when does a geometry-agnostic BBHT marked-set query reference remain informative?} The answer cannot be read from rarity alone. It depends jointly on the available information, threshold provenance, target geometry, predicate fidelity, and the cost of constructing coherent access. These quantities define the evidence gates used throughout \qcphast{}.

\begin{figure*}[!htbp]
\centering
\includegraphics[width=0.94\textwidth]{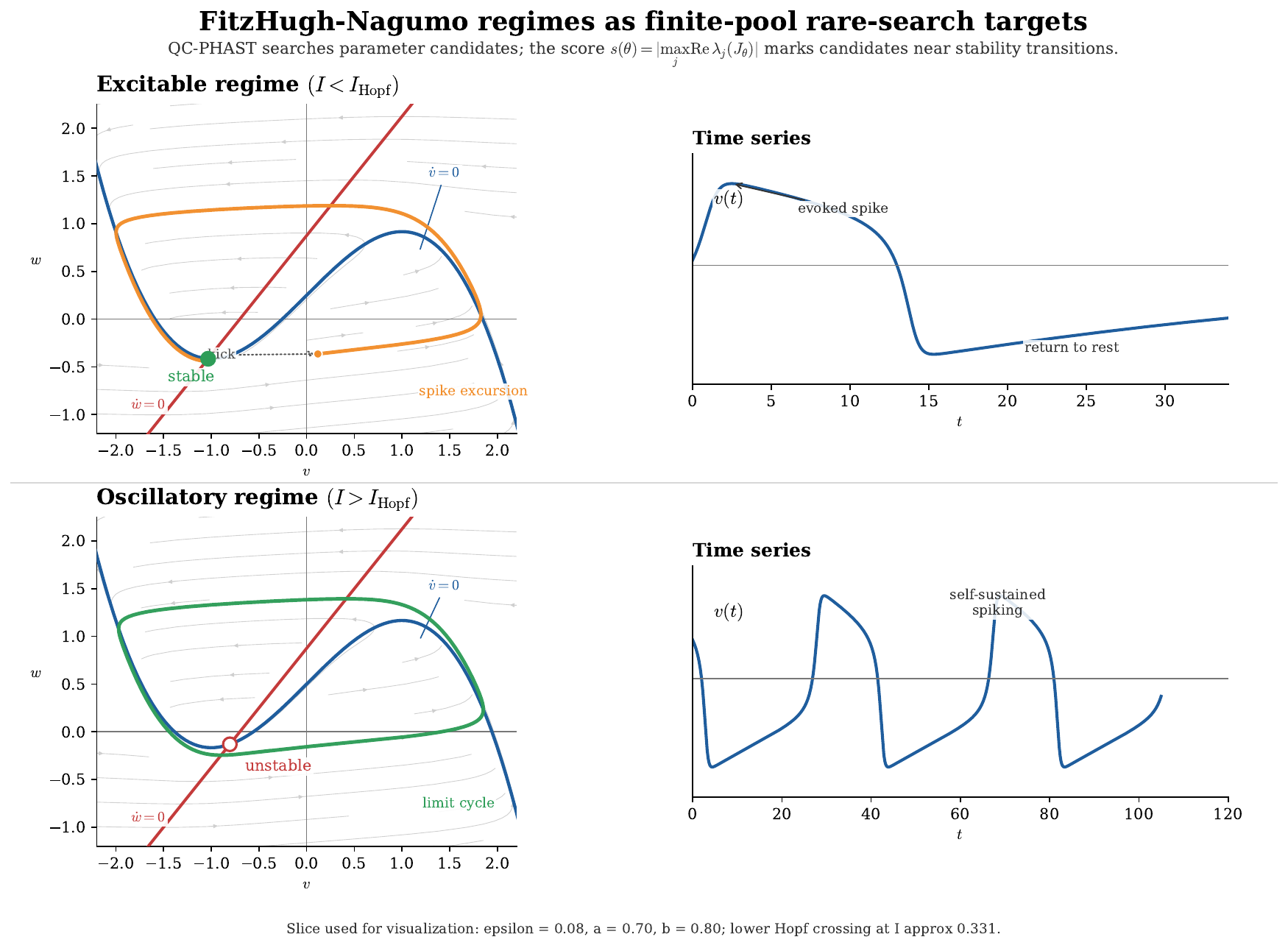}
\caption{FHN regime bridge. \qcphast{} searches parameter candidates, and each candidate induces a phase portrait, equilibrium stability, trajectory response, and voltage trace. In the excitable regime, a stable equilibrium returns to rest after a perturbation-driven spike; in the oscillatory regime, the equilibrium is unstable and trajectories approach a limit cycle. The benchmark score minimizes $|\max_j\operatorname{Re}\lambda_j(J_{\mathrm{FHN}})|$ over relevant equilibria; this slice visualizes one such stability transition and gives a concrete phase-space meaning to the finite-pool rare-regime predicate.}
\label{fig:fhn-regimes}
\end{figure*}

Figure~\ref{fig:fhn-regimes} gives a concrete example before the general framework. The FitzHugh-Nagumo (FHN) system is a canonical reduced model of excitable dynamics introduced for nerve-membrane and active-pulse phenomena \citep{fitzhugh1961,nagumo1962}. Standard phase-plane treatments and modern surveys emphasize that its nullclines, separatrices, limit cycles, and fast-slow structure make it a compact testbed for excitability and bifurcation analysis \citep{izhikevich2006scholarpedia,izhikevich2007neuroscience,cebrianlacasa2024sixdecades}. It is useful here because a small parameter change can move the phase portrait from a stable excitable rest state, where a perturbation creates one spike and returns to rest, to an oscillatory regime with an unstable equilibrium and a stable limit cycle. In the benchmark, FHN is not a clinical simulator; it is a controlled dynamical-system search target where the rare event has a physical meaning: proximity to a stability transition.

The machine-learning object in Figure~\ref{fig:fhn-regimes} is the active search policy, not a fitted FHN forecaster. After each online score or predicate query, a policy chooses which parameter candidate to evaluate next under a finite budget; in saved-score experiments, the identical interaction is replayed against an immutable score bank. Nonadaptive coverage rules ignore feedback; cross-entropy, subset-style, and GP active-search policies adapt from observed scores and uncertainty; noisy-label experiments test the boundary where learned triage is no longer safe as a marked predicate. Table~\ref{tab:fhn-anatomy} shows why even a two-state model can induce several qualitatively different dynamical objects. The current benchmark does not label all of those objects directly: its implemented FHN predicate is local stability-distance at equilibria. The broader phase portrait explains the scientific meaning of that boundary and identifies richer trajectory-level targets for future work. \qcphast{} uses evidence gates, rather than a trained meta-selector, to decide which search layer the measured access conditions support.
\begin{table*}[!htbp]
\centering
\caption{FHN phase-portrait anatomy and active-search implications. The table compresses the visual bridge in Figure~\ref{fig:fhn-regimes}: qualitative dynamical features become score geometry that an acquisition policy can exploit or fail to exploit.}
\label{tab:fhn-anatomy}
\footnotesize
\renewcommand{\arraystretch}{1.25}
\setlength{\tabcolsep}{4pt}
\begin{tabular}{p{0.18\textwidth} p{0.28\textwidth} p{0.40\textwidth}}
\toprule
Phase-space feature & Dynamical meaning & Active-search implication \\
\midrule
Spiral sink & Stable equilibrium; perturbations decay back to rest. & Usually not the target unless the leading eigenvalue is close to zero, but score feedback can reveal approach to the stability boundary. \\

Spiral source & Unstable equilibrium; nearby trajectories leave the fixed point. & Can occur after an equilibrium loses stability; the implemented leading-real-eigenvalue score detects proximity to the loss of stability, not the global attractor by itself. \\

Limit cycle & Closed attracting orbit and self-sustained spiking. & Gives the global dynamical interpretation of oscillatory behavior, but is not directly detected by the current local-stability predicate. A trajectory-level verifier would be required to mark it explicitly. \\

\shortstack[l]{Slow manifold /\\cubic branch} & Fast voltage motion approaches a nullcline branch, followed by slower recovery motion. & Can produce anisotropic parameter responses. The present benchmark uses equilibrium stability distance; it does not claim to reconstruct the slow manifold during search. \\

Fast jump & Rapid voltage excursion with recovery nearly fixed. & Illustrates why a small parameter displacement can change trajectory behavior sharply; direct jump detection would require a trajectory-level score. \\

\shortstack[l]{Saddle or\\separatrix} & Basin boundary separating qualitatively different outcomes. & Motivates fragmented or boundary-like stress tests, but separatrix geometry is not part of the implemented FHN label. \\
\bottomrule
\end{tabular}
\end{table*}

\begin{figure*}[!htbp]
\centering
\includegraphics[width=\textwidth]{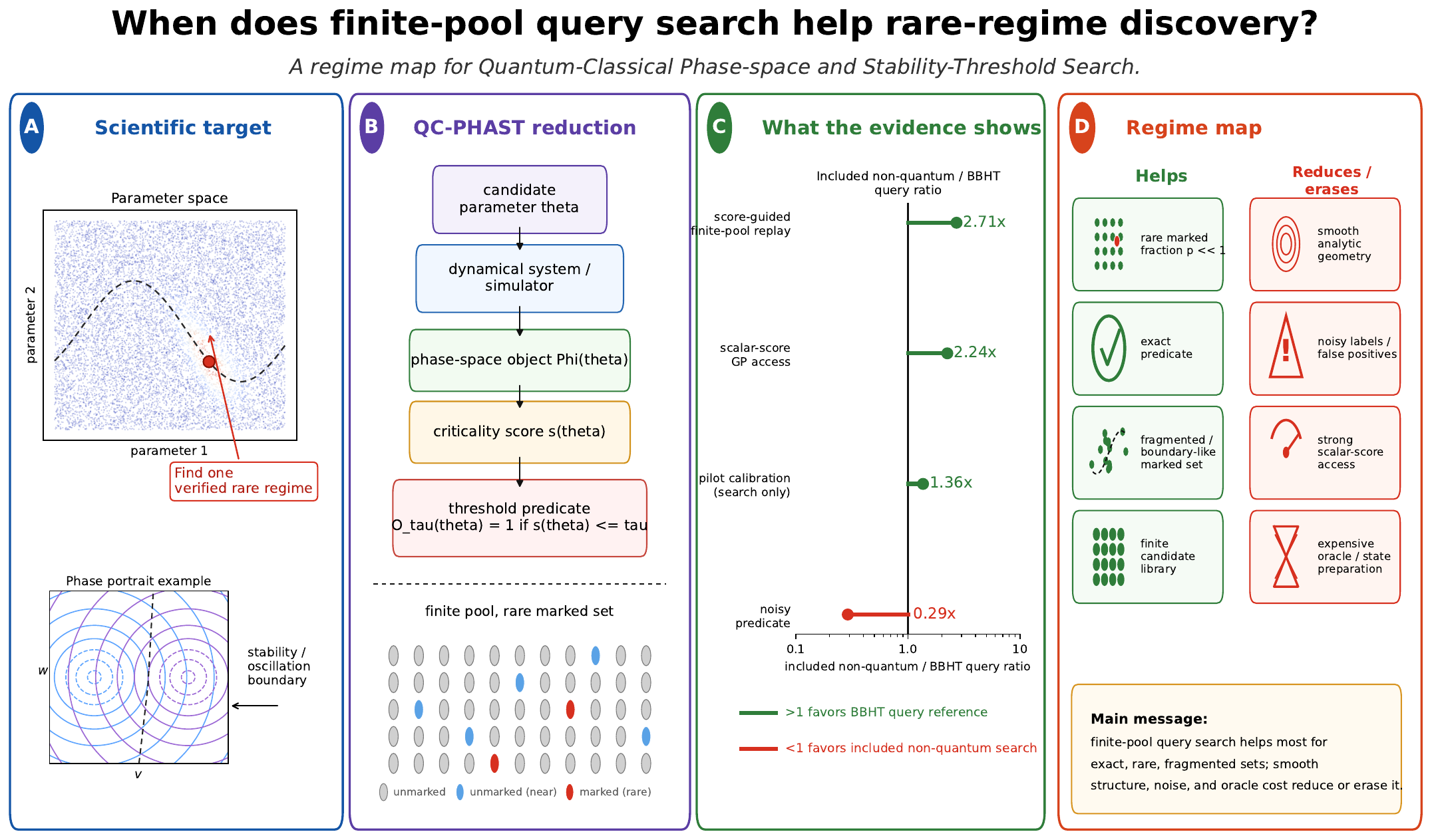}
\caption{A regime map for \qcphastexpanded{}. Panel A frames rare-regime discovery as finding one verified parameter near a phase-space transition. Panel B shows the reduction from candidate parameter to simulator, phase-space object, criticality score, threshold predicate, and finite marked set. Panel C reports selected included non-quantum / BBHT query-count ratios, where values above one favor the BBHT query reference and values below one favor included non-quantum search. The first row is the exact finite-pool replay with scalable score-guided baselines; BBHT itself has exact binary marked-oracle access. Its pilot point is a search-after-calibration diagnostic; Table~\ref{tab:pilot-all-in} separately charges the calibration budget. Panel D summarizes when the finite-pool marked-set layer helps and when smooth analytic structure, noisy labels, strong scalar-score access, or oracle/state-preparation cost reduces or erases the gain. \qcphast{} denotes \qcphastexpanded{}; BBHT denotes the \bbhtfull{} unknown-$M$ Grover-search schedule.}
\label{fig:teaser}
\end{figure*}

Figure~\ref{fig:teaser} summarizes the paper's search-selection view. Panel A shows the scientific objective: find one verified parameter near a qualitative phase-space transition. Panel B shows the \qcphast{} reduction from a candidate parameter to a simulator-induced phase-space object, criticality score, threshold predicate, and finite marked set. Panels C and D summarize the empirical message: exact rare finite predicates favor the BBHT query reference in query count, while scalar-score access, threshold uncertainty, noisy labels, smooth structure, and oracle/state-preparation costs narrow or reverse the margin.

\qcphast{} (\qcphastexpanded{}) studies this setting as a finite-pool active search and rare-regime discovery. Given candidates $\X=\{x_i\}_{i=1}^{N}$, a simulator-derived criticality score $s(x_i)$, and a threshold $\tau$, the marked set is $\A_\tau=\{x_i:s(x_i)\le\tau\}$ with marked fraction $p=M/N$. The objective is first verified hit: find any member of $\A_\tau$ with as few verified score or predicate checks as possible. This differs from continuous global optimization because the candidate set, threshold, access model, and verifier are fixed before the search policy is evaluated.

The name \qcphast{} reflects the two ingredients used throughout the framework: candidate parameters are interpreted through their induced phase-space objects, and rare regimes are defined by verified stability or criticality thresholds. The acronym names the finite-pool query-efficiency framework introduced here; it is not a port-Hamiltonian temporal-forecasting model and not a hardware quantum-runtime claim.

Quantum search gives a clean finite marked-set query reference. Grover search gives the square-root primitive for unstructured marked-set search; \bbht{} handles the more realistic case where the number of marked candidates $M$ is unknown; and amplitude amplification gives the general success-probability view \citep{grover1997,boyer1998,brassard2002}. \qcphast{} uses this theory only as a query-count layer. If the problem is reduced to exact finite marked-set search, BBHT's square-root behavior is expected. The empirical machine-learning question is thus instead whether this reference remains informative after the simulator score, threshold-estimation procedure, scalar-score access, continuous structure, noisy predicates, and oracle costs are made explicit.

This framing makes the machine-learning contribution explicit. \qcphast{} specifies an evidence-gated decision protocol and access-model evaluation framework for scientific simulators, not a claim that one search rule is best everywhere. Its value is the controlled routing logic and comparison of finite-pool search policies under matched first-hit objectives: smooth analytic or continuous-structure cases are routed to structure-aware non-quantum methods, scalar-score cases are tested with GP active-search diagnostics, noisy-predicate cases are routed away from a marked-oracle layer when label error is comparable to the rare-event density, and exact rare finite-pool cases are compared against the Grover/BBHT query reference. The protocol is a transparent scientific decision rule, not a learned per-instance algorithm selector; the paper evaluates its constituent evidence tracks rather than prospective selector regret on held-out simulator families.

\paragraph{Contributions.}
\begin{itemize}
    \item We formalize finite-pool rare-regime discovery for parameterized dynamical systems, including the map from searched parameters to phase-space structure, criticality scores, threshold predicates, verified first-hit queries, p90 budgets, and access models.
    \item We specify \qcphast{} as an evidence-gated decision protocol and access-model evaluation framework: pre-flight structure probes, finite-pool active-search comparisons, scalar-score diagnostics, noisy-predicate checks, and oracle break-even tests support an auditable routing interpretation under explicitly declared gates.
    \item We instantiate the protocol on seven systems and an 875-configuration controlled density sweep, then add 7,175 base configurations across fixed-threshold, all-in calibration, continuous-routing, and predicate-noise confirmations. The resulting regime guide shows that exact rare fragmented or boundary-like finite marked sets can favor the Grover/BBHT query reference, while scalar-score access, smooth continuous structure, threshold uncertainty, noisy predicates, and oracle/state-preparation costs define where classical or resource effects dominate.
\end{itemize}

Table~\ref{tab:claim-scope} states these claims in audit-ready form by pairing each claim with its evidence and scope boundary.
\begin{table*}[!htbp]
\centering
\caption{Claim-evidence-scope summary. The main result is a scoped search-selection guide for finite-pool rare-regime discovery.}
\label{tab:claim-scope}
\footnotesize
\renewcommand{\arraystretch}{1.22}
\setlength{\tabcolsep}{4pt}
\begin{tabular}{p{0.17\textwidth} p{0.48\textwidth} p{0.23\textwidth}}
\toprule
Claim & Evidence in this paper & Scope \\
\midrule

Evidence-gated protocol &
\qcphast{} (\qcphastexpanded{}) formalizes finite-pool rare-regime discovery through pre-flight structure probes, routing state $c$, $\X$, $s$, $\tau$, $\A_\tau$, $Q_{\mathrm{hit}}$, p90 budget, objective correctness, access-model accounting, and Algorithm~\ref{alg:qcphast}. &
Transparent rule-based decision/evaluation framework, not a prospectively validated learned meta-selector, continuous optimizer, or bifurcation solver. \\

Rarity improves BBHT query ratios &
875-configuration sweep; ratio point estimate 2.71 at $M/N=0.001$; paired hierarchical mean 2.71 [1.89, 3.68]. &
Exact BBHT marked-oracle reference versus the included scalable score-guided finite-pool baselines. \\

Stronger scalar-score access narrows the margin &
Scalar-score GP active-search access gives 2.24 [2.02, 2.47] ratio and 0.71 win rate at $M/N=0.001$. &
Finite-pool numerical scores are observed by the GP diagnostics. \\

Threshold and label realism matter &
The GP pilot stress is 1.36 [1.14, 1.58] after calibration only; a separate 1,750-configuration all-in $B=256$ ledger gives 2.62 [2.49, 2.77] under a restricted predicate-only portfolio. A 5\% noisy predicate gives 0.29 [0.27, 0.32] in the access-model ablation and 0.17 [0.15, 0.20] in the predicate-only confirmation. &
Pilot rows use distinct policy portfolios and are controlled calibration diagnostics, not deployment runtime claims. \\

Smooth structure can remove the need for the marked-set layer &
Across 425 fixed-threshold draws per system, direct FHN and Lorenz controls find a candidate in one evaluation; Duffing finds one in one evaluation with a 0.99 strict-local-certificate pass rate; smooth ball geometry gives ratio 0.78. &
Different access model, reported as routing controls and negative controls. \\

Resource headroom is necessary &
Oracle break-even at $M/N=0.001$ is about 2--3 classical score checks. &
Query model; hardware runtime requires separate resource estimates. \\

\bottomrule
\end{tabular}
\end{table*}

\FloatBarrier

\section{Related Work}

\paragraph{Quantum search and amplitude amplification.}
Grover's algorithm establishes a quadratic query improvement for unstructured search \citep{grover1997}. BBHT extends the setting to an unknown number of solutions \citep{boyer1998}, which is the relevant case for simulator-defined rare regimes because $M$ is generally not known before search. Amplitude amplification abstracts the same principle to arbitrary procedures with nonzero success probability \citep{brassard2002}. Approximate counting, fixed-point search, branch-and-bound speedups, and modern end-to-end quantum-algorithm surveys are related examples of how marked-state search primitives enter broader optimization and counting workflows \citep{aaronson2020approxcount,yoder2014fixedpoint,montanaro2020branch,dalzell2025quantumalgorithms}. Quantum speedups for Monte Carlo and randomized procedures provide additional query-model motivation \citep{montanaro2015}.

Modern quantum-algorithm work reinforces two points that shape our framing. First, amplitude-estimation variants have continued to reduce reliance on ideal phase-estimation subroutines and to clarify how amplitude information can be extracted in more practical settings \citep{suzuki2020,grinko2021}. Second, the broader NISQ and end-to-end algorithm literature emphasizes that near-term quantum algorithms must be evaluated with care: variational and noisy intermediate-scale algorithms are powerful research directions, but hardware noise, state preparation, circuit depth, data loading, and oracle construction remain central obstacles \citep{cerezo2021,bharti2022,dalzell2025quantumalgorithms}. Faulty-oracle work is especially relevant because even small oracle failure can remove Grover-style search advantage in standard noisy-query models \citep{regev2012faultyoracle,shenvi2003noisyoracle,lolck2024faultyoracle}. Resource-estimation work on qRAM, fault-tolerant implementations, arithmetic circuits, and magic-state factories likewise warns that loading classical candidate libraries into coherent access and evaluating numerical predicates can dominate a nominal query speedup \citep{dimatteo2019qram,duan2024qram,babbush2021beyondquadratic,cuccaro2004adder,haener2018arithmetic,remaud2024adders,gidney2019magic}. QC-PHAST uses BBHT as a finite-oracle query reference and separates query complexity from circuit implementation. Recent quantum algorithms and reviews for scientific computing and nonlinear differential equations also show that dynamical systems are a serious target for quantum scientific computing \citep{liu2021,auyeung2024quantumscientific,wu2024nonlineardynamics}; our contribution is complementary because we search over simulator parameters rather than quantum-solving the simulator dynamics.

\paragraph{Classical black-box and rare-event search.}
Random search is a competitive baseline in high-dimensional hyperparameter optimization because only some dimensions may matter \citep{bergstra2012}. Bayesian optimization models expensive black-box functions to select promising evaluations, with modern surveys and recent variants emphasizing constrained, high-dimensional, cost-aware, and grey-box settings \citep{snoek2012,wang2023bo,xie2024costaware,xu2023greybox}. Active search and level-set estimation are especially close to QC-PHAST because they prioritize discovering members of a target class or estimating a thresholded set rather than merely minimizing a scalar objective \citep{garnett2012active,gotovos2013levelset}. Cross-entropy methods and modern rare-event prediction surveys provide adaptive sampling and imbalanced-rare-event context \citep{rubinstein2004,shyalika2024rareevent}, while subset simulation addresses small failure probabilities by progressively exploring rarer sets \citep{au2001}. Reliability active learning and sequential failure-probability design further show how surrogate uncertainty can focus expensive limit-state evaluation \citep{echard2011akmcs,bect2012failure}. We include these classical perspectives to avoid comparing against weak strawmen.

QC-PHAST is related to rare-event estimation, but the objective is different. Subset simulation and related methods often estimate a small probability. Here the target is first discovery of a marked candidate under a shared thresholded score. Bayesian optimization is also adjacent but not identical: it usually optimizes or models a scalar response, while our final evidence asks how many marked-predicate queries are needed to find one candidate satisfying $s(x)\leq\tau$.

Modern Bayesian optimization has expanded well beyond a single expected-improvement-style template. Recent surveys emphasize noisy, constrained, batch, high-dimensional, cost-aware, grey-box, and multi-objective variants \citep{wang2023bo,xie2024costaware,xu2023greybox}; modern software, trust-region methods, and scalable constrained BO make off-grid continuous optimization a serious competitor rather than an afterthought \citep{balandat2020botorch,eriksson2019turbo,eriksson2021scbo}. High-dimensional BO methods can exploit sparse low-dimensional structure when it exists \citep{eriksson2021}, while recent active level-set work shows how prior information can be transferred robustly when the objective is threshold-set identification rather than minimization \citep{ngo2025levelset}. Derivative-free optimizers such as CMA-ES motivate rank-based continuous search controls \citep{hansen2016cma}. This matters for interpreting QC-PHAST. We do not treat classical adaptive search as weak. Bayesian LCB appears in the default and stress experiments as a finite-pool diagnostic, and the access-model ablation adds finite-pool GP active-search, level-set, and expected-threshold-improvement diagnostics on the full 875-base-configuration design. The paper's central metric is not improvement over random search alone; it is improvement over the included non-quantum baseline available in each configuration.

\paragraph{Algorithm selection and portfolios.}
Per-instance algorithm selection asks which member of a portfolio should solve a given problem instance. Recent work in automated algorithm selection emphasizes that the portfolio itself, the instance representation, and the cost of obtaining informative features all affect the selector's value \citep{pulatov2022algorithmselection,kostovska2023psaas}. \qcphast{} shares the portfolio view but does not train a meta-model to predict the winning search policy. Its routing variables are scientifically interpretable pilot quantities---access type, estimated rarity, predicate error, geometry, structure-probe success, and oracle-cost headroom---and its hard gates are designed to prevent invalid marked-oracle or runtime interpretations. Learning a calibrated router across a much larger family of simulators would be a separate meta-learning contribution. This distinction is important: the present contribution is an auditable decision protocol supported by a cross-regime benchmark, not a claim of learned algorithm-selection optimality.

Table~\ref{tab:method-family-comparison} makes the comparison boundary explicit. Several prior method families are excellent when their access assumptions match the problem: active search for finding many positives, level-set estimation for smooth threshold surfaces, reliability methods for failure-probability estimation, and continuation tools for tracing known solution branches \citep{govaerts2005matcont,liessi2025matcont}. QC-PHAST asks a narrower question: after a finite candidate set and a verified criticality predicate are defined, how many checks are needed to find a first rare candidate?

\begin{table*}[!htbp]
\centering
\caption{Method-family comparison. The table separates QC-PHAST's finite-pool first-hit objective from related classical and quantum objectives that use different access models.}
\label{tab:method-family-comparison}
\footnotesize
\renewcommand{\arraystretch}{1.18}
\setlength{\tabcolsep}{3pt}
\begin{tabular}{@{}p{0.16\textwidth}p{0.17\textwidth}p{0.17\textwidth}p{0.42\textwidth}@{}}
\toprule
Method family & Access model & Typical objective & Relation to QC-PHAST \\
\midrule
Active search \citep{garnett2012active} & Labels or scores over a pool & Find many positives & Closest classical objective; current paper includes first-hit finite-pool active-search diagnostics but does not optimize total-positive yield. \\
Algorithm selection \citep{pulatov2022algorithmselection,kostovska2023psaas} & Instance features and policy-performance data & Select a solver from a portfolio & Closest framework-level objective; QC-PHAST uses transparent evidence gates rather than a learned meta-selector. \\
GP level-set estimation \citep{gotovos2013levelset,shekhar2019multiscale,ngo2025levelset} & Noisy scalar evaluations & Estimate a thresholded set & Directly relevant for smooth score surfaces; QC-PHAST instead measures first verified hit under fixed finite candidates. \\
Bayesian optimization \citep{snoek2012,wang2023bo,eriksson2021,xie2024costaware} & Scalar objective feedback & Find optimum or high-value designs & Powerful when smooth surrogate assumptions hold; not identical to thresholded rare-regime first discovery. \\
Reliability active learning \citep{au2001,moustapha2021reliability} & Limit-state simulator and surrogate & Estimate failure probability or boundary & Shares rare-event motivation; QC-PHAST targets first discovery rather than probability estimation. \\
Continuation/bifurcation tools \citep{strogatz2015,liessi2025matcont} & Equations, derivatives, solution branch & Trace equilibria, periodic orbits, bifurcations & Best when a branch and differentiable model are available; outside the finite-pool black-box predicate setting. \\
Quantum variants \citep{aaronson2020approxcount,yoder2014fixedpoint,dalzell2025quantumalgorithms} & Quantum oracle access & Counting, fixed-point search, end-to-end quantum primitives & Useful references for future threshold-free, density-estimation, or lower-bound-known search extensions. \\
\bottomrule
\end{tabular}
\end{table*}

\paragraph{Dynamical-system benchmarks.}
FitzHugh-Nagumo models reduced excitable-neuron dynamics \citep{fitzhugh1961,nagumo1962}; Van der Pol dynamics are canonical relaxation oscillations \citep{vanderpol1926}; Duffing dynamics are a standard nonlinear oscillator benchmark in nonlinear-dynamics treatments \citep{strogatz2015}; and Lorenz flow is a canonical nonlinear chaotic system \citep{lorenz1963}. Latin hypercube and Sobol-style designs remain important space-filling candidate-generation baselines, while recent quasi-Monte Carlo software work emphasizes modern implementation, adaptive error control, and parallel evaluation support \citep{choi2023qmcsoftware,sorokin2023qmc}.

The broader scientific-machine-learning literature motivates why this problem is timely. Physics-informed and knowledge-guided machine learning aim to combine statistical learning with mechanistic structure rather than replacing models by unconstrained predictors \citep{karniadakis2021,willard2023}. In fluid mechanics and related dynamical systems, machine learning is often useful precisely because full mechanistic exploration is expensive and the relevant qualitative regimes may occupy restricted parts of parameter space \citep{brunton2020}. PDEBench and related benchmark efforts show the value of standardized simulation tasks for scientific ML, even though the present paper remains focused on rare-regime discovery rather than PDE-scale surrogate modeling \citep{takamoto2022pdebench}. Stiff ODE solvers and Radau-type methods provide the numerical background for the online stiff-chemistry stress test added here \citep{ekanathan2024radau}. QC-PHAST fits this scientific-computing pattern: it assumes a simulator, stability calculation, or score already exists, and asks how to find rare regimes in the resulting candidate space with fewer verified queries.

\section{Problem Formulation}

Let $\X=\{x_i\}_{i=1}^N\subset\R^d$ be a finite candidate set. Each $x_i$ is a parameter vector for a simulator or stability calculation. Let $s:\X\rightarrow\R_{\geq 0}$ be a criticality score, where smaller values indicate proximity to a qualitative transition. Given a threshold $\tau$, define the marked set
\begin{equation}
    \A_\tau=\{x_i\in\X:s(x_i)\leq \tau\},
    \qquad M=|\A_\tau|,\qquad p=M/N.
\end{equation}
The query objective is to find any $x\in\A_\tau$ while minimizing the number of evaluations of the predicate
\begin{equation}
    O_\tau(x)=\mathbf{1}\{s(x)\leq \tau\}.
\end{equation}

\paragraph{From dynamical systems to marked candidates.}
In the benchmark setting, a candidate $x$ is a parameter vector $\theta$ for a dynamical system
\begin{equation}
    \dot z = f(z;\theta),\qquad z\in\R^m,\qquad \theta\in\Theta\subset\R^d .
\end{equation}
The search is over parameter space, not over trajectory directions. Each $\theta$ induces a phase portrait, nullclines or equilibrium equations when they are available, and a local stability calculation. Let
\begin{equation}
    \Phi_\theta=\{f(\cdot;\theta),\,\mathcal E(\theta),\,J_\theta(z^\star),\,\lambda(J_\theta(z^\star))\}
\end{equation}
denote this induced phase-space object, where $\mathcal E(\theta)$ is the relevant set of equilibria and $J_\theta$ is the Jacobian at an equilibrium $z^\star$. The criticality score is therefore a composition
\begin{equation}
    s(\theta)=S(\Phi_\theta),
\end{equation}
where $S$ is a system-specific distance to a stability boundary, saddle-node boundary, conservative boundary, or Hopf-type boundary. For analytic FHN and Duffing scores, the relevant real equilibria are enumerated from the stated polynomial equations. For coupled FHN, by contrast, the operational equilibrium set is the finite set returned by three deterministic numerical root starts; that score is a stability proxy over found equilibria, not a proof that every coupled equilibrium has been enumerated. QC-PHAST then searches a finite candidate library $\X\subset\Theta$ for any parameter configuration whose induced phase portrait satisfies $s(\theta)\leq\tau$. In words, the objective is: find a parameter configuration that makes the model enter a rare critical regime, using as few verified score queries as possible.

The primary measurement is the first-hit query count
\begin{equation}
    Q_{\mathrm{hit}}=\min\{t\geq 1:O_\tau(x_t)=1\},
\end{equation}
where the returned candidate is counted only after simulator-score verification. The corresponding discovery record is not just a query number; it includes the returned score $s(x_{Q_{\mathrm{hit}}})$, the marked-set verification $O_\tau(x_{Q_{\mathrm{hit}}})=1$, empirical survival curves $\Pr(Q_{\mathrm{hit}}>b)$, budget-to-90\%-success values, and saved-score replay time. Thus a small query count is meaningful only when the run returns a verified marked candidate with high finite-budget success probability. If $c_s$ denotes the application-dependent cost of one simulator-score or oracle-predicate check, a deployment total must also charge any threshold-calibration pilot:
\begin{equation}
    T_{\mathrm{disc}}\approx (Q_{\mathrm{pilot}}+Q_{\mathrm{search}}+Q_{\mathrm{verification}})c_s+T_{\mathrm{policy}}.
\end{equation}
For a scientifically pre-specified threshold, $Q_{\mathrm{pilot}}=0$; for a pilot-estimated threshold it must be reported separately and included in the total. The main saved-score confidence sweep reports $Q_{\mathrm{hit}}$, survival/completeness diagnostics, and logged policy-replay time. Those replay seconds are not simulator wall-clock time or quantum-hardware runtime. Later online-simulator and QPU sanity diagnostics are explicitly labeled stress tests, not replacements for the primary query-count claim.

The phrase ``query efficient'' is therefore used in a constrained, measurable sense. For a search policy $\pi$, the evidence record is the tuple in \eqref{eq:evidence-record}:
\begin{equation}
    \label{eq:evidence-record}
    \mathcal R(\pi)=
    \left(
    \E[Q_{\mathrm{hit}}],
    \Pr_\pi(s(x_{\mathrm{hit}})\leq\tau),
    B_{0.9}(\pi),
    T_{\mathrm{policy}},
    T_{\mathrm{verify}}
    \right),
\end{equation}
where
\begin{equation}
    B_{0.9}(\pi)=\inf\{b:\Pr_\pi(Q_{\mathrm{hit}}\leq b)\geq 0.9\}.
\end{equation}
The first coordinate is the mean verification burden, the second is objective correctness, the third is a finite-budget completeness measure, and the last two record analysis time and final verification cost. A method is better in the sense claimed here only when it reaches the same thresholded scientific objective with fewer verified predicate checks and comparable or better finite-budget discovery probability. This is why the results section reports query means, objective-quality thresholds, survival curves, p90 budgets, replay time, and artifact logs rather than a single speedup number.

The finite-candidate formulation is deliberate. Continuous parameter spaces require discretization, candidate generation, or adaptive proposal mechanisms before a marked-set query model is well defined. Our claims apply to the generated candidate set and simulator-derived labels. A finite pool does not imply that every score must be evaluated before deployment: online policies may reveal scores lazily. Complete score banks are evaluation infrastructure used to replay many policies against identical latent landscapes.

\paragraph{Finite-pool scope versus continuous optimization.}
The finite candidate set is a modeling choice, not a claim that continuous optimizers are intrinsically weak. Many simulator workflows already operate through finite design libraries, batched parameter sweeps, archived simulation banks, or finite sets of feasible experimental conditions. In that setting the decision problem is which candidate to verify next. A continuous Bayesian optimizer that proposes new off-pool coordinates is solving a different problem: it changes the candidate set while searching. QC-PHAST therefore compares methods under a shared finite-pool access model. The results should not be read as a general defeat of continuous Bayesian optimization, adaptive mesh refinement, or continuation methods on smooth dynamical systems.

\paragraph{Query accounting.}
For exact non-quantum policies, one query means revealing one candidate's simulator-derived score and checking $O_\tau(x)$. In the final sweep the scores are precomputed and saved, but replay exposes them to each policy only in its query order; the counter represents how many online score evaluations would have been requested before the first verified hit. For BBHT, the counter includes each Grover-iteration marked-oracle call and one verification check for each measured observed mark. It does not include state preparation or a compiled oracle unless a later cost diagnostic adds them explicitly. For learned triage, model ranking is not counted as a substitute for the simulator: training labels and post-ranking verification checks are reported separately.

\paragraph{Why a thresholded score rather than pure optimization?}
Many simulator workflows are not asking for the global minimum of a scalar function. They ask whether the system has entered a scientifically meaningful regime: near a stability boundary, near an oscillatory onset, near a conservative limit, or near a bifurcation surface. A scalar criticality score is a convenient way to define such regimes, but the end task is first discovery under a threshold. This distinction matters. A global optimizer may spend many evaluations refining the very best score after it has already found a scientifically acceptable marked point. QC-PHAST instead measures the time to first verified hit.

\paragraph{Default thresholds and controlled density thresholds.}
The project uses three thresholding modes. The default and stress benchmarks use system-specific $\tau$ values chosen before policy evaluation to represent near-boundary regimes in each canonical system. These are concrete scientific benchmark instances. The final confidence sweep uses the complete replay score bank to set empirical quantiles and control $p=M/N$ directly. This is a controlled experimental instrument: it equalizes rarity across systems, sizes, and seeds, but it is not an online threshold-estimation procedure and gives every evaluated policy the same post-calibration labels. The pilot-threshold ablation estimates the same target quantile from 64, 256, or 1024 pilot scores and evaluates success against the intended full-pool threshold. Together the three modes separate scientific threshold definition, controlled density scaling, and threshold-estimation uncertainty.

\paragraph{Access models.}
It is helpful to distinguish five access tracks. In a full simulator-access setting, evaluating $s(x)$ may require running a numerical simulator or solving an equilibrium/stability problem, and unqueried scores remain unknown. In saved-score query-policy replay, an evaluator has already generated the latent score landscape so that many policies can be compared reproducibly, but each policy receives only the observations allowed by its declared access model. In a scalable score-guided finite-pool replay, cross-entropy and subset-style policies observe numerical scores while BBHT receives only the thresholded predicate. In the stronger GP scalar-score track, non-quantum methods fit finite-pool surrogates or active-search policies from numerical scores. In a quantum oracle setting, the binary predicate $O_\tau$ is assumed to be coherently available. The score bank is therefore not free side information given to BBHT or predicate-only policies; it is an offline evaluation device. These tracks are not interchangeable, and all claims are labeled by the access contract in Table~\ref{tab:access-contract}.

Table~\ref{tab:access-contract} defines the access-model contract used throughout the experiments, separating matched finite-pool comparisons from stronger scalar-score, continuous-structure, and hardware/runtime stress tests.
\begin{table*}[!htbp]
\centering
\caption{Evidence tracks and access-model contract. Each row defines what information is available to the included non-quantum policies and to the BBHT query reference, and therefore the only interpretation the comparison supports.}
\label{tab:access-contract}
\footnotesize
\renewcommand{\arraystretch}{1.22}
\setlength{\tabcolsep}{4pt}
\begin{tabular}{p{0.18\textwidth} p{0.30\textwidth} p{0.20\textwidth} p{0.24\textwidth}}
\toprule
Evidence track & Included non-quantum access & BBHT access & Permissible conclusion \\
\midrule

\shortstack[l]{Predicate-only\\finite pool} &
hit/miss labels from $O_\tau$ on fixed $\X$ &
binary marked oracle over the same $\X$ &
matched finite-pool query-count comparison \\

\shortstack[l]{Scalable score-guided\\finite pool} &
CE and subset-style policies observe numerical $s(x)$; coverage policies use no score feedback &
binary marked oracle over the same $\X$ &
875-configuration replay with scalable finite-pool policies \\

\shortstack[l]{GP scalar-score\\finite pool} &
numerical $s(x)$, GP active-search diagnostics, exact verification &
threshold predicate $O_\tau$ only &
stronger non-quantum access test \\

Continuous structure &
off-pool coordinates, analytic probes, adaptive mesh, continuation-style structure &
not an identical access model &
stress test and negative control \\

Hardware/runtime &
simulator wall time, score verification, policy overhead &
no compiled coherent oracle; break-even multiplier only &
query-accounting diagnostic only \\
\bottomrule
\end{tabular}
\end{table*}

\FloatBarrier

\section{Method: Evidence-Gated Search Selection and Grover/BBHT Query Reference}

\qcphast{} decouples scientific scoring from search-policy selection. First, each candidate parameter is mapped through a simulator or stability calculation to a criticality score. Second, a scientifically defined threshold, or a controlled benchmark calibration rule, turns this score into a verified rare-regime predicate. Third, the protocol specifies which search layer is admissible under the observed access model and pilot evidence. Classical policies may exploit coverage, score feedback, local elites, or GP uncertainty; the BBHT row receives only binary marked-oracle access and implements the unknown-$M$ Grover-search schedule. The method asks when to route to structure-aware non-quantum search, when to use finite-pool non-quantum active search, when the Grover/BBHT query reference is informative, and when noise or oracle cost should block the marked-oracle interpretation.

\subsection{Routing state, admissibility gates, and decision objective}

The search-selection state is an evidence vector
\begin{equation}
    c=\left(a,\widehat p,\widehat\eta_{\mathrm{fp}},\widehat\eta_{\mathrm{fn}},
    \widehat g,c_s,\widehat\mu,S_N,R_{\mathrm{reuse}}\right).
\end{equation}
Here $a$ records the available access type (predicate, scalar score, or continuous/equation structure); $\widehat p$ is the pilot marked-fraction estimate; $\widehat\eta_{\mathrm{fp}}$ and $\widehat\eta_{\mathrm{fn}}$ estimate false-positive and false-negative rates relative to exact verification; $\widehat g$ summarizes observed geometric coherence or fragmentation; $c_s$ is the measured score-verification cost; $\widehat\mu$ is the anticipated coherent-oracle cost in units of $c_s$; $S_N$ is the one-time state-preparation or loading cost in the same units; and $R_{\mathrm{reuse}}$ is the number of scientifically justified searches expected to reuse that preparation. These quantities need not all be known exactly. Their role is to expose what must be estimated before a method is declared admissible.

For each policy $\pi$ in a candidate portfolio $\Pi$, a pilot or replay study estimates the evidence record $\mathcal R(\pi)$ from Eq.~\eqref{eq:evidence-record}, augmented by an access label and censoring indicator. The framework then applies three hard gates. The \emph{structure gate} removes the marked-set layer when analytic inversion, root finding, continuation, adaptive mesh, or smooth continuous search already reaches a verified hit within the pilot budget. The \emph{fidelity gate} excludes a noisy marked predicate when its estimated true-positive purity is too low for reliable final verification; Proposition~\ref{prop:predicate-purity} makes this condition explicit. The \emph{resource gate} forbids an end-to-end quantum-runtime interpretation unless $S_N+\widehat\mu Q_B<Q_C$. These gates encode scientific validity before performance ranking.

Among the admissible policies $\Pi_{\mathrm{adm}}(c)$, the proposed operational choice is
\begin{equation}
    \pi^\star(c)\in
    \arg\min_{\pi\in\Pi_{\mathrm{adm}}(c)}
    \widehat C(\pi)
    \quad\text{subject to}\quad
    \widehat{\Pr}\{s(x_{\mathrm{hit}})\leq\tau\}\geq 1-\delta,
\end{equation}
where $\widehat C$ is either the p90 verified first-hit budget or the measured total-cost estimate $\widehat T_{\mathrm{disc}}$, chosen before comparison. This defines an evidence-gated decision rule, not a learned meta-selector. It uses pilot observations and access conditions while remaining transparent about why a policy was admitted, rejected, or retained only as a query-model reference. The paper does not claim a prospective regret evaluation for this rule.

Algorithm~\ref{alg:qcphast} gives the deployment workflow. It deliberately leaves unqueried scores unknown: complete score evaluation is legitimate only for the separate offline benchmark-construction protocol in Algorithm~\ref{alg:final-sweep}, not for an online discovery claim.

\begin{algorithm}[t]
\caption{\qcphastexpanded{} (\qcphast) online evidence-gated finite-pool search selection}
\label{alg:qcphast}
\begin{algorithmic}[1]
\Require Simulator score/verifier $s$, pre-specified threshold $\tau$ or pilot calibration rule $T$, candidate design $G$, calibration budget $B_{\mathrm{cal}}$, structure-probe budget $B_{\mathrm{struct}}$, policy portfolio $\Pi$, access mode $a$
\Ensure Routing decision and verified first-hit evidence tuple
\State Generate or load $\X=\{x_i\}_{i=1}^{N}\leftarrow G$
\If{$\tau$ is not scientifically pre-specified}
    \State Evaluate a disjoint pilot, charge every pilot score, and set $\tau\leftarrow T(\{s(x):x\in\X_{\mathrm{pilot}}\})$
\Else
    \State Record the external scientific basis for $\tau$
\EndIf
\State Run charged structure probes: analytic inversion, root finding, continuation, adaptive mesh, and smooth continuous search
\If{a probe finds a verified hit within $B_{\mathrm{struct}}$}
    \State Route to the structure-aware non-quantum method
    \State Record its evaluations and the finite-pool marked-set layer as unnecessary
    \State \Return routing decision and evidence tuple
\EndIf
\State Leave all nonpilot scores unknown; evaluate $s(x)$ lazily and cache each requested score
\State Estimate $\widehat p$, predicate error, geometry, score cost, and reuse assumptions from charged pilot evidence
\State Build routing state $c$ and apply structure, fidelity, and resource gates
\State Run admissible coverage, adaptive finite-pool, GP/active-search, and Grover/BBHT reference policies under their declared access models
\State Verify every claimed hit with the authoritative score $s(x)$
\State Report $Q_{\mathrm{pilot}}+Q_{\mathrm{search}}+Q_{\mathrm{verification}}$, p90 budgets, success, censoring, policy/verification time, and oracle break-even
\State Return the supported search layer and the evidence that admitted it
\end{algorithmic}
\end{algorithm}

\paragraph{Progressive phase-space discovery.}
The workflow is progressive in the sense used by scientific phase-space exploration: first define a finite representation of the parameter domain and an authoritative verifier, then reveal scores sequentially, update the search policy, and stop at the first verified threshold crossing. Saved-score replay emulates this sequence without rerunning the simulator; the policy sees only the information that its access model permits at each query, even though the evaluator holds the complete score bank. Classical baselines exploit geometry at different stages: low-discrepancy designs spread evaluations over parameter space, cross-entropy and subset-style methods adapt toward low-score regions, Bayesian and GP diagnostics model score structure, and learned triage ranks candidates before exact verification. QC-PHAST's marked-set layer is competitive only when the final set is sufficiently rare or fragmented that this exploitable geometry is limited. Smooth controls remain in the suite to identify when classical progressive search is already the supported route.

The phase-space view also explains why the benchmark examples are not interchangeable. FHN and Lorenz expose stability-boundary discovery; Duffing and Van der Pol test nonlinear oscillator thresholds; the pendulum mixes conservative and saddle-node boundary structure; spring-mass-damper is a smooth control case where adaptive classical samplers can compete. Table~\ref{tab:benchmark-suite} defines the score and role of each system, Table~\ref{tab:example-difficulty} quantifies the rare-search burden at the hardest density, Table~\ref{tab:system-rare} reports system-level query ratios, and Table~\ref{tab:geometry-ablation} separates dynamical-system details from marked-set shape.

We evaluate random search, Latin hypercube search, Sobol search, cross-entropy search, subset-style adaptive search, Bayesian lower-confidence-bound search on the default and stress runs, and Grover/BBHT-style marked-set query search. Learned triage models are evaluated as optional ranking surrogates only. They do not replace simulator verification.

\subsection{Search policies}

The compared policies are intentionally heterogeneous. They represent different assumptions about the candidate set rather than minor variants of one optimizer.

\paragraph{Random search.}
Random search samples a uniformly random permutation of the candidate set and queries until the first marked point. It is uncapped except by $N$ and therefore supplies the cleanest empirical reference for the exact $(N+1)/(M+1)$ first-hit expectation. This is the least structured baseline, but it is not a strawman: random search is known to be competitive when only a few dimensions matter or when model assumptions are unreliable \citep{bergstra2012}.

\paragraph{Latin hypercube and Sobol search.}
Latin hypercube and Sobol policies produce space-filling query sequences in the normalized candidate domain. Each proposed point is matched to the nearest unseen candidate. These baselines test whether low-discrepancy coverage alone can find rare regimes efficiently without using observed scores.

\paragraph{Cross-entropy search.}
The implemented cross-entropy policy maintains a diagonal Gaussian proposal in normalized parameter space. Each trial starts at mean $0.5$ and standard deviation $0.35$ in every coordinate, proposes batches of 24 points, and maps each proposal to its nearest unseen finite-pool candidate. After each batch, it retains the lowest-score $\max\{4,\lfloor n_{\mathrm{obs}}/5\rfloor\}$ observations, updates coordinate-wise means and standard deviations, and floors each standard deviation at $0.03$. Trials stop at the first verified hit or at 384 queries. This baseline is designed to be strong when low-score regions are geometrically coherent.

\paragraph{Subset-style search.}
The implemented subset-style policy begins with 32 uniformly sampled seed candidates. It retains the lowest-score $\max\{4,\lfloor n_{\mathrm{obs}}/4\rfloor\}$ observations, samples an elite center, adds an isotropic Gaussian perturbation with normalized standard deviation $0.08$, and queries the nearest unseen candidate. It stops at the first verified hit or at 384 queries. This is a simplified rare-event-inspired local proposal rule rather than a full probability-estimation implementation of subset simulation. It can exploit neighborhoods but may struggle when rare regions are fragmented or when early elites are misleading.

\paragraph{Bayesian LCB diagnostics.}
Bayesian lower-confidence-bound search fits a Gaussian-process surrogate to observed scores and queries candidates with low posterior lower confidence. The implementation uses 16 random initial observations, a fixed-amplitude Mat\'ern-$3/2$ kernel plus fixed $10^{-5}$ white noise, numerical jitter $10^{-6}$, an acquisition value $\mu-1.96\sigma$, batches of eight, a randomly subsampled acquisition pool capped at 1500 unseen candidates, a 192-query budget, and at most eight stochastic trials. Because repeated GP fitting is computationally heavier, this diagnostic is included in the default and stress benchmark tables but not in the original 875-configuration confidence sweep.

This finite-pool Bayesian LCB diagnostic is not intended to exhaust the Bayesian optimization literature. It asks how a GP-style acquisition rule behaves when restricted to the same finite candidate library as the other search policies. A fully continuous BO method with off-grid proposals, adaptive trust regions, constraints, and model-specific kernels may perform differently, especially on smooth low-dimensional systems. Such methods are better treated as continuous-optimization baselines, not as finite marked-set query policies.

\paragraph{Finite-pool GP active-search diagnostics.}
The access-model ablation adds three GP diagnostics that are closer to active search than to pure minimization. All use 16 random initial observations, the same fixed Mat\'ern-$3/2$ plus white-noise kernel, batches of eight, a 2500-candidate acquisition subsample, a 384-query budget, and eight stochastic trials per base configuration. The probability-of-hit rule maximizes $\Pr\{s(x)\leq\tau\}$ under the posterior. The level-set rule minimizes $|\mu(x)-\tau|-1.96\sigma(x)$, prioritizing candidates whose uncertainty intersects the threshold. Expected-threshold improvement maximizes $(\tau-\mu)\Phi(z)+\sigma\phi(z)$ with $z=(\tau-\mu)/\sigma$. These policies observe scalar scores, not only binary hits, and therefore represent a stronger non-quantum access model than a pure marked-predicate baseline. Their fewer stochastic trials are reported explicitly rather than silently equated with the 32-trial scalable methods.

\paragraph{Grover/BBHT-style marked-set query search.}
We simulate the \bbht{} unknown-$M$ variant of Grover marked-set search at the query-count level. For a candidate set with unknown marked count $M$, the policy samples Grover iteration counts from an expanding schedule and succeeds with the standard amplitude formula
\begin{equation}
\Pr(\mathrm{hit}\mid j)=\sin^2((2j+1)\theta),\qquad \sin^2\theta=M/N .
\end{equation}
Each attempted Grover schedule contributes $j$ marked-oracle calls plus one final verification check to the query total. In the exact-predicate simulation, a Bernoulli draw with the amplitude-amplification success probability represents whether measurement lands in the marked subspace; no statevector, circuit, or concrete candidate identity is materialized. The evaluator uses the true $M$ only to sample this success event, while the BBHT schedule that chooses $j$ is not given $M$. This row is therefore a binary marked-oracle query reference: it receives neither scalar scores, gradients, geometry, nor surrogate uncertainty.

\paragraph{Why these baselines are intentionally mixed.}
The baseline family spans three kinds of assumptions. Random, Latin hypercube, and Sobol search assume little or no score feedback and mostly test coverage. Cross-entropy and subset-style search assume that low-score regions have exploitable geometry and that early observations can guide later queries. Bayesian LCB and GP active-search diagnostics assume the score surface can be modeled by a surrogate with uncertainty estimates. BBHT assumes only a finite marked-set oracle, not smoothness. Because these assumptions are different, no single method should dominate all systems. A scientifically useful result should identify the regimes where each assumption is helpful.

\paragraph{Why the GP results are separated.}
Bayesian LCB is included in the default and stress experiments because it is an important classical comparator. The original 875-configuration confidence sweep used scalable adaptive baselines, cross-entropy and subset-style search, to keep the sweep focused on query-policy behavior rather than GP fitting cost. The later access-model ablation deliberately pays that cost for additional GP active-search diagnostics. This separation is useful: the main confidence sweep gives a large, balanced density trend, and the access-model ablation shows how the headline changes when stronger scalar-score classical methods and noisy-predicate stress tests are added.

\paragraph{Budget censoring and comparator selection.}
Cross-entropy, subset-style, and finite-pool GP trials that do not find a verified hit are assigned their 384-query cap; Bayesian LCB uses a 192-query cap. Their reported means are therefore budget-censored first-hit burdens, not conditional-on-success means. Assigning a failed classical trial the cap rather than an unknown larger time understates its uncensored burden and is conservative for an included non-quantum / BBHT ratio in which the classical count is the numerator. Random search and exact BBHT are capped only by $N$. For each configuration, the comparator is the minimum mean among the included non-quantum methods under the declared access model; aggregation is performed after this per-configuration selection. This distinction explains why the reported ratio is not generally equal to the ratio of two separately aggregated method means.

\section{Query-Model Theory}

\begin{assumption}[Finite marked-set oracle]
The search algorithm is given a finite candidate set $\X$ of size $N$ and oracle access to $O_\tau(x)=\mathbf{1}\{x\in\A_\tau\}$, where $|\A_\tau|=M>0$.
\end{assumption}

\begin{proposition}[Random search without replacement]
If candidates are queried in a uniformly random order without replacement, then the expected position of the first marked candidate is
\begin{equation}
    \E[T_{\mathrm{rand}}]=\frac{N+1}{M+1}.
\end{equation}
For rare marked sets, this scales as $\Theta(N/M)=\Theta(1/p)$.
\end{proposition}

\paragraph{Proof sketch.}
In a uniformly random permutation of $N$ candidates, the $M$ marked positions are an unordered sample without replacement from $\{1,\ldots,N\}$. The expected minimum of $M$ sampled positions is $(N+1)/(M+1)$ by the standard order-statistic identity for discrete samples without replacement. When $M\ll N$, this is proportional to $N/M$.

\begin{proposition}[Grover/BBHT query scaling]
Under the finite marked-set oracle and unknown $M$, the \bbhtfull{} schedule finds a marked item with expected query complexity $O(\sqrt{N/M})=O(1/\sqrt{p})$ up to constant factors \citep{boyer1998}.
\end{proposition}

\paragraph{Proof sketch.}
For known $M$, Grover search rotates the initial uniform state toward the marked subspace by angle $\theta=\arcsin\sqrt{M/N}$ and obtains constant success probability after $O(1/\theta)=O(\sqrt{N/M})$ oracle calls. BBHT removes the need to know $M$ by randomizing the Grover iteration count over a gradually expanding range. The expected query cost remains $O(\sqrt{N/M})$ up to the constant factors from the BBHT schedule.

\begin{proposition}[Oracle and state-preparation break-even]
\label{prop:oracle-break-even}
Let $Q_B$ be the measured BBHT oracle-call count and let $Q_C$ be the measured included non-quantum verified score-query count for the same finite candidate problem. Suppose one coherent marked-oracle call costs $\mu$ times one classical score verification, and suppose state preparation or coherent candidate loading contributes an additional cost $S_N$ per search instance. A quantum-query implementation can beat the classical verification count only if
\begin{equation}
    S_N+\mu Q_B < Q_C.
\end{equation}
Equivalently, the admissible per-oracle cost multiplier is
\begin{equation}
    \mu < \mu^\star=\frac{Q_C-S_N}{Q_B}.
\end{equation}
If no qRAM-like amortized loading is available and a single run pays $S_N=\Omega(N)$, then the finite-pool BBHT query advantage can be erased even when $Q_B\ll Q_C$.
\end{proposition}

\paragraph{Proof sketch.}
The claim is an accounting identity rather than a new quantum lower bound. The search layer uses $Q_B$ oracle calls, each carrying relative cost $\mu$, plus any one-time state-preparation or coherent-loading cost $S_N$. Parity with the classical verified-query baseline requires this total to be below $Q_C$. Without qRAM or an equivalent amortized state-preparation mechanism, loading a classical library into coherent access can cost order $N$; for single-instance rare-search tasks that cost can be as large as the full finite library and can dominate the square-root query term. Table~\ref{tab:oracle-break-even} reports $\mu^\star$ and the no-qRAM single-run penalty from the measured experiments.

\begin{proposition}[Codimension-to-density scaling]
\label{prop:codimension-density}
Let $\Theta\subset\R^d$ be a compact parameter domain with a smooth density bounded above and below near a regular transition set
\begin{equation}
    \mathcal B=\{\theta\in\Theta:g(\theta)=0\},\qquad g:\Theta\rightarrow\R^k,
\end{equation}
where $Dg$ has rank $k$ on $\mathcal B$. If the marked set is a small tube
\begin{equation}
    \A_\tau=\{\theta\in\Theta:\|g(\theta)\|\leq\tau\},
\end{equation}
then its probability mass scales as $p(\tau)=\Pr(\theta\in\A_\tau)=\Theta(\tau^k)$ for small $\tau$ away from domain-boundary degeneracies.
\end{proposition}

\paragraph{Proof sketch.}
The rank condition is the local transversality assumption: the parameter-to-transition map $g$ intersects the codimension-$k$ target $\{0\}\subset\R^k$ transversely. Standard parametric transversality makes such transverse intersections generic and stable under small smooth perturbations \citep{lee2013smoothmanifolds}. Near a regular codimension-$k$ transition set, the constant-rank theorem provides local coordinates $(u,v)\in\R^{d-k}\times\R^k$ in which the transition set is $v=0$ and the marked tube is $\|v\|\leq O(\tau)$. Integrating a bounded smooth density over the $k$ transverse directions gives a volume proportional to $\tau^k$ times the $(d-k)$-dimensional measure of the transition set. Thus codimension-one boundaries create rare bands with $p\propto\tau$, while codimension-two and codimension-three transition criteria shrink as $\tau^2$ and $\tau^3$. Proposition~\ref{prop:codimension-density} formalizes why multi-condition bifurcation discovery can naturally enter ultra-rare marked-fraction regimes. The present benchmarks use simpler score thresholds, but this scaling gives the physical reason to expect rare finite-pool discovery tasks in more constrained bifurcation searches.

\begin{proposition}[Finite-library coverage and conditional search burden]
\label{prop:finite-library-coverage}
Let $p_\tau=\Pr_{\theta\sim\rho}\{s(\theta)\leq\tau\}$, and draw a candidate library of size $N$ independently from $\rho$. Then
\begin{equation}
    \Pr(M=0)=(1-p_\tau)^N\leq \exp(-Np_\tau).
\end{equation}
Consequently, $N\geq p_\tau^{-1}\log(1/\delta)$ is sufficient for the library to contain at least one marked candidate with probability at least $1-\delta$. If the regular codimension-$k$ conditions of Proposition~\ref{prop:codimension-density} give $p_\tau=\Theta(\tau^k)$, then a sufficient high-probability library-size scale is
\begin{equation}
    N=\Theta\!\left(\tau^{-k}\log\frac{1}{\delta}\right)
\end{equation}
up to the constants in $p_\tau=\Theta(\tau^k)$ and for fixed target failure probability. When the realized fraction $M/N$ concentrates at $\Theta(p_\tau)$ and the declared access models hold, uniform random first-hit search has rare-event burden $\Theta(\tau^{-k})$, whereas the Grover/BBHT marked-oracle reference has burden $O(\tau^{-k/2})$ up to schedule constants.
\end{proposition}

\paragraph{Interpretation.}
The proposition separates two costs that must not be conflated. Candidate-library coverage determines whether the finite problem contains a discoverable regime at all; the first-hit query count begins only after that library has been defined. The square-root exponent therefore does not remove the potential $\Theta(\tau^{-k}\log(1/\delta))$ library-design or materialization burden when candidates must be explicitly generated or loaded. Procedurally reversible generation and amortized coherent access are separate implementation cases. The fixed-threshold experiment retains $M=0$ pools for precisely this reason, while the controlled density sweep conditions on $M>0$ to study search-policy scaling.

\begin{proposition}[Purity of a statically corrupted marked set]
\label{prop:predicate-purity}
Let the true marked fraction be $p$, let an unmarked candidate be mislabeled as marked with false-positive probability $\alpha$, and let a marked candidate be mislabeled as unmarked with false-negative probability $\beta$. For a fixed corrupted predicate, the expected observed marked fraction and the true-positive purity of its marked set are
\begin{equation}
    \widehat p_{\mathrm{obs}}=p(1-\beta)+(1-p)\alpha,
    \qquad
    \chi=\Pr\{O_\tau=1\mid \widehat O_\tau=1\}
    =\frac{p(1-\beta)}{p(1-\beta)+(1-p)\alpha}.
\end{equation}
If amplitude amplification targets the observed marked set and an exact verifier rejects false positives after measurement, the expected number of observed-marked measurements required for one true verified hit is $1/\chi$. The observed marked set has majority true positives only if
\begin{equation}
    \alpha < \frac{p(1-\beta)}{1-p},
\end{equation}
which reduces to the scale requirement $\alpha=O(p)$ in the rare-event limit.
\end{proposition}

\paragraph{Proof.}
The observed marked mass is the sum of retained true positives, $p(1-\beta)$, and false positives, $(1-p)\alpha$. Bayes' rule gives $\chi$. Starting from a uniform candidate superposition, Grover amplification of a fixed observed marked set preserves equal amplitudes within that subspace, so a measured observed mark passes exact verification with probability $\chi$. Independent restarts therefore require a geometric number of observed-marked measurements with expectation $1/\chi$. Solving $\chi>1/2$ gives the stated inequality. This proposition describes the static label-corruption model implemented in the dense noise phase diagram. It is not a theorem for stochastic nonunitary failure on every oracle call; those stronger faulty-oracle models are treated separately in the quantum-search literature \citep{regev2012faultyoracle,lolck2024faultyoracle}.

\begin{proposition}[Necessary pilot coverage for a rare quantile]
\label{prop:pilot-coverage}
Suppose a finite pool contains $M=pN$ candidates at or below the intended lower-tail threshold, and a pilot of size $n$ is drawn without replacement. If $K$ is the number of intended-tail candidates observed in the pilot, then
\begin{equation}
    \Pr(K=0)=\frac{\binom{N-M}{n}}{\binom{N}{n}}
    \approx (1-p)^n
\end{equation}
for $n\ll N$. A necessary condition for seeing at least one intended-tail candidate with probability at least $1-\delta$ is therefore
\begin{equation}
    n\gtrsim \frac{\log\delta}{\log(1-p)}
    \sim \frac{\log(1/\delta)}{p}.
\end{equation}
\end{proposition}

\paragraph{Interpretation.}
This is only a coverage condition, not a sufficient guarantee for accurate extreme-quantile estimation. It nevertheless explains why threshold calibration becomes unstable before search begins. At $p=0.001$, a pilot with $n=256$ contains no candidate from the intended lower $0.1\%$ tail with probability approximately $0.999^{256}=0.774$. Achieving even 90\% probability of one such observation requires roughly 2302 independent pilot samples under the large-pool approximation. The pilot-threshold ablation is therefore a stress test of a statistically data-starved calibration regime, not a substitute for a scientifically pre-specified threshold or a dedicated rare-tail estimator.

These propositions clarify the role of the final density sweep. If \qcphast{} is truly operating in a rare-regime search setting, the advantage over random search should grow as $p=M/N$ decreases. The empirical question is whether Grover/BBHT-style query counts also remain favorable against strong adaptive classical baselines.

The propositions themselves are not the empirical discovery of this paper. The square-root query separation is inherited from Grover/BBHT. The methodological contribution of \qcphast{} is the access-model study asking what remains of that reference advantage when the marked sets come from concrete simulator-derived score landscapes and when non-quantum baselines are allowed to exploit coverage, local elites, score feedback, surrogate uncertainty, and continuous structure. A final sweep that only compared BBHT against random search would largely restate the theorem. The nontrivial part of the benchmark is the comparison against geometry-aware policies across systems, densities, sizes, and seeds.

\subsection{What the theory does and does not predict}

The theory predicts a query-scaling separation between unstructured random search and BBHT under a finite marked oracle. It does not predict that BBHT must beat every structured classical heuristic on every finite benchmark. Cross-entropy, subset-style search, and Bayesian LCB use information that is absent from the unstructured marked-set model: observed scores, local geometry, smoothness, and surrogate uncertainty. If a marked region is large, smooth, and easy to model, a classical adaptive method can find it quickly. If a marked region is rare or poorly aligned with the adaptive method's assumptions, the square-root query scaling of BBHT becomes more visible.

This distinction is the reason the manuscript reports both random/BBHT and included non-quantum / BBHT ratios. The random ratio tests whether the rare-event trend matches the basic query-model intuition. The included non-quantum ratio tests whether the advantage survives a more practical scientific-computing comparison. A paper that reported only random-search improvement would overstate the result. A paper that ignored the random-search scaling would miss the cleanest connection to the query theory. \qcphast{} needs both.

\subsection{Pre-flight decision procedure}

Before using \qcphast{} on a new simulator, Algorithm~\ref{alg:qcphast} and Table~\ref{tab:regime-decision} should be applied as a gate sequence rather than as a post hoc explanation. First specify the scientific threshold and verifier. Next run cheap equation-aware probes and a pilot that is separate from final policy evaluation. If the pilot finds no marked candidate, report that $p$ is unresolved; Proposition~\ref{prop:pilot-coverage} shows that a small pilot can easily miss an ultra-rare tail. If labels are approximate, estimate false-positive and false-negative rates against exact verification and use Proposition~\ref{prop:predicate-purity} to test whether the observed marked set has adequate purity. Only after those checks should finite-pool policy comparisons be interpreted.

The final gate concerns cost rather than query count. Estimate the observed query ratio, the coherent-oracle cost multiplier, the one-time loading cost, and the number of legitimate reuses of the prepared library. Proposition~\ref{prop:oracle-break-even} then determines whether the evidence supports only a query-count statement or leaves any plausible total-cost headroom. The route is intentionally asymmetric: failure of a gate rejects the stronger interpretation, while passing a gate only permits the next analysis. It does not certify hardware advantage.

\begin{table*}[!htbp]
\centering
\caption{Pre-flight regime decision map. The recommended route is based on the finite-pool, geometry, access-model, noise, online-simulator, and break-even evidence in this manuscript.}
\label{tab:regime-decision}
\begingroup
\small
\setlength{\tabcolsep}{6pt}
\renewcommand{\arraystretch}{1.35}
\begin{tabular}{@{}p{0.23\textwidth}p{0.33\textwidth}p{0.36\textwidth}@{}}
\toprule
\textbf{Observed regime} & \textbf{Evidence to check} & \textbf{Recommended route} \\
\midrule
Smooth analytic or low-dimensional boundary & Closed-form score, monotone coordinate, continuation path, or analytic inversion succeeds in a few evaluations & Use analytic inversion, root finding, continuation, adaptive mesh, or continuous BO before QC-PHAST \\
Finite library, exact predicate, rare marked fraction & Pilot estimate $p\ll 1$, verified threshold labels, no obvious smooth boundary exploitable by classical baselines & Run finite-pool baselines and compare against the BBHT query reference \\
Fragmented, thin, boundary-like, or checkerboard marked set & Geometry ablation or pilot labels show disconnected, curved, or poorly aligned positives & QC-PHAST is a plausible query-efficiency layer if the predicate is verified \\
Noisy labels or surrogate-only positives & Estimated label error $\eta$ is comparable to or larger than $p$ & Do not use noisy labels as a marked oracle; add exact simulator verification or reject the quantum-query claim \\
Expensive or reversible predicate unclear & Oracle synthesis, state preparation, QRAM, or reversible simulator cost exceeds the break-even multiplier & Report query counts only; no hardware runtime speedup claim \\
Learned triage model available & Top-$k$ precision is useful but training labels are expensive & Use the model only to rank candidates, and count training plus verification labels \\
\bottomrule
\end{tabular}
\endgroup
\end{table*}

\subsection{Finite-size effects}

The asymptotic expressions $O(1/p)$ and $O(1/\sqrt{p})$ are not direct predictions of every observed mean query count. The experiments use finite candidate pools, finite marked counts, capped query budgets, stochastic trial counts, and adaptive methods whose behavior depends on candidate geometry. At the rarest fraction and smallest candidate pools, the marked count can be small. The final audit found a minimum marked count of two, which is valid but also means individual configurations can have high variance. For this reason, the main table aggregates 175 configurations per density and reports bootstrap confidence intervals.

\subsection{Noisy and uncertain predicates}

The clean BBHT proposition assumes an exact binary marked-set oracle. A practical simulator workflow may instead face numerical error, uncertain thresholds, surrogate labels, or stochastic simulator output. The predicate-robustness ablation therefore adds two diagnostics. In the noisy-predicate diagnostic, the observed predicate $\widehat O_\tau$ satisfies an empirical flip probability near $\eta$, so that $\Pr[\widehat O_\tau(x)\neq O_\tau(x)]\approx\eta$ before the search policy verifies candidates against the true saved label. In the threshold-uncertainty diagnostic, a pilot sample estimates $\widehat\tau$ before search, and final success is still judged against the intended target threshold.

These diagnostics are not a theorem for noisy Grover search. They are intentionally simpler and explicitly scoped: they ask whether the query-count story remains credible when the predicate seen by the search policy is imperfect. The answer is mixed. Threshold uncertainty weakens but does not eliminate the advantage in many settings. A 5\% noisy predicate can destroy the rare-density advantage because false marked items become common enough to dominate the observed oracle. This is exactly why the paper treats oracle fidelity as a first-order limitation rather than a footnote.

The relevant dimensionless quantity is the false-positive rate divided by the rare-event density. Proposition~\ref{prop:predicate-purity} shows why: under static false-positive rate $\alpha$ and false-negative rate $\beta$, the observed marked-set purity is $p(1-\beta)/[p(1-\beta)+(1-p)\alpha]$. When $\alpha\ll p$, most observed positives are true positives and verification adds a limited factor. When $\alpha$ becomes comparable to $p$, false targets become a first-order part of the observed oracle; when $\alpha$ is several times larger than $p$, amplitude amplification can concentrate probability on states that fail final verification. The dense phase diagram reports $\eta/p$ because its false-positive and symmetric models set $\alpha=\eta$. This static-corruption diagnostic is consistent with, but mathematically distinct from, stochastic faulty-oracle models in which the operation itself fails independently on repeated calls.

Threshold uncertainty has a separate source. Quantile calibration asks the pilot to resolve an extreme lower tail before the search starts. Proposition~\ref{prop:pilot-coverage} gives a necessary sample-size scale of order $1/p$ even to observe one intended-tail candidate with high probability. Consequently, the pilot-threshold rows should not be read as a minor implementation perturbation: at $p=0.001$, a 256-point pilot is usually too small to identify the intended tail reliably. A scientifically fixed threshold avoids this particular post hoc calibration problem, although it can still produce an unknown or even empty marked set in a finite candidate library.

\section{Dynamical-System Benchmarks}

The benchmark suite covers seven canonical systems grouped by scientific role.

FHN and coupled FHN are reduced excitable-dynamics benchmarks. They are not detailed neuron, synapse, tissue, or clinical disease models.

The suite is designed around diversity rather than biological or physical exhaustiveness. FHN and coupled FHN test excitable stability boundaries. Van der Pol tests a clean relaxation-oscillator slice. Duffing tests a nonlinear mechanical oscillator with equilibrium-dependent stability. Pendulum and spring-mass-damper test control-like boundaries, including an intentionally smooth case where adaptive classical sampling should do well. Lorenz tests a sensitive nonlinear system with classical stability thresholds. This mixture is useful because a query-efficiency claim should survive more than one geometry and should also expose cases where the claim weakens.

The score definitions are intentionally analytic or semi-analytic. This choice controls the benchmark so that search-policy behavior can be separated from simulator noise, numerical integration error, and uncertain physical calibration. It also limits the conclusion: the seven-system sweep is strongest as an access-model and geometry study, not as evidence that these canonical equations require an expensive black-box search in practice. The continuous-structure challenge deliberately exposes that limitation by solving several analytic controls almost immediately. A later application to a production biomedical, engineering, or climate simulator would need its own score validation, online-cost accounting, and structure-aware baselines.

Table~\ref{tab:benchmark-objectives-main} gives the compact search-policy view: searched coordinates, score construction, and why each example stresses a different route. Table~\ref{tab:objective-definitions} complements it with the state variables and transition conditions needed to interpret the score physically. The expanded reproducibility table, Table~\ref{tab:benchmark-suite}, preserves exact parameter bounds and source-library sizes in the appendix. These tables are intentionally not interchangeable. The common structure is that \qcphast{} does not search for a point in a plotted trajectory; it searches for a parameter vector whose induced vector field has a desired rare property.

\begin{table*}[!htbp]
\centering
\caption{Benchmark objectives used in the finite-pool rare-regime discovery protocol. Each row defines the searched parameter coordinates, the phase-space or stability quantity used to score a candidate, and the reason the example is informative for search selection.}
\label{tab:benchmark-objectives-main}
\small
\renewcommand{\arraystretch}{1.3}
\setlength{\tabcolsep}{5pt}
\begin{tabular}{p{0.14\textwidth} p{0.17\textwidth} p{0.35\textwidth} p{0.26\textwidth}}
\toprule
System & Searched parameters & Criticality score / rare predicate & Search-selection role \\
\midrule
FitzHugh-Nagumo & $(\epsilon,a,b,I)$ & $\min_{v^\star}|\max_j \operatorname{Re}\lambda_j(J_{\mathrm{FHN}})|$ near an equilibrium stability boundary. & Excitable-dynamics transition; representative phase-space bridge in Figure~\ref{fig:fhn-regimes}. \\
Coupled FHN & $(I_1,k)$ with fixed unit parameters & $\min_{z^\star\in\widehat{\mathcal E}(\theta)}|\max_j \operatorname{Re}\lambda_j(J_{\mathrm{cFHN}})|$, where $\widehat{\mathcal E}$ is returned by three fixed numerical root starts. & Coupled stability-proxy benchmark; not an exhaustive equilibrium enumeration. \\
Van der Pol & $(\mu,\omega)$ & $|\mu|$ for the simplified nonlinear-damping transition slice. & Smooth analytic control where structure-aware non-quantum search should be strong. \\
Duffing & $(\alpha,\beta,\gamma,d)$ & $\min_{x^\star}|\max_j \operatorname{Re}\lambda_j(J_{\mathrm{Duffing}})|$ across real cubic equilibria. & Nonlinear oscillator with moving equilibria and harder boundary geometry. \\
Windy pendulum & $(u,d,g)$ & $\min\{|\,|u|-g\,|,|d|\}$ for saddle-node or conservative-limit proximity. & Control-like nonlinear boundary with multiple ways to become rare. \\
Spring-mass-damper & $(k,c)$ & $\min\{|k|,|c|\}$ for near-zero stiffness or damping. & Smooth easy-regime negative control for the marked-set query layer. \\
Lorenz & $(\sigma,\beta,\rho)$ & $|\rho-\rho_H|$ near the nontrivial-equilibrium Hopf boundary. & Nonlinear stability-surface example with sensitive phase-space structure. \\
\bottomrule
\end{tabular}
\end{table*}

\begin{table*}[!htbp]
\centering
\caption{Dynamical-system search objectives. QC-PHAST searches parameter candidates, while each candidate induces a phase portrait, equilibria, and a criticality score.}
\label{tab:objective-definitions}
\footnotesize
\renewcommand{\arraystretch}{1.18}
\setlength{\tabcolsep}{3pt}
\begin{tabular}{@{}p{0.11\textwidth}p{0.10\textwidth}p{0.13\textwidth}p{0.20\textwidth}p{0.19\textwidth}p{0.20\textwidth}@{}}
\toprule
System & State & \shortstack[c]{Searched\\parameters} & \shortstack[c]{Criticality\\condition} & Score & \shortstack[c]{Role in\\difficulty ladder} \\
\midrule
FitzHugh-Nagumo & $(v,w)$ & $(\epsilon,a,b,I)$ & Equilibrium stability boundary; leading Jacobian real part near zero. & $\min_{v^\star}|\max_j \operatorname{Re}\lambda_j(J_{\mathrm{FHN}})|$ & Reduced excitable-dynamics benchmark; hard rare boundary in a four-dimensional parameter library. \\
Coupled FHN & $(v_1,w_1,v_2,w_2)$ & $(I_1,k)$ with fixed unit parameters & Four-state coupled equilibrium near a stability boundary. & $\min_{z^\star}|\max_j \operatorname{Re}\lambda_j(J_{\mathrm{cFHN}})|$ & Excitable interaction slice; tests whether the stability-search story survives coupling. \\
Van der Pol & $(x,\dot x)$ & $(\mu,\omega)$ & Zero nonlinear-damping slice in the simplified oscillator diagnostic. & $|\mu|$ & Clean relaxation-oscillator control with a simple analytic boundary. \\
Duffing & $(x,\dot x)$ & $(\alpha,\beta,\gamma,d)$ & Equilibrium-dependent linear stability boundary across real cubic roots. & $\min_{x^\star}|\max_j \operatorname{Re}\lambda_j(J_{\mathrm{Duffing}})|$ & Nonlinear mechanical oscillator; harder than smooth controls because equilibria depend on parameters. \\
Windy pendulum & $(\phi,\dot\phi)$ & $(u,d,g)$ & Torque-gravity saddle-node proximity or conservative zero-damping proximity. & $\min\{|\,|u|-g\,|,|d|\}$ & Control-like nonlinear boundary with both saddle-node and damping structure. \\
Spring-mass-damper & $(x,\dot x)$ & $(k,c)$ & Near-zero stiffness or damping. & $\min\{|k|,|c|\}$ & Smooth easy-regime control where adaptive classical search should compete. \\
Lorenz & $(x,y,z)$ & $(\sigma,\beta,\rho)$ & Near the nontrivial-equilibrium Hopf boundary $\rho_H$. & $|\rho-\rho_H|$ when $\sigma>\beta+1$ & Sensitive nonlinear stability-boundary benchmark. \\
\bottomrule
\end{tabular}
\end{table*}

The rarest-density objective-quality summary records the realized marked counts, score thresholds, and verified first-hit score quality for each system. It is included here to separate a small score from an unexplained binary label before the policy comparisons are interpreted.

\begin{table*}[!htbp]
\centering
\caption{Rarest-density objective quality at $M/N=0.001$. Scores are system-specific distances to the benchmark criticality condition, so values should be interpreted within each row rather than compared across systems. Query counts are first-hit simulator-score verifications from the final confidence sweep.}
\label{tab:objective-quality}
\footnotesize
\renewcommand{\arraystretch}{1.18}
\setlength{\tabcolsep}{3pt}
\begin{tabular}{@{}p{0.17\textwidth}c c c c c c c@{}}
\toprule
System & Configs & \shortstack[c]{Mean\\marked $M$} & \shortstack[c]{Mean\\$\tau$} & \shortstack[c]{Mean best\\score} & \shortstack[c]{Median marked\\score} & \shortstack[c]{Included non-quantum\\$Q$} & \shortstack[c]{BBHT\\$Q$} \\
\midrule
Coupled FHN & 25 & 14.80 & 3.33e-04 & 6.16e-05 & 1.98e-04 & 101.12 & 51.08 \\
Duffing & 25 & 31.80 & 0.025 & 0.020 & 0.025 & 119.69 & 52.98 \\
FitzHugh-Nagumo & 25 & 31.80 & 3.45e-04 & 6.57e-05 & 1.93e-04 & 262.99 & 50.77 \\
Lorenz Hopf Boundary & 25 & 31.80 & 0.013 & 9.44e-04 & 0.006 & 153.43 & 52.08 \\
Windy Pendulum & 25 & 7.40 & 2.06e-04 & 6.25e-05 & 1.36e-04 & 173.55 & 51.27 \\
Spring-Mass-Damper & 25 & 6.00 & 3.29e-04 & 6.16e-05 & 1.94e-04 & 65.57 & 50.92 \\
Van der Pol & 25 & 6.00 & 0.001 & 1.53e-04 & 6.85e-04 & 96.51 & 53.57 \\
\bottomrule
\end{tabular}
\end{table*}

\subsection{How the examples instantiate the search objective}

Each benchmark should be read through the same chain:
\begin{equation}
    \theta
    \longrightarrow
    \dot z=f(z;\theta)
    \longrightarrow
    \hbox{phase portrait/equilibria/Jacobian}
    \longrightarrow
    s(\theta)
    \longrightarrow
    O_\tau(\theta).
\end{equation}
The first object, $\theta$, is the searched parameter candidate. The middle objects are the dynamical-system interpretation induced by that candidate. The final object, $O_\tau$, is the rare-regime predicate queried by the search policy. This chain is the bridge between phase-space reasoning and the finite marked-set experiment. Figures~\ref{fig:fhn-objective-bridge} and~\ref{fig:duffing-objective-bridge} visualize two examples of this bridge, while Tables~\ref{tab:objective-definitions}, \ref{tab:objective-quality}, and~\ref{tab:example-difficulty} connect the mathematical objective to the query evidence.

FitzHugh--Nagumo is the main excitable-dynamics example. The state variables $(v,w)$ describe a voltage-like activation variable and a recovery variable. The four searched parameters $(\epsilon,a,b,I)$ change time-scale separation, nullcline geometry, and applied drive. For each candidate parameter vector, the $v$- and $w$-nullclines determine equilibria, and the Jacobian eigenvalues determine local stability. The marked candidates are not arbitrary low numerical scores; they are parameter settings whose induced phase portrait is close to a loss-of-stability boundary. In the generic finite-pool rare-density sweep, only about 32 candidates are marked on average across the size-seed configurations, and the per-configuration included non-quantum baseline needs 262.99 verified checks on average while BBHT needs 50.77 under the exact marked-predicate query model. That result profiles the generic policy portfolio, not the best equation-aware solver: the trace-zero construction in Table~\ref{tab:fixed-threshold-online} directly targets an FHN boundary and finds a fixed-threshold candidate in one evaluation.

Coupled FHN keeps the same excitable interpretation but adds interaction. The searched slice varies the drive of the first unit and the coupling strength. A candidate must first define a four-state coupled equilibrium, and only then can the stability score be evaluated. The target is still first discovery of a near-boundary parameter setting, but the boundary is warped by coupling rather than by a single isolated nullcline intersection. The improvement is smaller than single-unit FHN because the searched slice is only two-dimensional and more exploitable by adaptive samplers, but Table~\ref{tab:example-difficulty} still reports a 1.98 included non-quantum / BBHT ratio at the rarest density.

Duffing is the nonlinear mechanical oscillator example. The searched parameters $(\alpha,\beta,\gamma,d)$ change the potential well, tilt, nonlinearity, and damping. The rare predicate is not simply a threshold on one coordinate. It depends on the real roots of $\beta x^3+\alpha x-\gamma=0$ and on the leading real eigenvalue of the Jacobian at those roots. A generic classical adaptive sampler can exploit smooth regions of low score, but it must learn where the moving equilibria create near-critical behavior. Figure~\ref{fig:duffing-objective-bridge} therefore shows both parameter-space candidates and the induced potential/phase portrait. The rare-density result, 119.69 included non-quantum queries versus 52.98 BBHT queries, is evidence about generic finite-pool policies. It is not evidence that the equation is intrinsically hard: the direct saddle-node parameterization in Table~\ref{tab:fixed-threshold-online} finds a candidate in one evaluation, with a 0.99 strict local-certificate pass rate in Table~\ref{tab:transition-certificates}.

Lorenz supplies a sensitive nonlinear stability-boundary example. The score marks proximity to the Hopf boundary of the nontrivial equilibria, producing a stability-relevant surface in $(\sigma,\beta,\rho)$ space over the stated domain. The task is not to simulate chaos faster; it is to locate a finite-library parameter vector close to a known qualitative transition using fewer verified checks. The included non-quantum / BBHT ratio in Table~\ref{tab:example-difficulty} applies only to that finite-pool comparison. Once the analytic Hopf-surface formula is admitted as proposal structure, Table~\ref{tab:fixed-threshold-online} finds a verified point in one evaluation, correctly routing this case away from the marked-set layer.

The pendulum, Van der Pol, and spring examples act as calibration cases. The pendulum target combines torque-gravity saddle-node proximity with the conservative zero-damping boundary, so the marked set is a union of physically meaningful conditions. Van der Pol uses the clean $|\mu|$ zero-damping diagnostic, making it easier for geometry-aware methods than FHN. Spring-mass-damper is the smoothest control case: the score $\min\{|k|,|c|\}$ directly exposes near-zero stiffness or damping in two dimensions. Its 1.29 rare-density ratio and 0.88 BBHT win rate quantify a weak finite-pool margin, while the continuous analytic challenge finds the target in one to two evaluations. The supported routing decision is therefore classical structure-aware search.

This example ladder is central to the interpretation. \qcphast{} is not claiming that a quantum query reference defeats every classical scientific method. It is showing that once a physical transition is encoded as a verified finite-pool predicate, the query burden depends strongly on rarity and geometry. The method is most useful when the rare set is small, boundary-like, fragmented, or only implicitly defined through equilibria and stability calculations. It is less necessary when the target is a smooth low-dimensional surface that classical adaptive search can exploit directly.

\paragraph{Energy and Hamiltonian structure.}
Several examples have natural energy interpretations in conservative limits, including the spring, pendulum, and undamped Duffing oscillator. Table~\ref{tab:energy-structure} records those conservative-limit Hamiltonians and the associated dissipative terms. Other examples, including FHN, Van der Pol, and Lorenz, are dissipative systems. For those systems, forcing an ordinary Hamiltonian form would be misleading. The evidence in this paper therefore uses verified phase-space, equilibrium, and Jacobian/eigenvalue score maps. A port-Hamiltonian or energy-based reformulation would be useful for future structured-oracle design only when the corresponding formulation is derived and audited for the specific simulator \citep{desai2021porthamiltonian}.

\begin{table*}[!htbp]
\centering
\caption{Energy and Hamiltonian structure used for interpretation. Conservative Hamiltonians are written only where they naturally apply; dissipative benchmarks use stability-score maps instead of forced Hamiltonian forms.}
\label{tab:energy-structure}
\footnotesize
\renewcommand{\arraystretch}{1.18}
\setlength{\tabcolsep}{3pt}
\begin{tabular}{@{}p{0.15\textwidth}p{0.23\textwidth}p{0.24\textwidth}p{0.30\textwidth}@{}}
\toprule
Benchmark & Representative dynamics & Conservative energy/Hamiltonian & QC-PHAST relevance \\
\midrule
Spring-mass-damper & $\dot x=p,\; \dot p=-kx-cp$ & $H(x,p)=\frac{1}{2}p^2+\frac{1}{2}kx^2$ & The benchmark score $\min\{|k|,|c|\}$ marks near-zero stiffness or damping; damping gives $\dot H=-cp^2$ for $c\ge0$. \\
Windy pendulum & $\dot\phi=p,\; \dot p=-g\sin\phi-dp+u$ & $H(\phi,p)=\frac{1}{2}p^2+g(1-\cos\phi)$ for $d=u=0$ & The score $\min\{|\,|u|-g\,|,|d|\}$ marks torque-gravity saddle-node proximity or the conservative limit. \\
Duffing & $\dot x=y,\; \dot y=-dy-\alpha x-\beta x^3+\gamma$ & $H(x,y)=\frac{1}{2}y^2+\frac{1}{2}\alpha x^2+\frac{1}{4}\beta x^4-\gamma x$ for $d=0$ & The score uses the leading real eigenvalue at real equilibria; damping gives $\dot H=-dy^2$. \\
FHN, Van der Pol, Lorenz & Dissipative vector fields & Ordinary conservative Hamiltonian not used & The manuscript uses phase-space, equilibrium, and Jacobian/eigenvalue score maps rather than forcing a false Hamiltonian form. \\
\bottomrule
\end{tabular}
\end{table*}

\paragraph{Excitable and relaxation dynamics.}
FitzHugh-Nagumo (FHN) and coupled FHN test excitable stability boundaries. For FHN, equilibria are computed from the reduced model and the score is the smallest absolute leading real Jacobian eigenvalue across real equilibria. Coupled FHN solves a four-dimensional equilibrium system over a two-parameter slice and uses the same stability-boundary score. Van der Pol uses a relaxation-oscillator boundary score.

\paragraph{Nonlinear mechanics and control.}
Duffing evaluates a nonlinear oscillator stability boundary over stiffness, nonlinearity, forcing, and damping parameters. The pendulum benchmark marks proximity to saddle-node and conservative boundaries. The spring-mass-damper benchmark is a smooth control-like system and acts as a useful easy-regime control.

\paragraph{Chaotic and stability-boundary dynamics.}
Lorenz marks proximity to the Hopf boundary of the nontrivial equilibria. It tests whether the search story persists in a sensitive nonlinear system while retaining an analytic classical control.

\subsection{Criticality scores}

The score functions are chosen to be interpretable and cheap enough for repeated finite-candidate studies. They are not meant to be exhaustive physical simulators; they are canonical criticality tests.

\begin{figure*}[!htbp]
\centering
\includegraphics[width=0.94\textwidth]{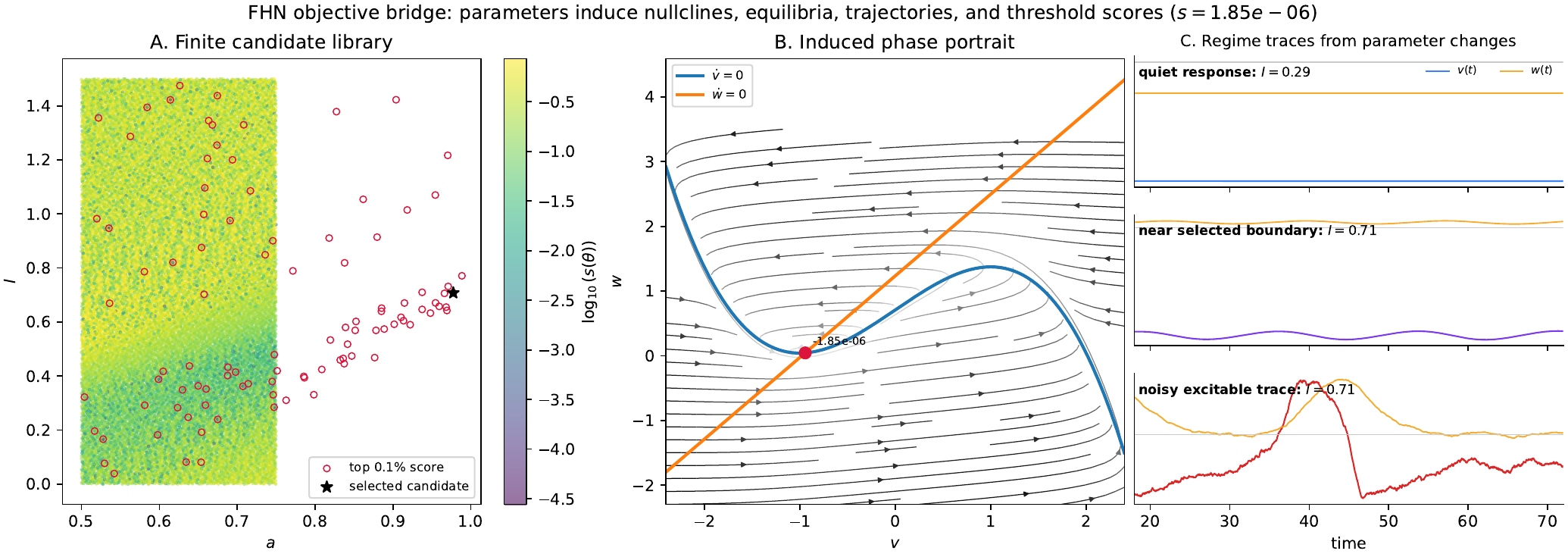}
\caption{FHN objective bridge generated from the saved stress dataset. Left: QC-PHAST searches parameter candidates; red outlines mark the top $0.1\%$ score set under the FHN stability-boundary score. Right: the selected parameter vector induces a phase portrait with $v$- and $w$-nullclines, an equilibrium, and a near-zero leading real Jacobian eigenvalue. The figure illustrates the mapping $\theta\mapsto\Phi_\theta\mapsto s(\theta)\mapsto O_\tau(\theta)$ used throughout the benchmark.}
\label{fig:fhn-objective-bridge}
\end{figure*}

\paragraph{FitzHugh-Nagumo.}
The single-unit FHN benchmark uses the reduced excitable-dynamics system \citep{fitzhugh1961,nagumo1962}
\begin{align}
    \dot v &= v-\frac{v^3}{3}-w+I,\\
    \dot w &= \epsilon(v+a-bw).
\end{align}
Here $v$ is the fast voltage-like activation variable, $w$ is the slow recovery variable, $I$ is an external drive, and $(\epsilon,a,b)$ control time-scale separation and nullcline geometry. The current experiments search the four-dimensional parameter vector $\theta=(\epsilon,a,b,I)$ over the bounds in Table~\ref{tab:benchmark-suite}. For a fixed $\theta$, the $v$-nullcline and $w$-nullcline are
\begin{equation}
    w=v-\frac{v^3}{3}+I,\qquad w=\frac{v+a}{b}.
\end{equation}
Their intersections define equilibria. Eliminating $w$ gives the cubic
\begin{equation}
    -\frac{1}{3}v^3+\left(1-\frac{1}{b}\right)v+I-\frac{a}{b}=0.
\end{equation}
For each real equilibrium $v^\star$, the Jacobian
\begin{equation}
J_{\mathrm{FHN}}=
\begin{bmatrix}
1-(v^\star)^2 & -1\\
\epsilon & -\epsilon b
\end{bmatrix}
\end{equation}
is evaluated. In this two-dimensional local linearization, the trace and determinant are
\begin{equation}
    T_{\mathrm{FHN}}=1-(v^\star)^2-\epsilon b,\qquad
    D_{\mathrm{FHN}}=\epsilon\left[1-b\left(1-(v^\star)^2\right)\right].
\end{equation}
A Hopf-type local stability boundary is indicated by $T_{\mathrm{FHN}}\approx 0$ with $D_{\mathrm{FHN}}>0$, while $D_{\mathrm{FHN}}\approx 0$ indicates proximity to a zero-eigenvalue boundary \citep{strogatz2015}. The implemented benchmark score is the direct eigenvalue-distance version,
\begin{equation}
s_{\mathrm{FHN}}(\theta)=
\min_{v^\star}\left|\max_j \operatorname{Re}\lambda_j(J_{\mathrm{FHN}})\right|.
\end{equation}
Small values mark proximity to a local stability transition. This score is not a complete bifurcation classifier; it is the verified criticality predicate used for finite-pool first discovery. In particular, the benchmark does not implement a full codimension-two Bogdanov--Takens detector; such a target would require a stricter condition such as simultaneous trace and determinant proximity near a double-zero eigenvalue. Figure~\ref{fig:fhn-objective-bridge} shows the corresponding parameter-space and phase-plane view for a data-derived rare FHN candidate.

\paragraph{Coupled FHN.}
The coupled benchmark uses two interacting FHN units over a two-parameter slice $(I_1,k)$:
\begin{align}
0 &= v_1-\frac{v_1^3}{3}-w_1+I_1+k(v_2-v_1),\\
0 &= \epsilon_1(v_1+a_1-b_1w_1),\\
0 &= v_2-\frac{v_2^3}{3}-w_2+I_2+k(v_1-v_2),\\
0 &= \epsilon_2(v_2+a_2-b_2w_2).
\end{align}
The experiments fix $\epsilon_1=\epsilon_2=0.1$, $a_1=a_2=0.7$, $b_1=b_2=0.8$, and $I_2=0.5$, then vary $I_1$ and coupling $k$. For each solved equilibrium, QC-PHAST evaluates the four-dimensional Jacobian and uses the same stability-boundary distance
\begin{equation}
s_{\mathrm{cFHN}}(\theta)=
\min_{z^\star}\left|\max_j \operatorname{Re}\lambda_j(J_{\mathrm{cFHN}}(z^\star))\right|.
\end{equation}

\paragraph{Van der Pol.}
The Van der Pol benchmark varies $(\mu,\omega)$ and uses the reduced score
\begin{equation}
s_{\mathrm{VDP}}(\mu,\omega)=|\mu|.
\end{equation}
The frequency parameter $\omega$ remains part of the candidate design, while the marked boundary is the zero-damping/nonlinearity slice $\mu=0$ in this simplified relaxation-oscillator diagnostic.

\begin{figure*}[!htbp]
\centering
\includegraphics[width=\textwidth]{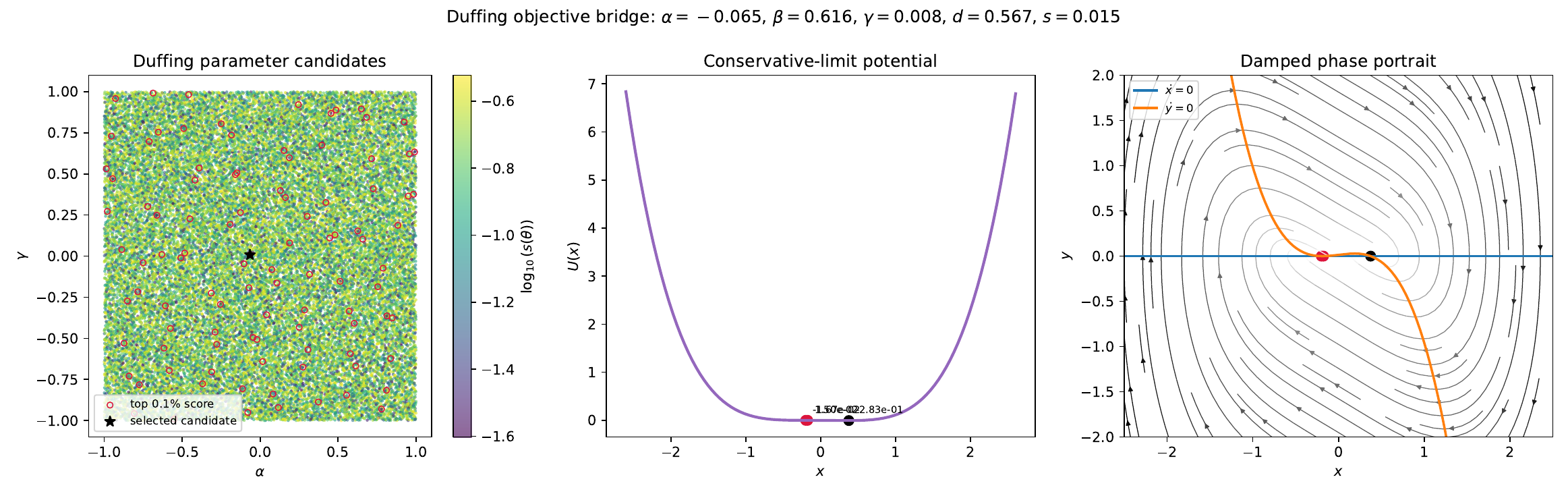}
\caption{Duffing objective bridge generated from the saved stress dataset. Left: parameter candidates projected onto $(\alpha,\gamma)$ with the top $0.1\%$ score set highlighted. Middle: conservative-limit potential for the selected parameter vector, with real equilibria annotated by leading real Jacobian eigenvalue. Right: damped phase portrait and nullclines. This second worked example shows the same QC-PHAST map as FHN: parameter candidate $\theta$, induced phase-space structure, stability score, and marked rare-regime predicate.}
\label{fig:duffing-objective-bridge}
\end{figure*}

\paragraph{Duffing.}
The Duffing benchmark uses the damped, tilted oscillator
\begin{align}
    \dot x &= y,\\
    \dot y &= -dy-\alpha x-\beta x^3+\gamma.
\end{align}
The searched parameter vector is $\theta=(\alpha,\beta,\gamma,d)$. In the conservative limit $d=0$, the Hamiltonian is
\begin{equation}
    H(x,y)=\frac{1}{2}y^2+\frac{1}{2}\alpha x^2+\frac{1}{4}\beta x^4-\gamma x,
\end{equation}
and for $d>0$ the energy derivative is $\dot H=-dy^2$. Thus damping makes the system dissipative while preserving a clear potential-energy interpretation for the equilibrium structure. Equilibria are roots of $\beta x^3+\alpha x-\gamma=0$. For each real equilibrium $x^\star$, the linearized Jacobian is
\begin{equation}
J_{\mathrm{Duffing}}=
\begin{bmatrix}
0 & 1\\
-(\alpha+3\beta (x^\star)^2) & -d
\end{bmatrix}.
\end{equation}
The score is the minimum absolute leading real eigenvalue across real equilibria,
\begin{equation}
    s_{\mathrm{Duffing}}(\theta)=
    \min_{x^\star}\left|\max_j\operatorname{Re}\lambda_j(J_{\mathrm{Duffing}}(x^\star))\right|.
\end{equation}
This is harder than the spring control because the equilibria themselves move nonlinearly with $(\alpha,\beta,\gamma)$; the target is not a fixed coordinate plane. Figure~\ref{fig:duffing-objective-bridge} shows the corresponding parameter map, conservative-limit potential, and damped phase portrait for the rarest saved-score candidate. The generated analysis recomputed the Duffing stability scores for all 100,000 stress candidates in parallel and matched the saved scores to floating-point precision.

\paragraph{Pendulum and spring controls.}
For torque $u$, damping $d$, and gravity $g$, the pendulum score is
\begin{equation}
s_{\mathrm{pend}}(u,d,g)=\min\{|\,|u|-g\,|, |d|\}.
\end{equation}
The first term marks proximity to a torque-gravity saddle-node boundary, and the second marks the conservative zero-damping boundary. For spring stiffness $k$ and damping $c$, the spring-mass-damper score is
\begin{equation}
s_{\mathrm{spring}}(k,c)=\min\{|k|, |c|\}.
\end{equation}
These systems are included because simple control-like boundaries should be favorable to adaptive classical methods.

\paragraph{Lorenz.}
The Lorenz benchmark scores proximity to the Hopf boundary
\begin{equation}
\rho_H=\frac{\sigma(\sigma+\beta+3)}{\sigma-\beta-1},
\end{equation}
when the denominator is valid. The score is
\begin{equation}
s_{\mathrm{Lorenz}}(\sigma,\beta,\rho)=
|\rho-\rho_H|.
\end{equation}
This gives a compact stability-boundary test in the Lorenz parameter space.

\FloatBarrier

\section{Experimental Protocol}

The evidence is organized into tracks because predicate-only replay, scalar-score active search, continuous equation-aware search, and hardware-resource accounting are different experiments. Table~\ref{tab:experiment-matrix} states the access and permissible conclusion for each track before the detailed implementation choices are given.

\begin{table*}[!htbp]
\centering
\caption{Evidence-track matrix. Each track answers a different question and uses only the access shown in its row; configuration counts are not pooled across tracks to manufacture a single estimator. The detailed protocol paragraphs below preserve budgets, trials, candidate sizes, and implementation choices.}
\label{tab:experiment-matrix}
\small
\renewcommand{\arraystretch}{1.25}
\setlength{\tabcolsep}{4pt}
\begin{tabular}{@{}p{0.16\textwidth}p{0.18\textwidth}p{0.24\textwidth}p{0.34\textwidth}@{}}
\toprule
Evidence track & Scale & Information and calibration & Scientific role and permissible conclusion \\
\midrule
Controlled density replay & 875 configurations; 32 scalable-policy trials & Saved scores construct a fixed hidden predicate at five controlled fractions; scalable non-quantum policies receive only their declared sequential feedback, while BBHT receives the binary marked predicate & Estimates conditional first-hit scaling after a nonempty finite instance is frozen; not online threshold discovery or simulator runtime. \\

Scalar-score access & Same 875 base configurations; 8 GP trials & GP policies observe numerical scores; BBHT remains a binary-oracle reference & Tests how much stronger finite-pool score access closes the gap to a geometry-agnostic reference. \\

Fixed-threshold replay & 2,975 configurations & System threshold fixed before policy exposure; $M=0$ pools retained & Fixed-objective candidate-library test and explicit no-target accounting; still saved-score replay rather than online runtime. \\

Pilot calibration & 1,750 configurations; $B=256$ & Disjoint exact-score pilot is charged before predicate-only search & Measures calibration overhead; reported separately from the scalar-score GP policy portfolio. \\

Continuous routing & 1,575 configurations & Off-pool proposals and equation-level structure are allowed & Determines whether finite-pool search should be bypassed; not an identical-access BBHT comparison. \\

Predicate noise & 875 base configurations across three noise models and eight rates & Static corrupted labels followed by exact post-measurement verification & Locates the $\eta/p$ fidelity boundary; not a per-call faulty-unitary theorem. \\

Geometry and dimension & Controlled synthetic shapes plus 4D--32D embedded FHN & Density is controlled while shape or nuisance representation changes & Diagnoses which target geometries help score-guided policies; does not establish intrinsic geometry of projected physical sets. \\

Online and resource sanity & Two propose--evaluate--update simulators plus break-even accounting & Classical simulator wall time is measured; quantum time remains a cost proxy without a compiled oracle & Tests workflow realism and the maximum oracle/state-preparation headroom; no hardware-speedup conclusion. \\
\bottomrule
\end{tabular}
\end{table*}

The main evidence uses the following experimental layers.

\paragraph{Default benchmark.}
Seven systems are evaluated with system-specific thresholds and baselines including grid, random, Latin hypercube, Sobol, Bayesian LCB, cross-entropy, subset-style search, and BBHT.

\paragraph{Stress benchmark.}
FHN, coupled FHN, Duffing, and Lorenz are scaled to larger candidate pools up to 100,000 candidates.

\paragraph{Density and dimension scaling.}
Density scaling controls the marked fraction at \\ $p\in\{0.1,0.03,0.01,0.003,0.001\}$. Dimension scaling embeds FHN from 4D to 32D at fixed rare density. The dimension result is diagnostic only; it is not a claim about arbitrary high-dimensional physics.

\paragraph{Final confidence sweep.}
The final sweep uses seven systems, five target marked fractions, five candidate-pool size levels, and five seeds, for 875 configurations. Each scalable stochastic search method uses 32 trials per configuration. Cross-entropy and subset-style trials are capped at 384 verified queries and failures contribute the cap; this censoring is conservative for the reported included non-quantum / BBHT ratio. The run used 72 workers and saved config, result, log, and manifest files.

\paragraph{Access-model ablation.}
The access-model run reuses the 875-base-configuration design and evaluates exact scalar-score access, a 5\% noisy-predicate stress test, and a pilot-threshold-uncertainty setting. It also adds finite-pool GP active-search, GP level-set, and expected-threshold-improvement diagnostics. The regenerated run used 72 workers, 32 trials for scalable stochastic methods, and 8 GP trials per base configuration.

\paragraph{Fixed-threshold and calibration accounting controls.}
The expanded fixed-threshold replay uses the default benchmark thresholds before policy exposure to the sampled pool and retains no-target pools rather than adding positives. It contains 2,975 configurations: seven systems, five size levels, and 85 finite-pool resampling seeds. It pairs generic finite-pool policies with direct equation-aware constructions only as routing controls. A separate 1,750-configuration pilot-calibration ledger uses a disjoint $B=256$ score pilot and charges every calibration query before search; its restricted predicate-only portfolio is reported separately from the stronger scalar-score GP access track.

\paragraph{Hierarchical and simulator validation.}
Paired hierarchical resampling nests systems, finite-pool resampling seeds, and candidate sizes to avoid treating related configurations as independent. The saved source score bank is fixed within each system, so this hierarchy does not represent independent simulator-source landscapes. A 500-case amplitude-subspace validation compares the BBHT probability simulator with the exact two-dimensional marked/unmarked recurrence before the query counts are used in the manuscript.

\paragraph{Dense noise phase diagram.}
The full-capacity noise phase-diagram run uses 875 base configurations and expands the single 5\% noisy-predicate stress test to $\eta\in\{0,10^{-4},3\cdot10^{-4},10^{-3},3\cdot10^{-3},10^{-2},3\cdot10^{-2},5\cdot10^{-2}\}$ under false-positive, false-negative, and symmetric noise models. The printed rare-density table focuses on false-positive and symmetric noise because these models introduce false marked targets that directly corrupt amplitude amplification; the full false-negative rows remain in the reproducibility artifact. The diagnostic reports performance as a function of $\eta/p$, the noise rate divided by the true marked fraction.

\paragraph{Geometry and oracle-resource ablations.}
The geometry ablation replaces system-specific score landscapes by controlled synthetic marked-set geometries: smooth balls, islands, boundary bands, curved manifolds, active subspaces, checkerboards, and noisy boundaries. This separates density from shape. The oracle-resource ablation is an illustrative fixed-point arithmetic proxy for simple analytic threshold predicates; it is neither a compiled circuit estimate nor a BBHT oracle-query count.

\paragraph{Oracle break-even analysis.}
The break-even run converts each query-count ratio into a maximum allowable quantum-oracle cost multiplier. It also adds a no-qRAM single-run state-preparation penalty and an amortized-run count needed before a one-time $O(N)$ candidate-loading cost stops dominating the BBHT query term.

\paragraph{Continuous and structure-aware classical challenge.}
The full-capacity continuous challenge contains 1,575 configurations: seven systems, five target fractions, three candidate-size levels, and 15 resampling seeds. It allows classical methods to leave the finite pool or use analytic structure when the score permits it. It includes analytic probes for the smooth systems, differential evolution, adaptive mesh, a rank-Gaussian proposal rule, and continuous GP-LCB diagnostics under a 128-evaluation budget. This is intentionally not the same query model as BBHT; it is a stress test for the criticism that finite-pool evaluation can handicap continuous optimizers.

\paragraph{Online simulator challenge.}
The online challenge adds two true propose-evaluate-update loops: a stiff Robertson-type chemical-kinetics ODE solved with an implicit stiff solver and a compact two-dimensional Kolmogorov-flow transition proxy. Classical methods pay actual simulator wall time. The BBHT row is reported only as a query-count time proxy obtained by multiplying measured simulator-evaluation time by BBHT query count.

\paragraph{Learned-triage accounting.}
The learned-oracle diagnostic is rerun with explicit accounting for training, validation, test, model-selection, and ranking-verification labels. Learned models are evaluated as candidate-ranking tools only; their training labels are counted as simulator-derived labels rather than free information.

The target fractions are $p\in\{0.1,0.03,0.01,0.003,0.001\}$ and the seeds are 20260801 through 20260805. Candidate-size levels are source-limited by system: FHN, Duffing, and Lorenz use $N\in\{2048,5414,14311,37830,100000\}$; coupled FHN uses $N\in\{2048,4305,9051,19027,40000\}$; pendulum uses $N\in\{2048,3424,5724,9570,16000\}$; and Van der Pol and spring use \\ $N\in\{2048,3186,4957,7713,12000\}$.

\begin{algorithm}[t]
\caption{Offline controlled confidence-sweep construction for query-policy replay}
\label{alg:final-sweep}
\begin{algorithmic}[1]
\Require Source candidate sets and their offline simulator-derived score banks
\Require Systems $S$, target fractions $P$, candidate-size levels $L$, seeds $R$, trials $K$
\For{$s\in S$}
  \For{$N\in L_s$}
    \For{$p\in P$}
      \For{$r\in R$}
        \State Sample a size-$N$ candidate pool from the saved source dataset using seed $r$
        \State Set $\tau$ to the empirical score quantile of the already generated pool score bank that targets marked fraction $p$
        \State Freeze $O_\tau$ and record actual $M/N$; this controlled benchmark excludes $M=0$ rows by design
        \For{each search method $\pi$}
          \State Hide unqueried labels from $\pi$ and run $K=32$ stochastic trials or deterministic query sequences
          \State Store first-hit query counts and method metadata
        \EndFor
      \EndFor
    \EndFor
  \EndFor
\EndFor
\State Aggregate by target fraction, system, and included non-quantum comparator
\end{algorithmic}
\end{algorithm}

Algorithm~\ref{alg:final-sweep} is important because it prevents a common ambiguity. The final sweep is not seven isolated demonstrations. It is a Cartesian evaluation design: seven systems, five target densities, five size levels, five seeds, and 32 stochastic trials where applicable. The analysis audit confirms 875 configurations and no zero-positive rows. The access-model ablation keeps the same base design so that stress tests can be compared against the original headline without changing the underlying system-density-size-seed population. This construction gives the paper enough replication to discuss density effects rather than relying on a single favorable system or a single favorable threshold.

\subsection{Aggregation and confidence intervals}

For every final-sweep row, each method returns an observed first-hit query count or its declared censoring cap over 32 trials. The analysis stores the mean, standard deviation, median, minimum, maximum, and raw query list. The included non-quantum comparator is selected separately within each configuration from the available method means; only then are ratios grouped by target marked fraction or by system and fraction. Table~\ref{tab:final-by-fraction} reports ordinary configuration-level bootstrap 95\% intervals. Table~\ref{tab:hierarchical-density} additionally performs paired hierarchical resampling by system, finite-pool resampling seed, and candidate size; it is the conservative uncertainty summary used for cross-system interpretation. Win rates compare BBHT with that per-configuration included comparator. This hierarchy keeps 32 repeated policy trials from being mistaken for 32 independent simulator landscapes or related candidate pools from being mistaken for independent systems; it does not create independent source-score landscapes where only one saved source bank was used.

\subsection{Why the final sweep uses saved simulator-score datasets}

The final confidence sweep is a query-policy study over simulator-derived scores. Source landscapes are generated once by the default and stress benchmark runs. The final sweep resamples finite candidate pools from those immutable landscapes, applies controlled quantile thresholds to obtain target rare fractions, and replays search policies while revealing only the observations allowed by each access model. This separation keeps the scientific score definition fixed while supporting many replicated policy configurations. It does not measure end-to-end simulator savings, because source-landscape generation has already been paid; the online simulator challenge is the separate check in which proposals trigger actual solver calls.

\FloatBarrier

\section{Results}

\paragraph{Main finding.}
The fairest ML-facing result is the stronger scalar-score comparison: when GP active-search diagnostics observe numerical scores, the configuration-level included non-quantum / BBHT ratio at $M/N=0.001$ is 2.24 [2.02, 2.47], with BBHT favorable in 0.71 of configurations. The exact finite-pool replay with scalable score-guided baselines gives a ratio point estimate of 2.71 and win rate 0.98; paired hierarchical resampling gives a conservative 2.71 [1.89, 3.68] mean-ratio interval and 2.39 [1.76, 3.31] geometric-mean interval. Under 5\% noisy predicates at the same rare density, the ratio falls below one to 0.29 [0.27, 0.32], reversing the comparison. The evidence therefore supports a scoped regime claim: the marked-set query reference is informative when the predicate is exact, the target is rare, and the geometry remains hard for the included non-quantum policies; stronger access, imperfect predicates, and exposed equation structure materially narrow or remove that result.

\begin{figure*}[!htbp]
\centering
\includegraphics[width=0.83\textwidth]{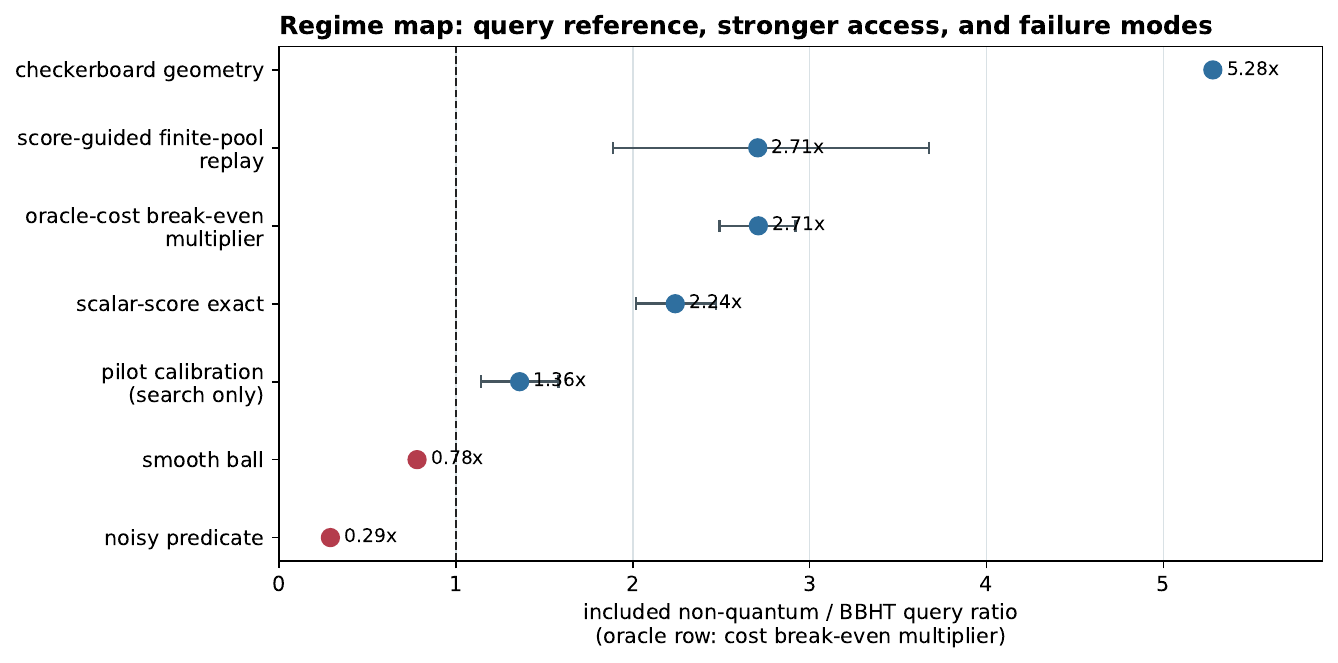}
\caption{Forest view of the main regime map. Values above one favor the \bbhtfull{} (BBHT) marked-set query reference in verified first-hit query count, except for the oracle-cost row, which reports the maximum per-oracle cost multiplier that preserves parity with the included non-quantum baseline. The figure shows query-count and break-even diagnostics only, not simulator or quantum-hardware runtime. Geometry and access-model rows explain when the finite-pool query reference is meaningful and when it should be bypassed.}
\label{fig:regime-forest}
\end{figure*}

Table~\ref{tab:compact-results} condenses the main positive and negative controls at the rarest density, while Figure~\ref{fig:regime-forest} visualizes the same regime map as query-count ratios.
\begin{table*}[!htbp]
\centering
\caption{Compact result table at rare density. BBHT denotes the \bbhtfull{} unknown-$M$ Grover-search schedule. Ratios above one mean that the included non-quantum baseline used more verified first-hit queries than the BBHT marked-set query reference. The oracle-cost row is a per-configuration break-even multiplier, so its interval can differ slightly by rounding from the ideal query-ratio row. Full method-specific counts and confidence sweeps appear below.}
\label{tab:compact-results}
\footnotesize
\renewcommand{\arraystretch}{1.2}
\setlength{\tabcolsep}{4pt}
\begin{tabular}{p{0.21\textwidth} c c c p{0.30\textwidth}}
\toprule
Setting & $M/N$ & \shortstack{Incl.\ non-Q /\\BBHT} & \shortstack{BBHT\\win rate} & Message \\
\midrule
Exact finite-pool replay & 0.001 & 2.71 [1.89, 3.68] & 0.98 [0.93, 1.00] & BBHT binary-oracle reference versus scalable score-guided baselines. \\

Scalar-score exact & 0.001 & 2.24 [2.02, 2.47] & 0.71 & Stronger non-quantum access reduces the margin. \\

\shortstack[l]{Pilot-calibrated GP\\stress, search only} & 0.001 & 1.36 [1.14, 1.58] & 0.43 & Calibration cost excluded here; stronger scalar-score portfolio. \\

\shortstack[l]{Pilot-calibration\\control, all in} & 0.001 & 2.62 [2.49, 2.77] & 0.94 & $B=256$ is charged; restricted predicate-only portfolio. \\

Noisy predicate, $\eta=0.05$ & 0.001 & 0.29 [0.27, 0.32] & 0.00 & False positives dominate rare marked sets. \\

Smooth ball geometry & 0.001 & 0.78 & 0.11 & Smooth coherent rare sets favor adaptive non-quantum search. \\

Checkerboard geometry & 0.001 & 5.28 & 1.00 & Fragmented finite sets favor marked-set query search. \\

\shortstack[l]{Oracle-cost break-even\\multiplier} & 0.001 & 2.71 [2.49, 2.92] & -- & Coherent oracle must fit within about 2--3 classical checks. \\
\bottomrule
\end{tabular}
\end{table*}

\subsection{Question 1: How does query burden scale with rarity?}

\paragraph{Example difficulty and verified discovery.}

A query-count table alone is not enough to establish a useful discovery result. The evidence must say what makes each example hard, whether the returned point is a verified marked point, how much finite-budget completeness is obtained, and how much saved-score replay time was used to evaluate the policy. Table~\ref{tab:example-difficulty} gives that view for the rarest target fraction in the final sweep.

The table is organized around the practical discovery question rather than around a theorem alone. The columns $N$ and $M$ describe the candidate-library scale and the number of acceptable rare regimes. The random-search column measures the burden of unguided finite-pool discovery. The per-configuration included non-quantum column measures how much structure the included adaptive classical baselines can exploit. The BBHT column measures the marked-predicate query reference. The ratio and win-rate columns compare BBHT against the strongest non-quantum method available in each configuration, not just against random search. The replay-time column records how long the saved-score policy evaluation took per configuration; it is included to keep query efficiency separate from analysis runtime and from any future simulator or quantum-hardware runtime.

\begin{table*}[!htbp]
\centering
\caption{Example difficulty and rare-discovery burden at the rarest target fraction, $M/N=0.001$. Query counts are first-hit verified predicate queries averaged over the recorded size-seed configurations for each system. The included non-quantum column selects the best included finite-pool method separately in each configuration. Replay time is saved-score query-policy evaluation time per configuration, not simulator or quantum-hardware runtime.}
\label{tab:example-difficulty}
\footnotesize
\renewcommand{\arraystretch}{1.20}
\setlength{\tabcolsep}{3pt}
\begin{tabular}{@{}p{0.15\textwidth}c c c r r p{0.14\textwidth}p{0.15\textwidth}c c@{}}
\toprule
System & $d$ & \shortstack[c]{Sweep\\$N$} & \shortstack[c]{Marked\\$M$} & \shortstack[c]{Random\\$q$} & \shortstack[c]{BBHT\\$q$} & \shortstack[c]{Per-config included\\non-quantum $q$} & \shortstack[c]{Included non-quantum\\ / BBHT} & \shortstack[c]{BBHT\\win} & \shortstack[c]{Replay\\s/config} \\
\midrule
Coupled FHN & 2 & 2,048--40,000 & 2--40 & 870.15 & 51.08 & 101.12 & 1.98 & 1.00 & 1.15 \\
Duffing & 4 & 2,048--100,000 & 2--100 & 907.86 & 52.98 & 119.69 & 2.26 & 1.00 & 2.37 \\
FitzHugh-Nagumo & 4 & 2,048--100,000 & 2--100 & 851.34 & 50.77 & 262.99 & 5.20 & 1.00 & 2.22 \\
Lorenz Hopf Boundary & 3 & 2,048--100,000 & 2--100 & 921.92 & 52.08 & 153.43 & 2.96 & 1.00 & 1.62 \\
Windy Pendulum & 3 & 2,048--16,000 & 2--16 & 791.06 & 51.27 & 173.55 & 3.44 & 1.00 & 1.55 \\
Spring-Mass-Damper & 2 & 2,048--12,000 & 2--12 & 848.75 & 50.92 & 65.57 & 1.29 & 0.88 & 1.19 \\
Van der Pol & 2 & 2,048--12,000 & 2--12 & 794.85 & 53.57 & 96.51 & 1.81 & 1.00 & 1.15 \\
\bottomrule
\end{tabular}
\end{table*}

For each system at $M/N=0.001$, the threshold $\tau$ is the system-specific distance-to-criticality cutoff used to define the marked set. Thus a reported first hit is not merely a table row: it is accepted only after the authoritative simulator-derived score satisfies $s(x_{\mathrm{hit}})\leq\tau$. Scores are not comparable across systems because each score has its own physical units and criticality definition. The direct transition certificates introduced below provide the complementary dynamical validation for systems where a local boundary construction is available.

At $M/N=0.001$, the sweep contains only 2--100 marked candidates inside candidate pools of size 2,048--100,000. This is the core generic finite-pool difficulty: a method that spreads queries without exploiting the rare set should require hundreds of checks before first discovery. The observed random-search means are 791--922 verified predicate queries across the seven systems. Grover/BBHT-style marked-set query search stays near 51--54 queries because the simulated unknown-$M$ schedule depends on the actual marked count through the marked-oracle success probability, not on a manually assigned constant. The per-configuration included non-quantum baseline remains strong and system dependent, ranging from 65.57 queries in the smooth spring control case to 262.99 queries in FHN, but the included non-quantum / BBHT ratio remains above one for every system at this rare density. These counts do not establish that the underlying equations are intrinsically difficult: Table~\ref{tab:fixed-threshold-online} separately shows that direct structure-aware constructions solve several systems immediately. Table~\ref{tab:system-rare} reports a complementary selected-method summary, where the FHN row uses the aggregate subset-search mean rather than the per-configuration best-method mean.

The table also clarifies what ``better'' means in this paper. FHN, Lorenz, pendulum, and Duffing have larger gaps for the included generic finite-pool policies because those policies do not fully exploit their equation-level boundary descriptions. Spring-mass-damper is intentionally easier because the score is a smooth two-dimensional control boundary; it has the smallest ratio and the only rare-density system-level BBHT win rate below one. This is not an inconsistency. It is evidence that QC-PHAST is not merely defeating weak baselines, and that classical geometry-aware or equation-aware search can be the right tool when the transition is smooth or analytically exposed.

Completeness is reported separately from mean query count. Every configuration in the final sweep has at least one verified marked candidate, and all policies are evaluated against the same simulator-derived labels. Adaptive trials that do not hit before their method cap contribute that cap to budget-censored means but are not reclassified as successful discoveries. Table~\ref{tab:survival} therefore separates verified success by the 384-query cap from an estimable p90 budget. At $M/N=0.001$, BBHT reaches 90\% empirical first-hit success by 86 queries and random search by 1872 queries. Cross-entropy, subset-style search, GP probability-of-hit, and GP level-set search reach verified-hit probabilities of 0.87, 0.85, 0.80, and 0.79, respectively, by their 384-query cap, so their p90 budgets are reported as greater than 384 rather than as successful hits at the cap. The replay-time column and logs report saved-score policy-evaluation time; they are not simulator or quantum-hardware runtimes.

\begin{figure*}[!htbp]
\centering
\includegraphics[width=0.76\textwidth]{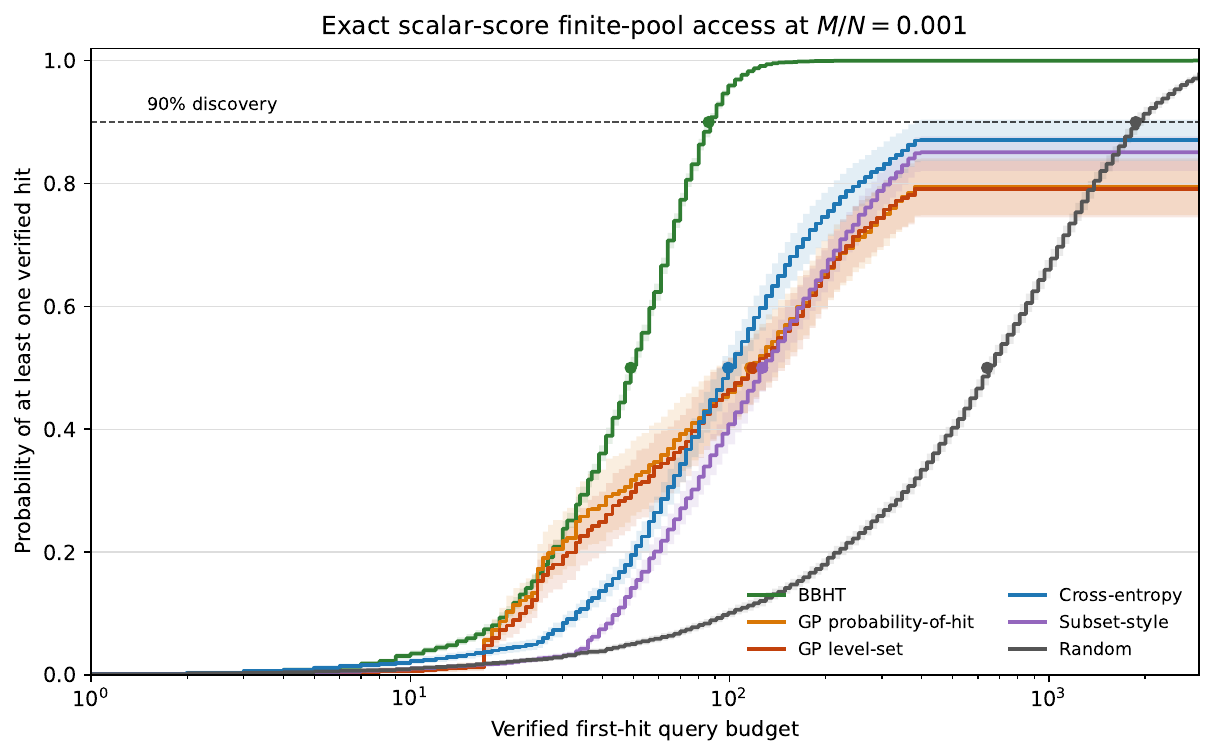}
\caption{Empirical verified-hit cumulative distributions for exact scalar-score access at $M/N=0.001$, with configuration-bootstrap uncertainty bands. Higher curves indicate faster discovery. Capped adaptive failures remain right-censored, so their curves plateau at the observed verified-hit probability rather than jumping to one; median and p90 markers are shown only when those probabilities are reached.}
\label{fig:access-survival}
\end{figure*}

This is the direct connection between query count, objective accuracy, completeness, and time. Query count measures the number of verified predicate checks before the first acceptable parameter vector. Objective correctness is enforced by the condition $s(x_{\mathrm{hit}})\leq\tau$; for direct structure-aware controls, the transition-certificate checks below additionally test the relevant local crossing conditions. Completeness is represented by the empirical survival and p90 budgets in Table~\ref{tab:survival} and Figure~\ref{fig:access-survival}. Time is represented only as saved-score policy replay time in Table~\ref{tab:example-difficulty} and in the logs, while simulator and quantum-hardware runtime remain outside the claim. The paper's conclusion relies on this combined evidence, not on the asymptotic square-root expression by itself.

\paragraph{Density scaling and hierarchical uncertainty.}

Table~\ref{tab:final-by-fraction} is the central density-sweep result. As the target marked fraction decreases from 0.1 to 0.001, the included non-quantum / BBHT ratio increases from 1.51 to 2.71. At $M/N=0.001$, BBHT requires 51.81 queries on average, while the selected included non-quantum baseline is cross-entropy at 144.74 queries and random search requires 855.13 queries.

The table should be read in two ways. First, random search follows the expected rare-event burden: the random query count grows rapidly as $M/N$ decreases. Second, adaptive classical methods close much of that gap. Cross-entropy and subset search reduce query counts by factors of several relative to random search. The meaningful comparison is therefore not BBHT versus random alone, but BBHT versus the included non-quantum method available for that configuration. On this stricter comparison, BBHT remains favorable in 93.94\% of final-sweep configurations. Figure~\ref{fig:final-speedups} visualizes the same included non-quantum / BBHT trend across systems.

The row-by-row pattern is more important than any single number. At $p=0.1$, the marked set is not very rare, and all methods have many opportunities to encounter a positive candidate. BBHT still has a lower mean query count, but the included non-quantum ratio is modest. As the target fraction moves to $0.03$ and $0.01$, adaptive classical methods remain useful but the gap grows. At $0.003$ and $0.001$, the search problem becomes closer to the rare-regime setting that motivated the framework. The included non-quantum / BBHT ratio reaches 2.41$\times$ and 2.71$\times$, and the BBHT win rate reaches 93.71\% and 98.29\%. This is a conditional density trend for the included finite-pool policy portfolio, not a claim that all system-specific solvers scale this way.

The absolute BBHT counts also behave plausibly. They vary with the actual marked counts and are generated by the BBHT query simulation rather than by assigning a closed-form constant to each density row. Table~\ref{tab:bbht-validation} independently validates the simulator against the exact two-dimensional amplitude recurrence over 500 finite-pool cases and 32,000 probability checks. The configuration-level intervals in Table~\ref{tab:final-by-fraction} are narrow because each density row aggregates many related system-size-seed configurations; the paired hierarchical intervals below are wider and should be used for cross-system uncertainty.

\begin{table*}[!htbp]
\centering
\caption{Expanded offline controlled confidence sweep by target marked fraction. BBHT denotes the Boyer--Brassard--H{\o}yer--Tapp unknown-$M$ Grover-search schedule. Each row aggregates the recorded system, candidate-size, and finite-pool resampling-seed configurations. Entries are mean verified first-hit query burdens with configuration-level bootstrap 95\% confidence intervals. Table~\ref{tab:hierarchical-density} provides paired hierarchical uncertainty across systems, finite-pool resampling seeds, and sizes. The right columns compare the binary-oracle BBHT reference with the per-configuration included non-quantum finite-pool baseline; CE and subset policies use their declared scalar-score feedback.}
\label{tab:final-by-fraction}
\footnotesize
\renewcommand{\arraystretch}{1.22}
\setlength{\tabcolsep}{3pt}
\begin{tabular}{@{}c c c c c c c c c@{}}
\toprule
\shortstack[c]{Target\\$M/N$} & Random & LHS & Sobol & CE & Subset & BBHT & \shortstack[c]{Included non-quantum\\ / BBHT} & \shortstack[c]{BBHT\\win rate} \\
\midrule
0.001 & \cientry{855.13}{831.62}{878.79} & \cientry{995.70}{957.68}{1032.86} & \cientry{998.11}{959.76}{1038.01} & \cientry{144.74}{133.30}{156.21} & \cientry{169.82}{159.39}{180.63} & \cientry{51.81}{51.11}{52.54} & \cientry{2.71}{2.49}{2.93} & 0.98 \\
0.003 & \cientry{308.40}{300.69}{316.02} & \cientry{370.90}{357.71}{385.12} & \cientry{360.04}{347.98}{372.85} & \cientry{79.09}{71.70}{86.31} & \cientry{87.53}{82.72}{92.60} & \cientry{29.84}{29.45}{30.21} & \cientry{2.41}{2.24}{2.59} & 0.94 \\
0.010 & \cientry{95.91}{93.55}{98.37} & \cientry{111.44}{107.50}{115.48} & \cientry{101.86}{98.23}{105.55} & \cientry{37.95}{35.39}{40.57} & \cientry{43.04}{41.73}{44.38} & \cientry{16.13}{15.85}{16.40} & \cientry{2.19}{2.06}{2.32} & 0.91 \\
0.030 & \cientry{32.98}{32.10}{33.87} & \cientry{35.36}{34.32}{36.37} & \cientry{29.19}{28.27}{30.15} & \cientry{20.19}{19.11}{21.28} & \cientry{24.59}{24.13}{25.06} & \cientry{8.89}{8.75}{9.02} & \cientry{2.13}{2.03}{2.23} & 0.90 \\
0.100 & \cientry{9.86}{9.64}{10.08} & \cientry{10.64}{10.37}{10.90} & \cientry{7.02}{6.78}{7.26} & \cientry{8.42}{8.12}{8.73} & \cientry{9.86}{9.65}{10.07} & \cientry{4.31}{4.25}{4.37} & \cientry{1.51}{1.46}{1.56} & 0.97 \\
\bottomrule
\end{tabular}
\end{table*}

\begin{table}[!htbp]
\centering
\caption{Paired hierarchical uncertainty for the exact finite-pool replay density sweep. Resampling is nested by system, finite-pool resampling seed, and candidate-size level, preserving the paired included non-quantum and BBHT configuration values. The saved source score bank is fixed within each system, so this hierarchy does not represent independently generated simulator landscapes. BBHT has binary marked-oracle access, while the scalable finite-pool baselines use their declared score feedback.}
\label{tab:hierarchical-density}
\footnotesize
\renewcommand{\arraystretch}{1.18}
\setlength{\tabcolsep}{4pt}
\begin{tabular}{lrrr}
\toprule
\shortstack[c]{Target\\$M/N$} & Mean ratio & Geometric mean & \shortstack[c]{BBHT\\win rate} \\
\midrule
0.001 & 2.71 [1.89, 3.68] & 2.39 [1.76, 3.31] & 0.98 [0.93, 1.00] \\
0.003 & 2.41 [1.72, 3.28] & 2.17 [1.61, 2.95] & 0.94 [0.79, 1.00] \\
0.010 & 2.19 [1.64, 2.77] & 2.01 [1.45, 2.64] & 0.91 [0.71, 1.00] \\
0.030 & 2.13 [1.67, 2.52] & 1.99 [1.48, 2.45] & 0.90 [0.69, 1.00] \\
0.100 & 1.51 [1.35, 1.69] & 1.48 [1.32, 1.67] & 0.97 [0.91, 1.00] \\
\bottomrule
\end{tabular}
\end{table}

\begin{table}[!htbp]
\centering
\caption{Fitted density exponents from $\log Q=\alpha+\beta\log p$ over paired system-size-seed sweeps. Negative values indicate a larger first-hit burden as targets become rarer.}
\label{tab:density-exponents}
\footnotesize
\renewcommand{\arraystretch}{1.18}
\setlength{\tabcolsep}{5pt}
\begin{tabular}{lrr}
\toprule
Method & Mean $\beta$ & Median $\beta$ \\
\midrule
bbht & -0.54 [-0.56, -0.52] & -0.54 \\
cross entropy & -0.59 [-0.67, -0.52] & -0.58 \\
lhs & -0.99 [-1.04, -0.93] & -0.99 \\
random & -0.97 [-1.01, -0.93] & -0.97 \\
sobol & -1.08 [-1.14, -1.01] & -1.08 \\
subset & -0.59 [-0.66, -0.51] & -0.60 \\
\bottomrule
\end{tabular}
\end{table}

\begin{figure*}[!htbp]
\centering
\includegraphics[width=0.72\textwidth]{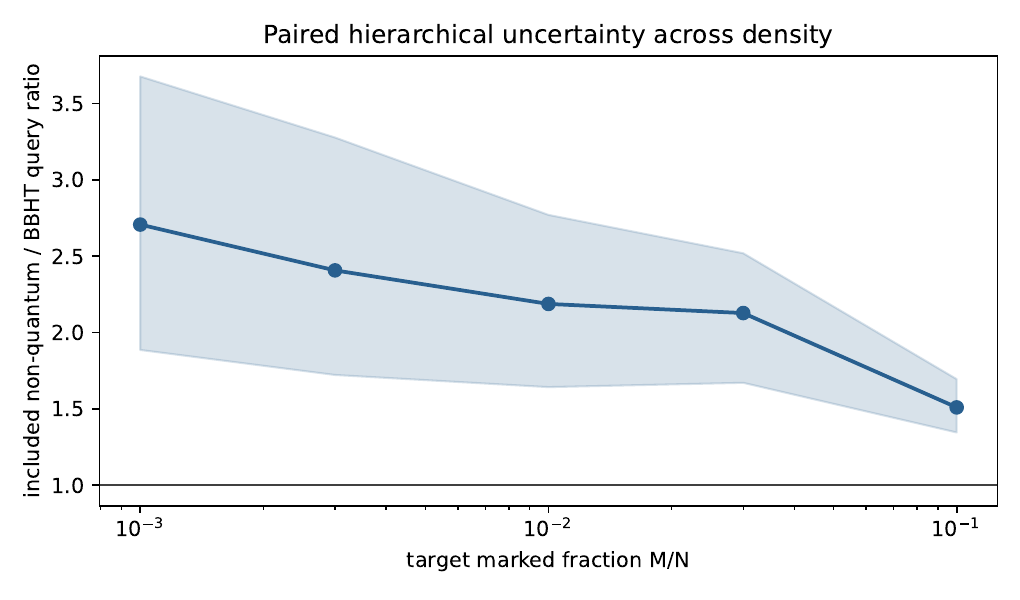}
\caption{Paired hierarchical uncertainty across target marked fractions. Resampling is nested by system, finite-pool resampling seed, and candidate size and preserves each configuration's included non-quantum / BBHT pairing. The wider intervals relative to ordinary configuration bootstrap reflect between-system heterogeneity rather than an additional source of query noise; they do not represent independent source-score landscapes.}
\label{fig:hierarchical-density}
\end{figure*}

The hierarchical result in Figure~\ref{fig:hierarchical-density} preserves the point estimate at $p=0.001$ but broadens its uncertainty to 2.71 [1.89, 3.68]; the corresponding geometric mean is 2.39 [1.76, 3.31]. The interval remains above one in this benchmark, but it is appropriately less precise than a flat resampling of related configurations. The fitted density exponents in Table~\ref{tab:density-exponents} provide a second check on the intended query-model trend: BBHT has mean exponent $-0.54$ [-0.56, -0.52], close to the expected $-1/2$, while random has $-0.97$ [-1.01, -0.93], close to $-1$. Cross-entropy and subset-style policies have intermediate, system-dependent slopes. These fits describe the implemented finite-pool policies; they are not universal dynamical-system exponents.

\begin{figure*}[!htbp]
\centering
\includegraphics[width=0.78\textwidth]{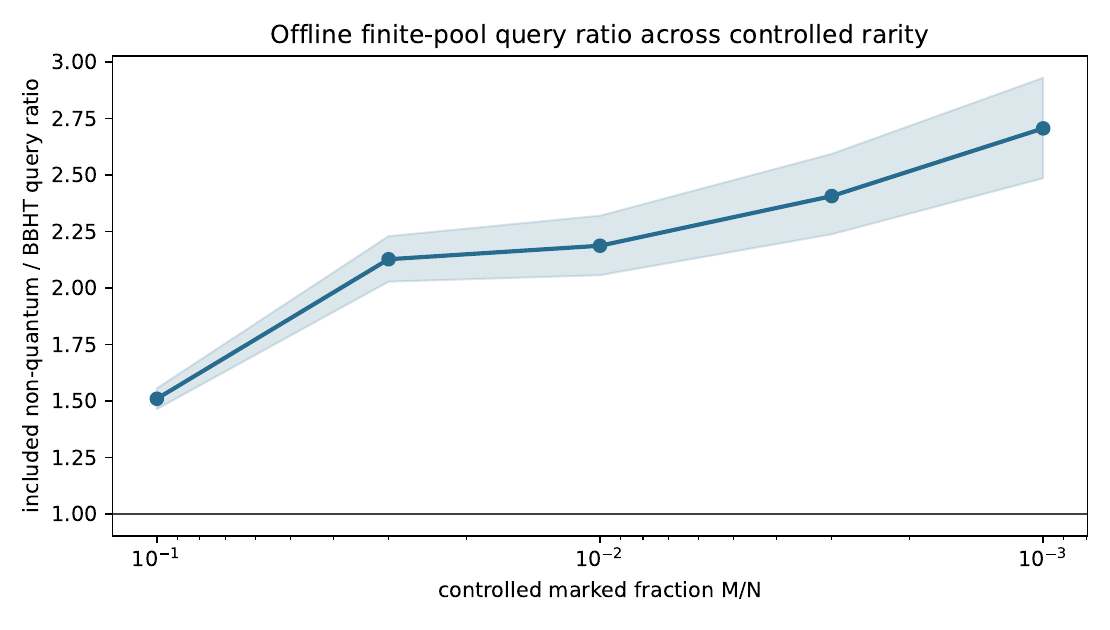}
\caption{System-resolved ratios from the final confidence sweep. Values above one indicate fewer BBHT queries than the included non-quantum baseline selected per configuration. This view exposes system heterogeneity across densities; Figure~\ref{fig:hierarchical-density} is the corresponding paired hierarchical uncertainty summary. Smooth or analytically exposed systems can reduce or eliminate the margin.}
\label{fig:final-speedups}
\end{figure*}

\subsection{Question 2: How do access, calibration, fidelity, and geometry change the result?}

The final confidence sweep answers the large replicated density question, but it still leaves two methodological concerns: whether the more expressive GP scalar-score diagnostics narrow the margin relative to the scalable score-guided baseline portfolio, and whether the advantage is merely an artifact of discretizing smooth continuous systems. The access-model and geometry ablations address both concerns directly.

Table~\ref{tab:access-ablation} separates exact scalar-score access, a 5\% noisy-predicate stress test, and pilot-threshold uncertainty over the same 875 base configurations. The exact scalar-score row is the fairest strengthened comparison because GP active-search diagnostics observe numerical scores and can model proximity to the threshold. At $M/N=0.001$, BBHT requires 51.62 [50.86, 52.36] queries and the included non-quantum / BBHT ratio is 2.24 [2.02, 2.47]. This is lower than the original scalable finite-pool replay point estimate of 2.71 because the GP diagnostics are stronger than the scalable baselines in some configurations, but it remains favorable in 0.71 of configurations.

The noisy-predicate rows provide the opposite result. With symmetric static corruption at $\eta=0.05$, the rarest-density verified BBHT count rises to 4405.30 [3924.00, 4926.50] and the included non-quantum / BBHT ratio falls to 0.29 [0.27, 0.32]. At $p=0.001$, Proposition~\ref{prop:predicate-purity} predicts observed-mark purity of only $0.001(0.95)/[0.001(0.95)+0.999(0.05)]\approx0.0187$ before finite-sample variation: roughly one true target per 53 observed marks. The measured collapse is therefore a mechanistic consequence of false-positive mass, not an unexplained sensitivity. Practical use requires an exact predicate or a verification architecture whose repeated rejection cost remains acceptable.

\begin{figure}[!htbp]
\centering
\includegraphics[width=0.82\columnwidth]{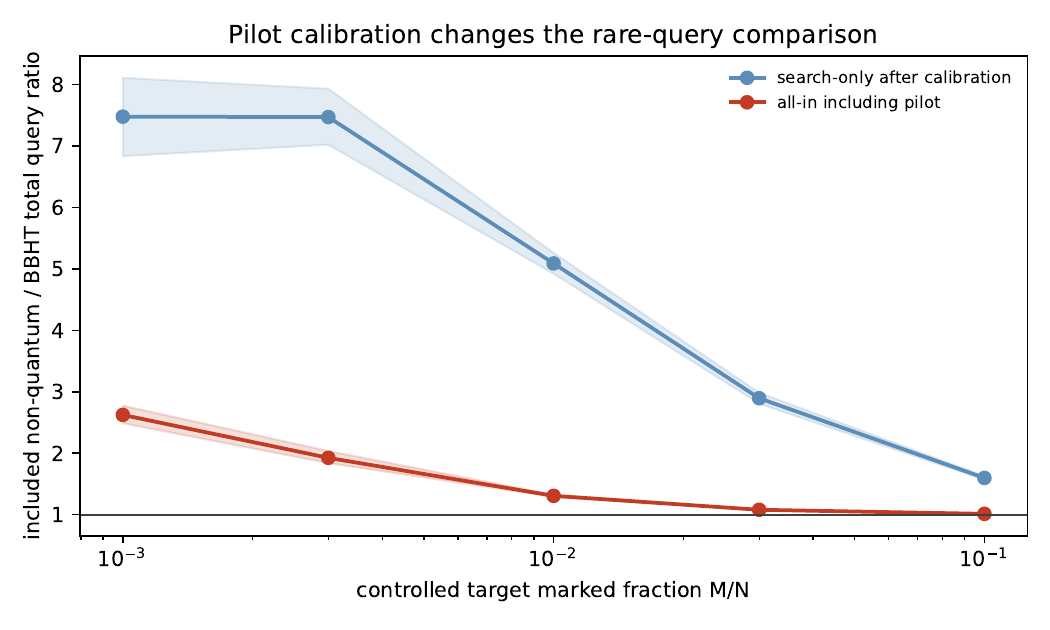}
\caption{Fully charged pilot-calibration control. The blue curve excludes the calibration budget after the pilot has set $\hat\tau$; the red curve charges all $B=256$ pilot scores before search. This restricted predicate-only control is displayed separately from the scalar-score GP access track because it answers an accounting question, not a direct policy-ranking question.}
\label{fig:pilot-all-in}
\end{figure}

The pilot-threshold GP row is also more severe than its label may suggest. It reports search after calibration only: with $n=256$ and intended $p=0.001$, Proposition~\ref{prop:pilot-coverage} gives approximately 77.4\% probability that the pilot contains no candidate from the intended target tail. The resulting threshold often marks a broader surrogate tail and sends search policies toward candidates that fail the authoritative target. Its search-only ratio is 1.36 [1.14, 1.58] and win rate 0.43. Table~\ref{tab:pilot-all-in} and Figure~\ref{fig:pilot-all-in} report a separate fully charged 1,750-configuration calibration control: every $B=256$ pilot evaluation is added to both policies before the finite-pool search, yielding an all-in ratio of 2.62 [2.49, 2.77] at $p=0.001$ for a restricted predicate-only portfolio. This does not contradict the GP stress row because the policy portfolios and information models differ. Together they show that calibration cannot be treated as free and that a small pilot has poor intended-tail coverage. In an application, a scientifically fixed threshold or a separately validated tail model is preferable to inspecting the full score distribution.

\begin{table*}[!htbp]
\centering
\caption{Access-model and predicate-robustness ablation on the same 875 offline controlled base configurations. BBHT has binary marked-oracle access; scalar-score GP diagnostics receive numerical simulator scores. Ratios above one favor the BBHT query reference. The pilot-threshold row reports search after calibration only; Table~\ref{tab:pilot-all-in} separately charges the calibration pilot.}
\label{tab:access-ablation}
\footnotesize
\renewcommand{\arraystretch}{1.20}
\setlength{\tabcolsep}{4pt}
\begin{tabular}{@{}p{0.17\textwidth}p{0.13\textwidth}c p{0.18\textwidth}p{0.23\textwidth}c@{}}
\toprule
Access track & Setting & \shortstack[c]{Target\\$M/N$} & \shortstack[c]{BBHT\\queries} & \shortstack[c]{Included non-quantum\\/ BBHT} & \shortstack[c]{BBHT\\win rate} \\
\midrule
Predicate-only & Exact & 0.001 & \shortstack[r]{52.02\\[-1pt]\scriptsize [51.31, 52.72]} & \shortstack[r]{15.16\\[-1pt]\scriptsize [14.61, 15.70]} & 1.00 \\
Scalar-score & Exact & 0.001 & \shortstack[r]{51.62\\[-1pt]\scriptsize [50.86, 52.36]} & \shortstack[r]{2.24\\[-1pt]\scriptsize [2.02, 2.47]} & 0.71 \\
Pilot-calibrated threshold & $B=256$ & 0.001 & \shortstack[r]{237.35\\[-1pt]\scriptsize [197.42, 286.80]} & \shortstack[r]{1.36\\[-1pt]\scriptsize [1.14, 1.58]} & 0.43 \\
Noisy predicate & $\eta=0.05$ & 0.001 & \shortstack[r]{4405.30\\[-1pt]\scriptsize [3924.00, 4926.50]} & \shortstack[r]{0.29\\[-1pt]\scriptsize [0.27, 0.32]} & 0.00 \\
\bottomrule
\end{tabular}
\end{table*}

\begin{table}[t]
\centering
\caption{All-in pilot-threshold accounting over 1750 controlled-replay configurations. A calibration pilot of $B=256$ exact score evaluations is charged before finite-pool search. The pilot estimates $\hat\tau$; the evaluator retains the full-pool controlled threshold only to assess whether a returned candidate is truly marked. Search-only values exclude the pilot solely to show the calibration effect; all-in values are the honest totals for this controlled calibration experiment, not an end-to-end runtime claim.}
\label{tab:pilot-all-in}
\footnotesize
\renewcommand{\arraystretch}{1.18}
\setlength{\tabcolsep}{4pt}
\begin{tabular}{@{}c p{0.20\textwidth}p{0.18\textwidth}c p{0.19\textwidth}@{}}
\toprule
\shortstack[c]{Target\\$M/N$} & \shortstack[c]{Search-only\\ratio} & \shortstack[c]{All-in\\ratio} & \shortstack[c]{BBHT\\win rate} & \shortstack[c]{Pilot\\true-hit rate} \\
\midrule
0.001 & 7.48 [6.83, 8.11] & 2.62 [2.49, 2.77] & 0.94 & 0.22 \\
0.003 & 7.47 [7.02, 7.93] & 1.92 [1.84, 2.03] & 0.99 & 0.56 \\
0.010 & 5.09 [4.92, 5.26] & 1.31 [1.29, 1.32] & 1.00 & 0.94 \\
0.030 & 2.89 [2.81, 2.98] & 1.08 [1.08, 1.08] & 1.00 & 1.00 \\
0.100 & 1.60 [1.57, 1.63] & 1.01 [1.01, 1.01] & 0.99 & 1.00 \\
\bottomrule
\end{tabular}
\end{table}

\begin{figure*}[!htbp]
\centering
\includegraphics[width=0.82\textwidth]{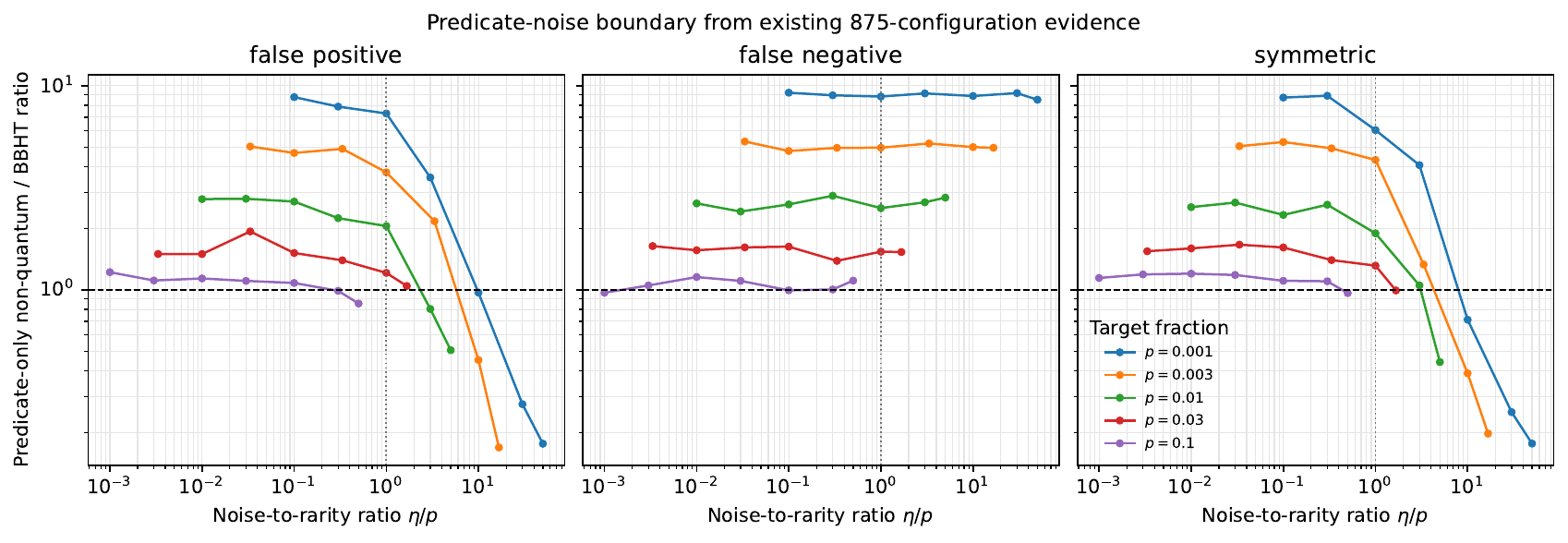}
\caption{Noisy-oracle phase diagram indexed by the physically relevant ratio $\eta/p$. The plotted value is the predicate-only included non-quantum / BBHT query ratio under verified final labels, not the final-sweep included non-quantum ratio. The exact $\eta=0$ control remains in Table~\ref{tab:noise-phase-diagram}; the log-scale figure resolves the nonzero-noise transition and marks ratio one, where the comparison reverses.}
\label{fig:noise-phase-diagram}
\end{figure*}

The dense noise phase diagram in Table~\ref{tab:noise-phase-diagram} and Figure~\ref{fig:noise-phase-diagram} refines that single 5\% stress row with 875 base configurations. The degradation is governed by $\eta/p$ rather than by $\eta$ alone. The table should not be read as a third estimate of the final-sweep headline. Its $\eta=0$ control row uses a predicate-only comparator restricted to random, Latin-hypercube, and Sobol-style query orders, so the 9.02$\times$ rare-density ratio is larger than both the 2.71$\times$ final-sweep ratio, which includes cross-entropy and subset-style adaptive search, and the 2.24$\times$ scalar-score access ratio, which adds stronger GP active-search diagnostics. Its purpose is to isolate noisy-predicate behavior under a common predicate-only access model. The printed table focuses on false-positive and symmetric noise because these are the regimes that can make the observed marked set contain many false targets; false-negative-only rows are included in the artifact summary. When false-positive or symmetric noise is small relative to the true marked fraction, verification overhead is limited. At the rarest density, the noisy BBHT procedure is still favorable at $\eta/p=1$, weakens by $\eta/p=3$, and reverses around $\eta/p=10$ for the false-positive and symmetric models. By $\eta/p=30$--$50$, noisy BBHT loses decisively. This is one of the strongest practical boundaries in the paper: a rare-regime quantum query layer requires either an exact predicate or an observed marked set whose purity remains high relative to $p$ and whose rejected candidates can be verified at acceptable cost.

\begin{table*}[t]
\centering
\caption{Noise phase-diagram summary at the rarest marked fraction from 875 base configurations. The printed rows focus on false-positive and symmetric noise because those regimes create false marked targets; false-negative-only rows are included in the reproducibility artifact. The comparator is restricted to predicate-only included non-quantum orders, so the $\eta=0$ control row is not the same estimator as the final sweep or scalar-score access ablation. Values below one indicate that noisy BBHT no longer beats the included predicate-only non-quantum baseline.}
\label{tab:noise-phase-diagram}
\footnotesize
\renewcommand{\arraystretch}{1.14}
\setlength{\tabcolsep}{4pt}
\begin{tabular}{@{}p{0.16\textwidth}c c p{0.25\textwidth}c@{}}
\toprule
Noise model & $\eta$ & $\eta/p$ & \shortstack[c]{Predicate-only\\non-quantum / BBHT} & \shortstack[c]{BBHT\\win rate} \\
\midrule
false positive & 0 & 0.00 & 9.02 [8.12, 9.92] & 0.91 \\
false positive & 0.0001 & 0.10 & 9.03 [8.20, 9.97] & 0.92 \\
false positive & 0.0003 & 0.30 & 7.28 [6.55, 8.00] & 0.90 \\
false positive & 0.001 & 1.00 & 6.72 [6.04, 7.43] & 0.90 \\
false positive & 0.003 & 3.00 & 3.31 [2.96, 3.68] & 0.79 \\
false positive & 0.01 & 10.00 & 0.90 [0.79, 1.00] & 0.38 \\
false positive & 0.03 & 30.00 & 0.26 [0.23, 0.29] & 0.01 \\
false positive & 0.05 & 50.00 & 0.17 [0.15, 0.20] & 0.00 \\
symmetric & 0 & 0.00 & 8.93 [8.07, 9.83] & 0.92 \\
symmetric & 0.0001 & 0.10 & 8.63 [7.79, 9.53] & 0.91 \\
symmetric & 0.0003 & 0.30 & 8.45 [7.61, 9.29] & 0.90 \\
symmetric & 0.001 & 1.00 & 6.57 [5.88, 7.27] & 0.87 \\
symmetric & 0.003 & 3.00 & 3.58 [3.19, 4.02] & 0.81 \\
symmetric & 0.01 & 10.00 & 0.75 [0.67, 0.83] & 0.34 \\
symmetric & 0.03 & 30.00 & 0.24 [0.21, 0.27] & 0.00 \\
symmetric & 0.05 & 50.00 & 0.18 [0.15, 0.20] & 0.00 \\
\bottomrule
\end{tabular}
\end{table*}

The first-hit survival summary in Table~\ref{tab:survival} gives a different view of the same rare-density exact-access case. At $M/N=0.001$, BBHT has median 49 and p90 86 queries over 5600 trials, while random search has median 641 and p90 1872. Cross-entropy and subset-style search have medians of 99 and 126.5 but reach only 0.87 and 0.85 verified-hit probability by their 384-query cap. The GP probability-of-hit and level-set diagnostics reach 0.80 and 0.79. Their unobserved p90 budgets are therefore reported as $>384$, not imputed at the cap. These scalar-score methods improve materially over random search but do not match the BBHT tail budget in this diagnostic.

\begin{table}[!htbp]
\centering
\caption{First-hit survival summary for exact scalar-score finite-pool access at target marked fraction $M/N=0.001$. Adaptive trials recorded at their 384-query cap are treated as right-censored rather than as verified hits. The success column therefore reports empirical verified-hit probability by query 384, and a p90 value is shown only when 90\% verified discovery is reached.}
\label{tab:survival}
\footnotesize
\renewcommand{\arraystretch}{1.18}
\setlength{\tabcolsep}{4pt}
\begin{tabular}{lrrrr}
\toprule
Method & Trials & Median & \shortstack[c]{Success by\\query 384} & \shortstack[c]{Verified-hit\\p90 budget} \\
\midrule
BBHT & 5600 & 49.0 & 1.00 & 86 \\
Cross-entropy & 5600 & 99.0 & 0.87 & $>384$ \\
GP level-set & 1400 & 118.0 & 0.79 & $>384$ \\
GP probability-of-hit & 1400 & 116.0 & 0.80 & $>384$ \\
Random & 5600 & 641.0 & 0.32 & 1872 \\
Subset-style & 5600 & 126.5 & 0.85 & $>384$ \\
\bottomrule
\end{tabular}
\end{table}

\begin{figure*}[!htbp]
\centering
\includegraphics[width=0.76\textwidth]{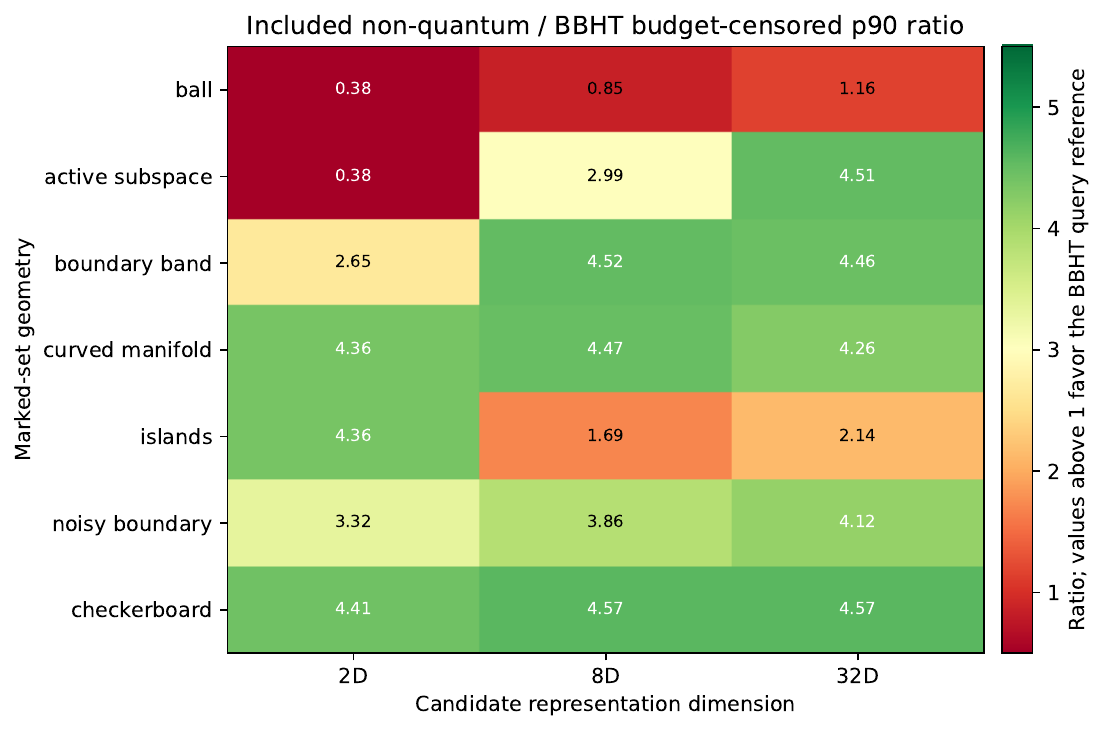}
\caption{Synthetic-geometry heatmap at $M/N=0.001$. Rows are marked-set constructions, columns are ambient dimensions, and cells report the budget-censored p90 included non-quantum / BBHT ratio; values above one favor the BBHT query reference. The smooth ball is the clearest classical-friendly control, while fragmented or boundary-like constructions produce larger ratios. These categories are controlled diagnostics, not estimates of intrinsic geometry in the projected dynamical-system plots.}
\label{fig:geometry-p90}
\end{figure*}

Table~\ref{tab:geometry-ablation} and Figure~\ref{fig:geometry-p90} isolate marked-set shape from dynamical-system details. Smooth balls are classical-friendly: the mean included non-quantum / BBHT ratio is 0.78 and the BBHT win rate is 0.11. This is the expected behavior for an easy coherent target set. Boundary bands, checkerboards, curved manifolds, islands, and noisy boundaries are much more favorable to BBHT, with mean ratios from 2.46 to 5.28. The geometry ablation therefore refines the central claim: \qcphast{} is strongest for rare finite sets whose geometry is fragmented, thin, boundary-like, or poorly matched to a simple adaptive classical proposal; it is not a replacement for continuation or smooth active learning on easy level sets.

\begin{table}[!htbp]
\centering
\caption{Synthetic geometry ablation at target marked fraction $M/N=0.001$, averaged over dimensions and candidate-pool sizes. The smooth ball is the clearest classical-friendly case, while fragmented or boundary-like marked sets show larger Grover/BBHT query advantages.}
\label{tab:geometry-ablation}
\footnotesize
\renewcommand{\arraystretch}{1.18}
\setlength{\tabcolsep}{4pt}
\begin{tabular}{lrrr}
\toprule
Geometry & Groups & \shortstack[c]{Mean included\\non-quantum / BBHT} & \shortstack[c]{Mean\\win rate} \\
\midrule
active\_subspace & 3 & 2.51 & 0.67 \\
ball & 3 & 0.78 & 0.11 \\
boundary\_band & 3 & 3.53 & 1.00 \\
checkerboard & 3 & 5.28 & 1.00 \\
curved\_manifold & 3 & 4.13 & 1.00 \\
islands & 3 & 2.46 & 0.96 \\
noisy\_boundary & 3 & 2.65 & 1.00 \\
\bottomrule
\end{tabular}
\end{table}

Finally, Table~\ref{tab:oracle-resource} records a deliberately modest arithmetic proxy for simple analytic predicates. The spring and Van der Pol rows report fixed-point register and operation proxies at $N=32768$. The table's final column is an illustrative $(2j+1)$ sequence factor multiplied by a per-call arithmetic proxy; it is not the BBHT oracle-query counter, which counts $j$ marked-oracle calls per sampled Grover schedule plus final verification. These values are not a compiled circuit, do not include error correction, and do not justify a hardware runtime claim. Their role is only to make the oracle issue explicit: even when the query count is favorable, predicate synthesis and data loading can dominate practical cost.

\begin{table}[!htbp]
\centering
\caption{Illustrative fixed-point arithmetic proxy for simple analytic threshold predicates at 16-bit precision and $N=32768$. The final column is $(2j+1)$ times the per-call proxy, where $j$ is the nominal Grover-iteration reference. It is a sequence arithmetic proxy, not a compiled circuit count or the BBHT oracle-query counter.}
\label{tab:oracle-resource}
\footnotesize
\renewcommand{\arraystretch}{1.18}
\setlength{\tabcolsep}{3pt}
\begin{tabular}{@{}p{0.16\textwidth}c c p{0.15\textwidth}p{0.19\textwidth}p{0.20\textwidth}@{}}
\toprule
System & $M/N$ & \shortstack[c]{Nominal\\$j$} & \shortstack[c]{Logical-qubit\\proxy} & \shortstack[c]{Per-call arithmetic\\proxy} & \shortstack[c]{Sequence arithmetic\\proxy} \\
\midrule
spring & 0.01000 & 8 & 100 & 191 & 3247 \\
spring & 0.00300 & 15 & 100 & 191 & 5921 \\
spring & 0.00100 & 25 & 100 & 191 & 9741 \\
vdp & 0.01000 & 8 & 67 & 95 & 1615 \\
vdp & 0.00300 & 15 & 67 & 95 & 2945 \\
vdp & 0.00100 & 25 & 67 & 95 & 4845 \\
\bottomrule
\end{tabular}
\end{table}

\subsection{Question 3: When should direct structure or online classical search bypass the finite-pool layer?}

\begin{figure}[!htbp]
\centering
\includegraphics[width=0.86\columnwidth]{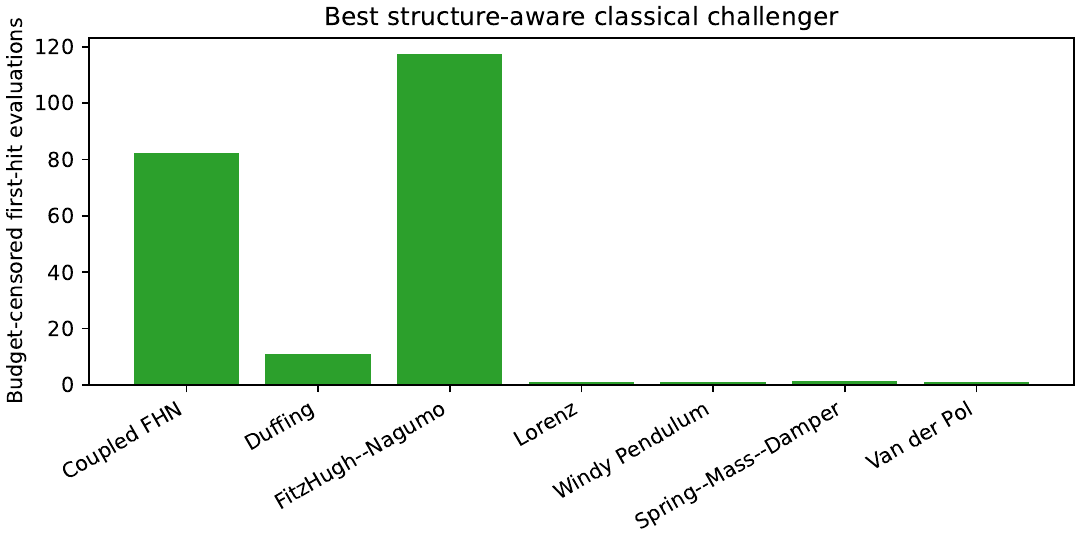}
\caption{Best implemented structure-aware classical challenger at $M/N=0.001$ in the 1,575-configuration continuous-routing study. The bar height is the budget-censored mean number of online evaluations. Analytic probes solve exposed smooth boundaries immediately, Duffing is favorable to a continuous rank-Gaussian proposal, and the implemented generic off-pool challengers remain difficult on FHN and coupled FHN. These are routing controls with stronger access than the finite-pool BBHT reference.}
\label{fig:classical-structure-challenge}
\end{figure}

Table~\ref{tab:classical-structure-challenge} and Figure~\ref{fig:classical-structure-challenge} directly address the concern that finite-pool search can handicap continuous classical optimization. The full-capacity study contains 1,575 configurations and is intentionally mixed. Smooth analytic systems are routed to analytic probes and classical methods win immediately: Van der Pol, spring, pendulum, and Lorenz require only one or two evaluations in this challenge. That result strengthens rather than weakens the paper because it identifies where \qcphast{} should bypass its marked-set branch. At $M/N=0.001$ under the 128-evaluation budget, Duffing is substantially easier under a rank-Gaussian continuous proposal (11.05 [10.40, 11.66] evaluations), whereas generic off-pool coupled FHN and FHN remain difficult for the implemented challengers (82.36 [79.99, 84.71] and 117.30 [115.34, 119.12] evaluations for differential evolution). The reported first-hit means are budget-censored, not conditional-on-success means: when a trial fails to find a verified hit, it contributes the full evaluation budget. This is why low-success rows such as generic FHN remain interpretable rather than being averaged only over the few successful trials. The conclusion is therefore not ``BBHT beats continuous optimization''; it is that the finite-pool marked-set branch is relevant only after an appropriate structure-aware route has failed or is unavailable.

\begin{table*}[t]
\centering
\caption{Continuous and structure-aware classical challenge at the rarest marked fraction from 1575 configurations. These methods are allowed to propose off-pool points or use analytic structure, so the table is a stress test of the finite-pool conclusion rather than the same query model as BBHT. First-hit means are budget-censored: trials without a verified hit contribute the full evaluation budget.}
\label{tab:classical-structure-challenge}
\footnotesize
\renewcommand{\arraystretch}{1.18}
\setlength{\tabcolsep}{4pt}
\begin{tabular}{@{}p{0.19\textwidth}p{0.22\textwidth}p{0.25\textwidth}c c@{}}
\toprule
System & Best method & \shortstack[c]{Budget-censored mean\\first-hit evaluations} & \shortstack[c]{Success\\rate} & \shortstack[c]{Mean wall\\time (s)} \\
\midrule
Coupled FHN & differential evolution & 82.36 [79.99, 84.71] & 0.93 & 0.023 \\
Duffing & rank gaussian & 11.05 [10.40, 11.66] & 1.00 & $<0.001$ \\
FitzHugh--Nagumo & differential evolution & 117.30 [115.34, 119.12] & 0.14 & 0.007 \\
Lorenz & analytic probe & 1.00 [1.00, 1.00] & 1.00 & $<0.001$ \\
Windy Pendulum & analytic probe & 1.07 [1.00, 1.16] & 1.00 & $<0.001$ \\
Spring--Mass--Damper & analytic probe & 1.29 [1.16, 1.42] & 1.00 & $<0.001$ \\
Van der Pol & analytic probe & 1.00 [1.00, 1.00] & 1.00 & $<0.001$ \\
\bottomrule
\end{tabular}
\end{table*}

\paragraph{Fixed thresholds, no-target handling, and direct transition checks.}

The controlled density sweep is intentionally an offline instrument. To make the deployment distinction concrete, Table~\ref{tab:fixed-threshold-online} uses the fixed default threshold for each system before a policy sees the sampled pool scores. The expanded replay contains 2,975 draws and retains pools with $M=0$ rather than repairing the pool or forcing a first hit. Five of 425 Lorenz draws (1.2\%) are no-target pools; the other six systems have no empty pools in this particular benchmark design. Thus the table validates retention and reporting behavior, not a claim that no-target cases are common for the chosen benchmark thresholds. A real online deployment must stop at its specified budget and report ``no verified target found in this library'' rather than infer that a target exists.

The finite-pool ratios in Table~\ref{tab:fixed-threshold-online} are compared only on nonempty pools and use the generic finite-pool policy portfolio. The structure-aware controls intentionally change the access model. FHN is parameterized directly on a trace-zero, positive-determinant local stability boundary; Duffing is parameterized on a zero-eigenvalue equilibrium boundary; Lorenz uses its analytic Hopf surface; coupled FHN uses equilibrium root following. These controls reach FHN, Duffing, and Lorenz fixed-threshold candidates in one score evaluation on the reported draws. FHN meets the stated local certificate on all 425 constructions; Lorenz is analytically verified against the declared Hopf surface; and the Duffing marked candidate meets the stricter local certificate in 0.99 of constructions. The remaining rows are authoritative score-verification or routing controls, not complete bifurcation-certificate claims. They are not apples-to-apples BBHT competitors, and they establish that those systems should not support a claim of intrinsic continuous-search difficulty when their governing equations are available.

\begin{table*}[t]
\centering
\caption{Expanded fixed-threshold deployment-style replay over 2975 configuration draws. Thresholds are specified before a policy sees the sampled pool scores; pools with no qualifying candidate are retained rather than repaired. The finite-pool ratio is reported only on pools with at least one true hit. The evaluator uses a saved immutable score bank, so this is a lazy-information policy replay rather than an end-to-end simulator-runtime measurement. The structure-aware control is allowed continuous equation access and is therefore a routing control rather than an identical-access comparator.}
\label{tab:fixed-threshold-online}
\footnotesize
\renewcommand{\arraystretch}{1.20}
\setlength{\tabcolsep}{3pt}
\begin{tabular}{@{}p{0.17\textwidth}c c c c p{0.21\textwidth}c c c@{}}
\toprule
System & Configs & \shortstack[c]{$M=0$\\rate} & Eligible & \shortstack[c]{Included non-quantum\\ / BBHT} & \shortstack[c]{Structure-aware\\control} & \shortstack[c]{Hit\\rate} & Evaluations & \shortstack[c]{Certificate\\pass} \\
\midrule
Coupled FHN & 425 & 0.00 & 425 & \cientry{2.19}{2.15}{2.22} & root following & 1.00 & \cientry{32.35}{30.43}{34.26} & 1.00 \\
Duffing & 425 & 0.00 & 425 & \cientry{2.22}{2.18}{2.26} & analytic saddle node parameterization & 1.00 & \cientry{1.00}{1.00}{1.00} & 0.99 \\
FitzHugh-Nagumo & 425 & 0.00 & 425 & \cientry{4.48}{4.39}{4.57} & analytic hopf parameterization & 1.00 & \cientry{1.00}{1.00}{1.00} & 1.00 \\
Lorenz Hopf Boundary & 425 & 0.01 & 420 & \cientry{2.70}{2.63}{2.76} & analytic hopf surface & 1.00 & \cientry{1.00}{1.00}{1.00} & 1.00 \\
Windy Pendulum & 425 & 0.00 & 425 & \cientry{1.95}{1.91}{1.99} & analytic control & 1.00 & \cientry{1.00}{1.00}{1.00} & 1.00 \\
Spring-Mass-Damper & 425 & 0.00 & 425 & \cientry{0.91}{0.89}{0.94} & analytic control & 1.00 & \cientry{1.00}{1.00}{1.00} & 1.00 \\
Van der Pol & 425 & 0.00 & 425 & \cientry{1.78}{1.76}{1.81} & analytic control & 1.00 & \cientry{1.00}{1.00}{1.00} & 1.00 \\
\bottomrule
\end{tabular}
\end{table*}

Table~\ref{tab:transition-certificates} records the stricter local checks for the two direct constructions most relevant to the phase-space narrative. The FHN certificate requires a small equilibrium residual, trace near zero, positive determinant, a nonzero imaginary pair, and a transverse sign change. The Duffing certificate requires a small equilibrium residual, a simple near-zero eigenvalue, a nondegeneracy condition, and a changed equilibrium-root count on either side of the candidate. These are local transition certificates, not complete normal-form calculations or a claim to classify every global bifurcation.

\begin{table}[t]
\centering
\caption{Transition-certificate checks for direct structure-aware controls. Values aggregate 850 successful fixed-threshold constructions; a certificate requires the stated local equilibrium and crossing checks.}
\label{tab:transition-certificates}
\footnotesize
\renewcommand{\arraystretch}{1.18}
\setlength{\tabcolsep}{4pt}
\begin{tabular}{lrrr}
\toprule
System & \shortstack[c]{Successful\\constructions} & \shortstack[c]{Certificate\\pass rate} & \shortstack[c]{Median verified\\score} \\
\midrule
FitzHugh--Nagumo & 425 & 1.00 & 1.41e-16 \\
Duffing & 425 & 0.99 & 7.22e-16 \\
\bottomrule
\end{tabular}
\end{table}

\begin{figure*}[!htbp]
\centering
\includegraphics[width=0.94\textwidth]{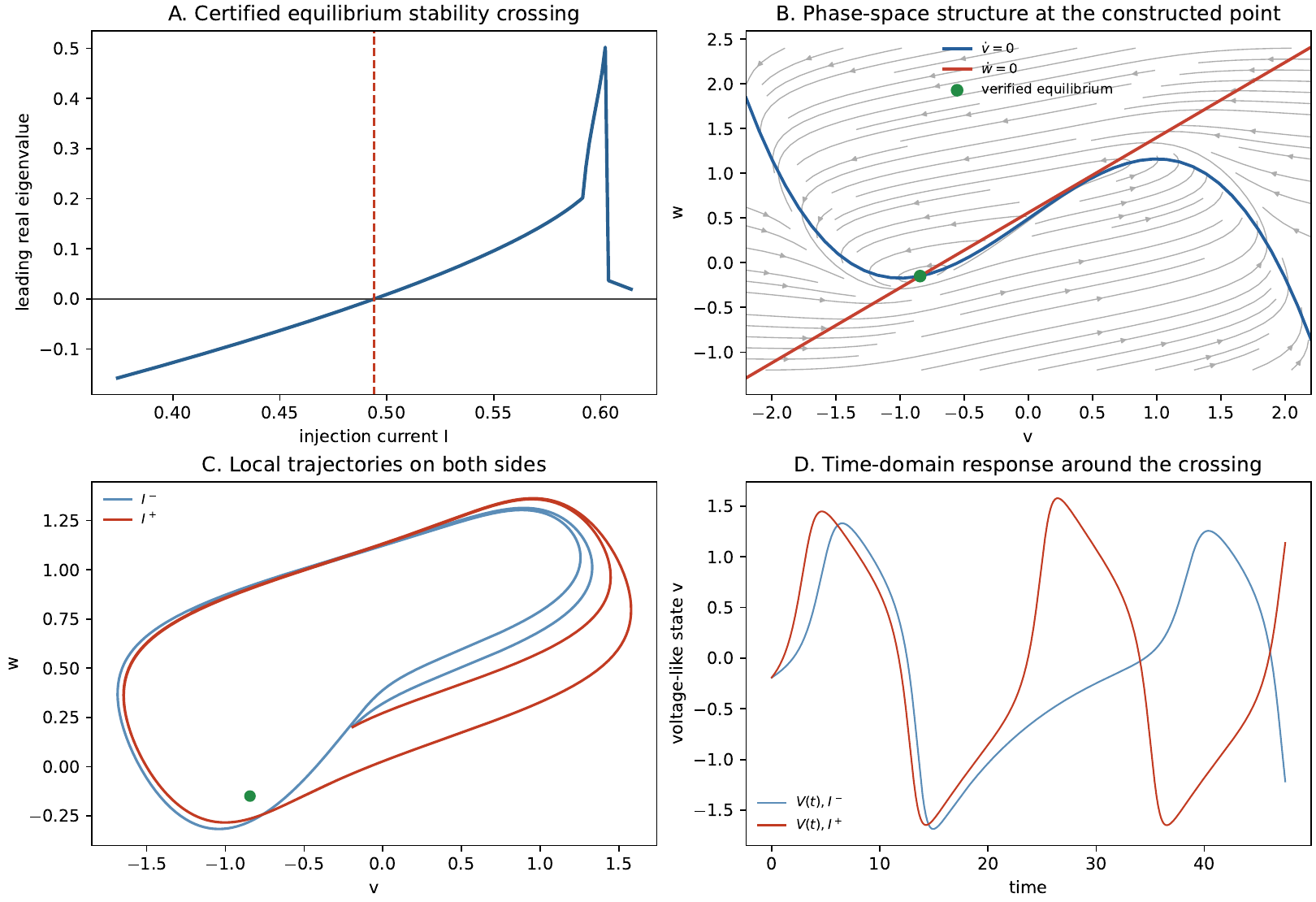}
\caption{FHN fixed-threshold transition certificate for a direct structure-aware construction. Panel A shows a transverse leading-real-eigenvalue crossing under a local change in injection current. Panel B shows the induced state-phase-plane streamlines, nullclines, and verified equilibrium at the constructed candidate. Panels C and D show local trajectories and voltage-like responses on the two sides of the crossing. This is dynamical validation of a constructed boundary point; it is not a geometric trajectory of the BBHT query reference through parameter space.}
\label{fig:fhn-certificate}
\end{figure*}

Table~\ref{tab:online-simulator-challenge} adds two true online propose-evaluate-update simulator loops. The stiff Robertson-type ODE and Kolmogorov-flow proxy are not saved-score replay in the classical rows: each classical proposal is evaluated by the simulator before the policy updates. The table is an online simulator challenge rather than a production costly-simulator benchmark; the measured wall times are sub-second and should not be overstated. Its value is that stiffness, solver failure, and boundary-like flow-transition penalties create non-smooth regions that can trap or disrupt continuous surrogate policies while a verified binary predicate remains well-defined. The BBHT time column remains a proxy, computed as BBHT query count multiplied by measured simulator-evaluation time, so it is not a quantum hardware runtime. At the rarest target density, the included non-quantum / BBHT query ratios are 2.89 [2.56, 3.24] for the stiff ODE and 3.81 [3.46, 4.21] for the flow proxy. These online results support the search-layer story while preserving the same oracle-realism caveat.

\begin{table*}[!htbp]
\centering
\caption{Online simulator challenge at the rarest target density. Classical rows are true online propose-evaluate-update loops. The BBHT time entry is a query-count proxy multiplied by measured simulator-evaluation time, not a quantum hardware runtime.}
\label{tab:online-simulator-challenge}
\small
\renewcommand{\arraystretch}{1.24}
\setlength{\tabcolsep}{4pt}
\begin{tabular}{@{}p{0.20\textwidth}c c c c@{}}
\toprule
Simulator & \shortstack[c]{Included non-quantum / BBHT\\query ratio} & \shortstack[c]{Included non-quantum / BBHT\\time-proxy ratio} & \shortstack[c]{Non-quantum\\wall time (s)} & \shortstack[c]{BBHT proxy\\time (s)} \\
\midrule
kolmogorov flow & \readableci{3.81}{3.46}{4.21} & \readableci{3.83}{3.48}{4.24} & \readableci{0.67}{0.65}{0.69} & 0.193 \\
stiff robertson & \readableci{2.89}{2.56}{3.24} & \readableci{2.94}{2.56}{3.35} & \readableci{0.49}{0.45}{0.53} & 0.184 \\
\bottomrule
\end{tabular}
\end{table*}

Table~\ref{tab:qpu-runtime-sanity} converts the online challenge into a wall-clock sanity bound for a hypothetical fault-tolerant quantum processing unit. If a logical oracle has depth $D_{\mathrm{oracle}}$ cycles and the logical clock is $f_{\mathrm{clock}}$, the quantum runtime proxy is $Q_{\mathrm{BBHT}}D_{\mathrm{oracle}}/f_{\mathrm{clock}}$. Aggregating the per-configuration bounds, matching the measured CPU wall time in the online rows would require only about 434--591 logical cycles per oracle at 10 kHz, 4340--5910 cycles at 100 kHz, or 43,401--59,101 cycles at 1 MHz. These thresholds are far below what one should assume for a reversible simulator, stability calculation, uncomputation, and error-corrected implementation without a dedicated resource estimate. The online challenge therefore strengthens the query-count story but does not imply end-to-end QPU wall-clock speedup.

\begin{table*}[!htbp]
\centering
\caption{QPU wall-clock sanity check for the online simulator challenge at the rarest target density. The maximum logical oracle depth is computed per configuration as $T_{\mathrm{classical}} f_{\mathrm{clock}} / Q_{\mathrm{BBHT}}$ and then aggregated, so the displayed cycle values need not equal the quotient of the rounded table means. These are not resource estimates for a compiled oracle.}
\label{tab:qpu-runtime-sanity}
\small
\renewcommand{\arraystretch}{1.24}
\setlength{\tabcolsep}{2.5pt}
\begin{tabular}{@{}p{0.18\textwidth}c c c c c c c@{}}
\toprule
Simulator & Configs & \shortstack[c]{BBHT\\queries} & \shortstack[c]{Classical\\wall time (s)} & \shortstack[c]{BBHT proxy\\time (s)} & \shortstack[c]{10 kHz\\cycles} & \shortstack[c]{100 kHz\\cycles} & \shortstack[c]{1 MHz\\cycles} \\
\midrule
kolmogorov flow & 50 & \readableci{12.50}{11.54}{13.53} & \readableci{0.67}{0.65}{0.69} & \readableci{0.19}{0.18}{0.21} & 591 & 5,910 & 59,101 \\
stiff robertson & 50 & \readableci{12.50}{11.40}{13.55} & \readableci{0.49}{0.45}{0.53} & \readableci{0.18}{0.17}{0.20} & 434 & 4,340 & 43,401 \\
\bottomrule
\end{tabular}
\end{table*}

\subsection{Question 4: When do oracle cost and state preparation erase the query margin?}

\begin{figure*}[!htbp]
\centering
\includegraphics[width=0.88\textwidth]{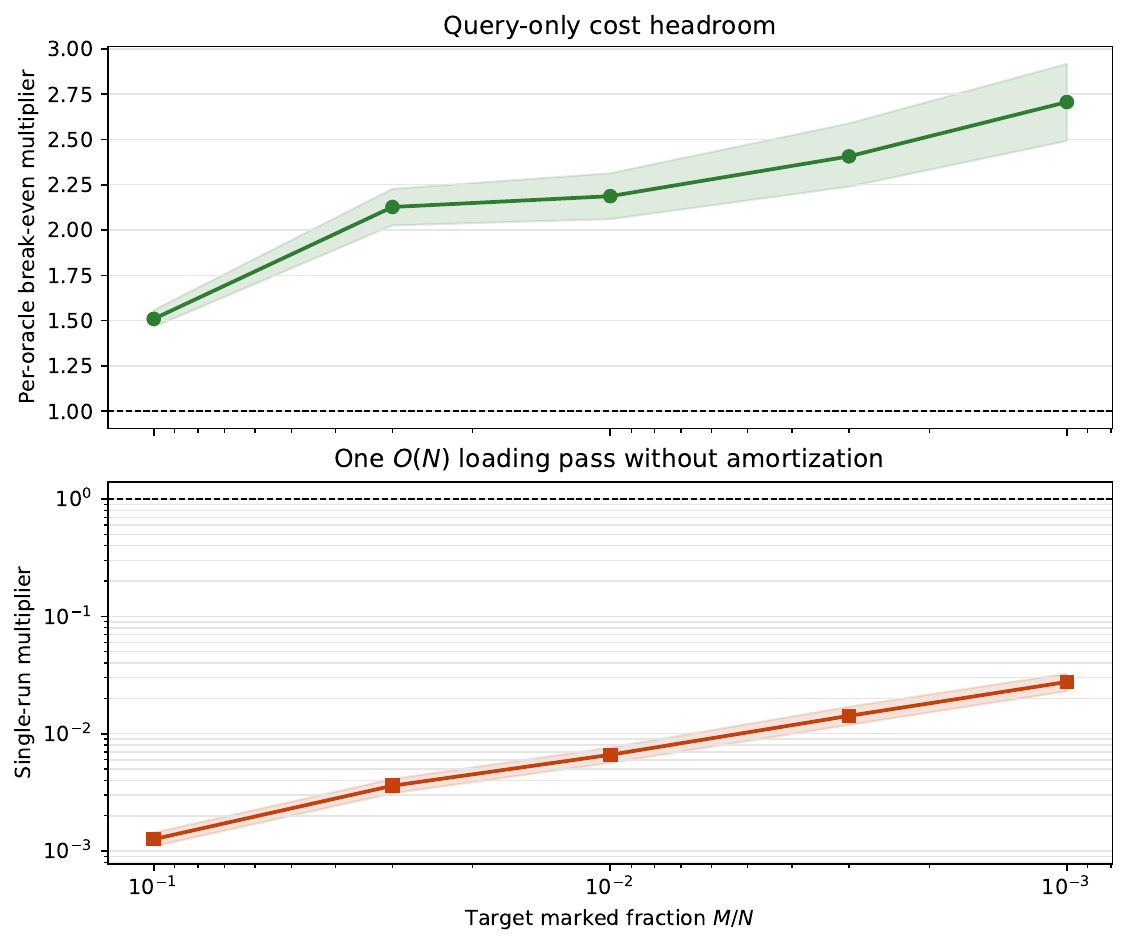}
\caption{Oracle break-even multipliers from the final confidence sweep. The upper panel reports the maximum per-oracle cost multiplier compatible with parity against the included non-quantum baseline. The lower panel adds one $O(N)$ candidate-loading pass to a single BBHT run. Values are cost-accounting boundaries, not simulator or quantum-hardware runtime measurements.}
\label{fig:oracle-break-even}
\end{figure*}

Table~\ref{tab:oracle-break-even} converts the main query ratios into an oracle-cost budget. At the rarest density in the final finite-pool sweep, the measured break-even multiplier is 2.71 [2.49, 2.92]: if a coherent quantum oracle call costs more than about 2.7 classical verified score checks, the reported query advantage no longer represents a total-cost advantage. The no-qRAM single-run multiplier is only 0.03 [0.02, 0.03], meaning that a one-time $O(N)$ candidate-loading cost would erase the single-run advantage. The amortization column reports how many repeated searches over the prepared candidate library would be needed before that one-time loading cost stops dominating. Figure~\ref{fig:oracle-break-even} displays the query-only and no-qRAM diagnostics separately so that the state-preparation penalty remains legible.

\begin{table*}[!htbp]
\centering
\caption{Oracle break-even interpretation of the final finite-pool sweep. The break-even multiplier is the maximum per-call quantum oracle cost, measured relative to one classical verified score check, that preserves parity with the included non-quantum baseline. The no-QRAM column adds one $O(N)$ state-preparation pass to the BBHT query count for a single run.}
\label{tab:oracle-break-even}
\footnotesize
\renewcommand{\arraystretch}{1.18}
\setlength{\tabcolsep}{4pt}
\begin{tabular}{@{}c c p{0.20\textwidth}p{0.24\textwidth}p{0.17\textwidth}@{}}
\toprule
$M/N$ & Configs & \shortstack[c]{Break-even\\multiplier} & \shortstack[c]{No-QRAM single-run\\multiplier} & \shortstack[c]{Median amortized\\runs} \\
\midrule
0.001 & 175 & 2.71 [2.49, 2.92] & 0.03 [0.02, 0.03] & 134.0 \\
0.003 & 175 & 2.41 [2.24, 2.59] & 0.01 [0.01, 0.02] & 247.9 \\
0.01 & 175 & 2.19 [2.06, 2.31] & 0.01 [0.01, 0.01] & 418.5 \\
0.03 & 175 & 2.13 [2.03, 2.23] & 0.00 [0.00, 0.00] & 788.1 \\
0.1 & 175 & 1.51 [1.46, 1.56] & 0.00 [0.00, 0.00] & 4438.3 \\
\bottomrule
\end{tabular}
\end{table*}

\subsection{Supporting diagnostics retained for completeness}

The following system, default/stress, learned-triage, and nuisance-dimension diagnostics preserve the broader evidence developed for the long arXiv version. They support interpretation and reproducibility, but they are not additional independent estimates of the four main questions above.

\paragraph{System-level behavior.}

Table~\ref{tab:system-rare} shows generic finite-pool policy behavior at the rarest target fraction. FHN has the largest included non-quantum / BBHT ratio, 5.20$\times$, while Lorenz and pendulum also show substantial gaps for this policy portfolio. These values profile how much the included methods exploit the saved score landscape; they do not measure intrinsic continuous-search difficulty. In particular, the direct constructions in Table~\ref{tab:fixed-threshold-online} route FHN and Lorenz away from the marked-set layer when their equations are exposed. Spring-mass-damper is intentionally retained as a smooth control case: BBHT still wins at $M/N=0.001$, but the margin is smaller, and at less rare fractions cross-entropy can match or beat BBHT. This boundary case is scientifically useful because it identifies where adaptive classical structure exploitation is competitive. Figure~\ref{fig:fhn-certificate} provides the corresponding phase-space validation for the FHN direct control.

This system-level behavior is important for the paper's interpretation. The result is not that quantum query search erases the value of modeling structure. Rather, the result is that the rare-set query model gives a robust advantage when the positive set is small and not fully exploited by adaptive classical geometry. Smooth boundaries can still be found efficiently by classical adaptive samplers. That observation is a feature of the analysis, not a defect.

The system table also identifies where future work should focus. FHN shows the largest rare-fraction gap, demonstrating that the implemented generic adaptive baselines do not fully exploit the score landscape at low density. The direct FHN parameterization makes the complementary point: a system-specific boundary solver can remove the need for generic search entirely. Lorenz and pendulum have similarly strong finite-pool reductions but exposed structure. Coupled FHN is less easily reduced and therefore remains a useful root-following case. Spring-mass-damper is the methodological control case. Its rare-fraction result remains favorable to BBHT, but the smaller margin warns against treating all stability-boundary search as unstructured search. In a smooth, low-dimensional system, adaptive classical search can be excellent.

\begin{table*}[!htbp]
\centering
\caption{System-level offline controlled sweep behavior at $M/N=0.001$. The named non-quantum method is the most frequent per-configuration finite-pool winner, while the reported non-quantum burden remains the per-configuration best included value. This is a query-policy replay table, not an end-to-end simulator-runtime comparison.}
\label{tab:system-rare}
\footnotesize
\renewcommand{\arraystretch}{1.20}
\setlength{\tabcolsep}{3pt}
\begin{tabular}{@{}p{0.18\textwidth}c p{0.17\textwidth}c c c@{}}
\toprule
System & \shortstack[c]{BBHT\\queries} & \shortstack[c]{Most frequent\\included method} & \shortstack[c]{Included non-quantum\\queries} & \shortstack[c]{Included non-quantum\\ / BBHT} & \shortstack[c]{BBHT\\win rate} \\
\midrule
Coupled FHN & \cientry{51.08}{49.52}{52.72} & cross entropy & \cientry{101.12}{88.14}{117.38} & \cientry{1.98}{1.74}{2.28} & 1.00 \\
Duffing & \cientry{52.98}{51.28}{54.60} & cross entropy & \cientry{119.69}{107.69}{133.46} & \cientry{2.26}{2.04}{2.51} & 1.00 \\
FitzHugh-Nagumo & \cientry{50.77}{49.20}{52.27} & subset & \cientry{262.99}{252.78}{272.73} & \cientry{5.20}{4.97}{5.44} & 1.00 \\
Lorenz Hopf Boundary & \cientry{52.08}{50.12}{54.12} & cross entropy & \cientry{153.43}{141.21}{165.45} & \cientry{2.96}{2.73}{3.20} & 1.00 \\
Windy Pendulum & \cientry{51.27}{49.01}{53.44} & subset & \cientry{173.55}{146.33}{203.74} & \cientry{3.44}{2.88}{4.09} & 1.00 \\
Spring-Mass-Damper & \cientry{50.92}{49.08}{52.88} & cross entropy & \cientry{65.57}{59.22}{72.06} & \cientry{1.29}{1.17}{1.40} & 0.88 \\
Van der Pol & \cientry{53.57}{51.76}{55.51} & cross entropy & \cientry{96.51}{90.96}{101.70} & \cientry{1.81}{1.70}{1.93} & 1.00 \\
\bottomrule
\end{tabular}
\end{table*}

\paragraph{Default and stress results.}

The default and stress runs include Bayesian LCB in addition to scalable baselines. Table~\ref{tab:default-stress} shows that BBHT is favorable across most systems, while the spring benchmark demonstrates that a smooth control boundary can favor cross-entropy. Bayesian LCB is retained here as an expensive adaptive comparator. The later access-model ablation extends this scalar-score comparison with GP active-search and level-set diagnostics over the full 875-base-configuration design.

The default and stress tables serve a different purpose from the controlled density sweep. They show concrete benchmark instances with fixed system thresholds. These runs are closer to how a researcher would initially define a criticality criterion for a particular simulator. The stress runs then enlarge the source candidate pools for FHN, coupled FHN, Duffing, and Lorenz. They are not the main statistical claim, but they protect the paper from relying only on small default candidate sets.

\begin{table*}[!htbp]
\centering
\caption{Validated default and stress benchmark summaries. Query counts are means to first marked configuration. The default run includes Bayesian LCB; the final 875-configuration confidence sweep omits Bayesian LCB for scalability and uses CE/subset as the scalable adaptive non-quantum comparators.}
\label{tab:default-stress}
\footnotesize
\renewcommand{\arraystretch}{1.16}
\setlength{\tabcolsep}{3pt}
\begin{tabular}{@{}p{0.09\textwidth}p{0.15\textwidth}r c r r p{0.19\textwidth}r@{}}
\toprule
Run & System & $N$ & $M/N$ & Random & \shortstack[c]{Bayesian\\LCB} & \shortstack[c]{Selected scalable\\non-quantum} & BBHT \\
\midrule
Default & FHN & 20,000 & 0.00585 & 159.6 & 106.1 & 81.2 Subset & 20.2 \\
Default & Coupled FHN & 10,000 & 0.01350 & 78.4 & 62.2 & 35.9 Subset & 13.9 \\
Default & Van der Pol & 12,000 & 0.00392 & 221.5 & 70.2 & 46.2 CE & 25.9 \\
Default & Duffing & 20,000 & 0.01815 & 47.7 & 17.1 & 23.3 CE & 12.0 \\
Default & Pendulum & 16,000 & 0.03413 & 27.7 & 15.4 & 11.7 CE & 8.6 \\
Default & Spring & 12,000 & 0.01242 & 76.1 & 24.4 & 8.5 CE & 13.5 \\
Default & Lorenz & 16,000 & 0.00256 & 337.4 & 124.9 & 69.2 CE & 32.2 \\
\midrule
Stress & FHN & 100,000 & 0.00349 & 251.0 & 156.2 & 121.6 Subset & 26.1 \\
Stress & Coupled FHN & 40,000 & 0.00817 & 108.7 & 69.9 & 42.7 CE & 16.7 \\
Stress & Duffing & 100,000 & 0.01820 & 51.5 & 15.0 & 23.4 CE & 11.1 \\
Stress & Lorenz & 100,000 & 0.00139 & 693.6 & 159.5 & 106.8 CE & 41.1 \\
\bottomrule
\end{tabular}
\end{table*}

\paragraph{Learned triage diagnostics.}

\begin{figure*}[!htbp]
\centering
\includegraphics[width=0.70\textwidth]{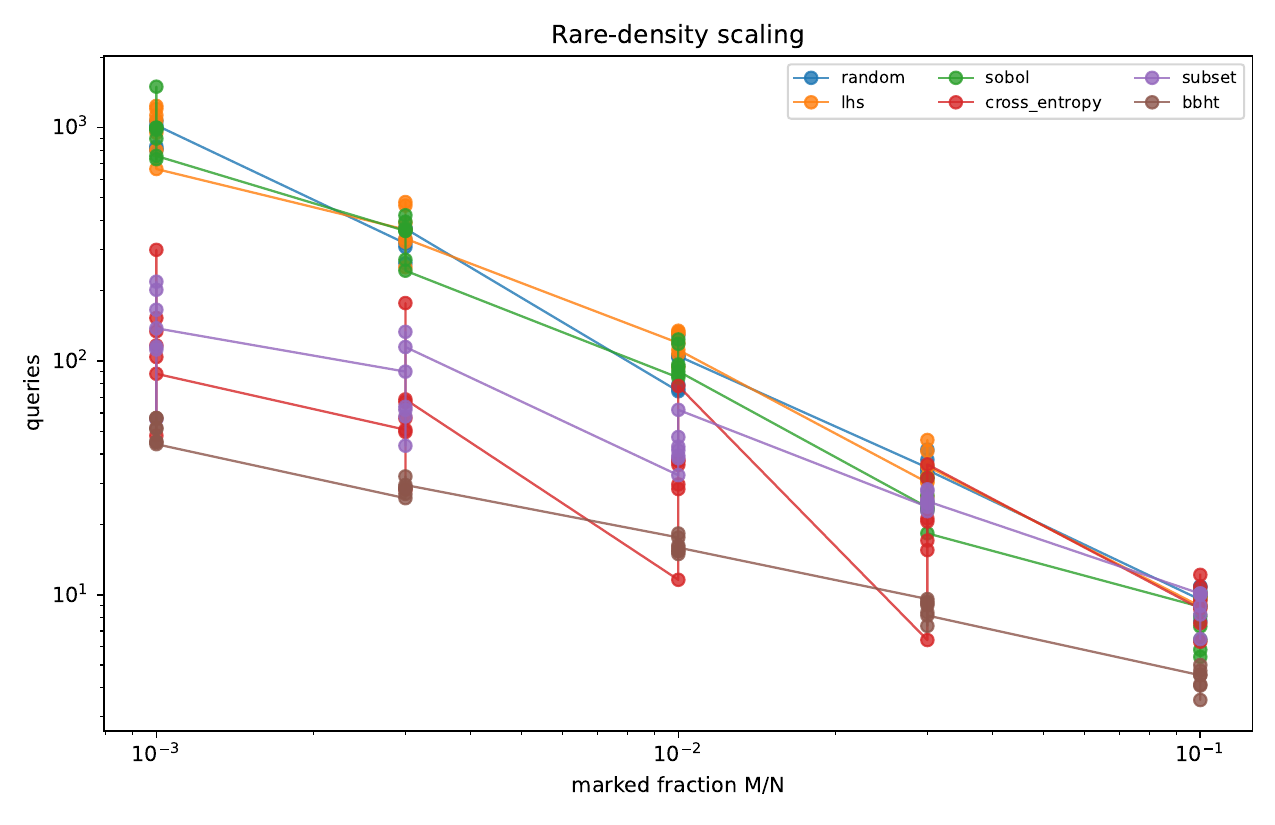}
\caption{Raw method-level query counts from the earlier scaling study, retained as a supporting diagnostic. Unlike the system-resolved ratios in Figure~\ref{fig:final-speedups} and the paired uncertainty in Figure~\ref{fig:hierarchical-density}, this view shows how the individual search policies' absolute query burdens change with rarity. Table~\ref{tab:final-by-fraction} remains the primary replicated estimate.}
\label{fig:density}
\end{figure*}

Learned models are not central to the proof of query advantage. Table~\ref{tab:learned-triage} reports MLP learned-triage diagnostics on the default datasets with explicit accounting for training, validation, test, and ranking-verification labels. Several systems show strong ranking, but FHN and Lorenz have weak classifier precision/recall despite good ability to surface a verified positive early. The total train-plus-verify column is therefore the honest cost view: a model that finds a positive after one ranked verification may still have consumed thousands of simulator-derived labels to train and select the ranking function. We treat learned models as ranking and triage tools followed by simulator verification, not as replacements for exact criticality scoring. Figure~\ref{fig:density} provides an additional visualization of rare-density scaling from the earlier scaling study.

This conservative interpretation matters because rare-event classifiers can be misleading. A classifier may have high AUROC while still producing poor precision at the rare threshold. A ranking model may surface one true positive early while still being poorly calibrated as a binary oracle. In a quantum setting, false labels are especially dangerous because an oracle error is not just a harmless prediction mistake; it changes the marked set. QC-PHAST therefore treats learned models as triage and proposal tools only. The verified simulator-derived score remains the authority.

\begin{table*}[!htbp]
\centering
\caption{Learned triage accounting for the default benchmark datasets. These models are ranking diagnostics only; the final marked predicate remains the simulator-derived score. Training, validation, and test labels are counted as simulator-derived labels rather than treated as free.}
\label{tab:learned-triage}
\footnotesize
\renewcommand{\arraystretch}{1.18}
\setlength{\tabcolsep}{3pt}
\begin{tabular}{@{}p{0.16\textwidth}r r r r r r r r r@{}}
\toprule
System & Positives & Train & Val & Test & Precision & Recall & AUPRC & Verify & \shortstack[c]{Train +\\verify} \\
\midrule
FHN & 117 & 13000 & 3500 & 3500 & 0.200 & 0.476 & 0.222 & 1 & 16501 \\
Coupled FHN & 135 & 6500 & 1750 & 1750 & 0.957 & 0.917 & 0.980 & 1 & 8251 \\
Van der Pol & 47 & 7800 & 2100 & 2100 & 0.667 & 0.750 & 0.897 & 1 & 9901 \\
Duffing & 363 & 13000 & 3500 & 3500 & 0.984 & 0.938 & 0.994 & 1 & 16501 \\
Pendulum & 546 & 10400 & 2800 & 2800 & 0.515 & 0.698 & 0.648 & 1 & 13201 \\
Spring & 149 & 7800 & 2100 & 2100 & 0.828 & 0.923 & 0.962 & 1 & 9901 \\
Lorenz & 41 & 10400 & 2800 & 2800 & 0.500 & 0.429 & 0.401 & 3 & 13203 \\
\bottomrule
\end{tabular}
\end{table*}

\paragraph{Dimension-scaling diagnostic.}

\begin{figure*}[!htbp]
\centering
\includegraphics[width=0.70\textwidth]{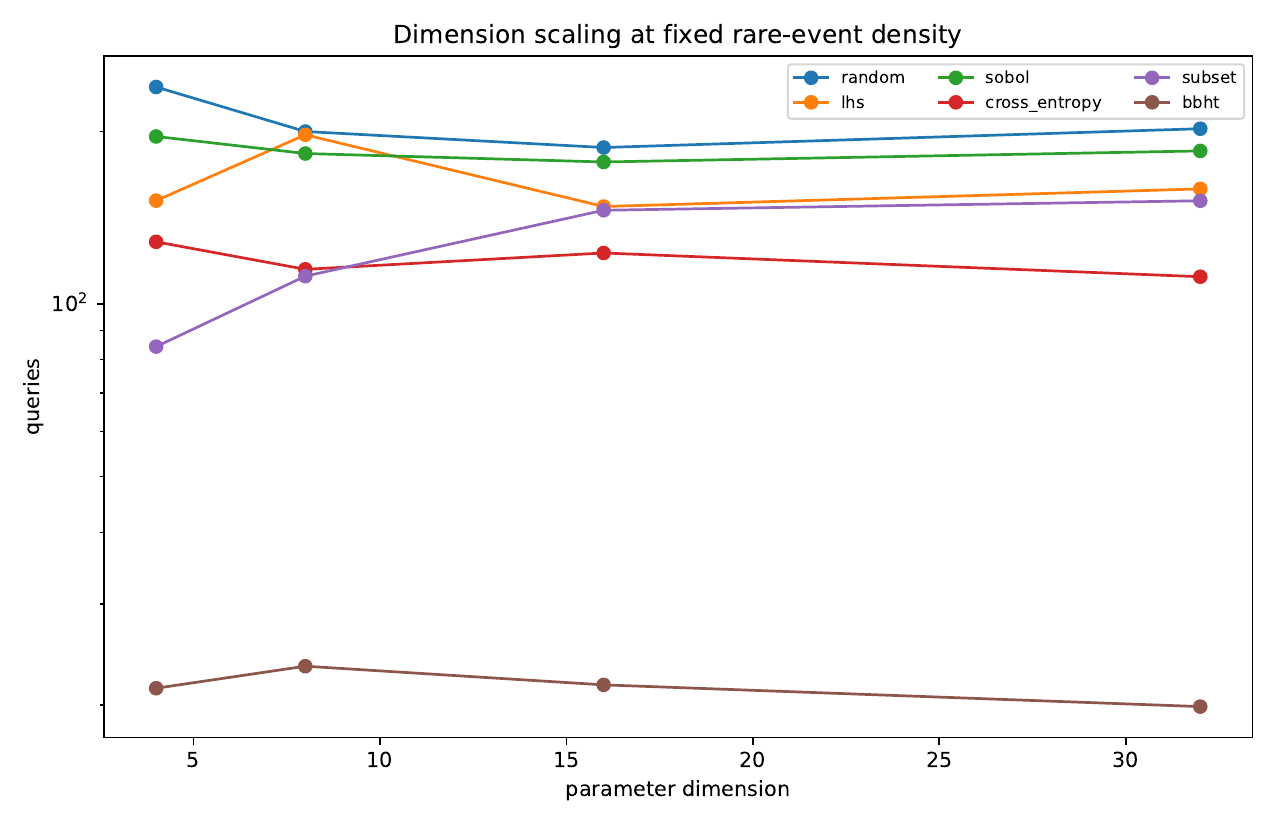}
\caption{Embedded FHN dimension-scaling diagnostic. The added dimensions are nuisance coordinates; the purpose is to check query-policy robustness under wider finite candidate representations.}
\label{fig:dimension}
\end{figure*}

The dimension-scaling study embeds FHN into wider candidate representations while keeping the score tied to the original excitable-dynamics variables. Table~\ref{tab:dimension-scaling} and Figure~\ref{fig:dimension} show that the BBHT query count remains near 20--23 queries at fixed marked fraction $M/N=0.005$, while classical baselines vary with the widened representation. This is a diagnostic, not a claim that arbitrary high-dimensional scientific simulators are solved.

The correct interpretation is that irrelevant dimensions can dilute simple space-filling strategies, while adaptive methods may or may not recover the active structure. Because the score still depends on the embedded FHN variables, this is not a proof of arbitrary high-dimensional success. It is a controlled nuisance-dimension test that belongs in the long arXiv version because it clarifies the limits of the final confidence sweep.

\begin{table*}[!htbp]
\centering
\caption{Embedded FHN dimension-scaling diagnostic at fixed target density. The added coordinates are nuisance dimensions; this table checks whether the finite-candidate query comparison remains stable when the candidate representation is widened.}
\label{tab:dimension-scaling}
\footnotesize
\renewcommand{\arraystretch}{1.18}
\setlength{\tabcolsep}{3pt}
\begin{tabular}{lrrrrrrrrr}
\toprule
Case & $N$ & $M$ & $M/N$ & Random & LHS & Sobol & CE & Subset & BBHT \\
\midrule
Embedded FHN 4D & 32,768 & 164 & 0.00500 & 239.2 & 151.5 & 196.1 & 128.4 & 84.3 & 21.4 \\
Embedded FHN 8D & 32,768 & 164 & 0.00500 & 200.0 & 197.4 & 183.1 & 115.0 & 111.9 & 23.4 \\
Embedded FHN 16D & 32,768 & 164 & 0.00500 & 187.6 & 148.0 & 177.0 & 122.8 & 145.8 & 21.7 \\
Embedded FHN 32D & 32,768 & 164 & 0.00500 & 202.3 & 158.9 & 185.0 & 111.6 & 151.5 & 19.9 \\
\bottomrule
\end{tabular}
\end{table*}

Table~\ref{tab:embedded-fhn-32d-continuous} adds the corresponding off-grid continuous stress test for the 32-dimensional embedded FHN diagnostic. The run uses 100 independent threshold calibrations across five marked fractions, giving 500 configurations, and lets continuous challengers propose points directly in the widened unit cube. At the rarest fraction, the best implemented continuous challenger is rank-adaptive Gaussian search with a budget-censored mean of 113.99 evaluations and 0.30 success rate, compared with 51.00 BBHT reference queries. This result does not prove that all high-dimensional continuous optimizers fail; it shows that the implemented off-grid challengers did not erase the finite-pool query reference on this embedded rare-boundary diagnostic. Production BoTorch/TuRBO and CMA-ES packages remain distinct comparators unless they are installed and run as named methods.

\begin{table*}[!htbp]
\centering
\caption{32-dimensional embedded-FHN continuous stress test. Continuous challengers propose off-grid points in the widened unit cube and are compared with a finite-pool BBHT reference on the same target marked fraction. First-hit means are budget-censored, and method names report the implemented algorithms rather than unrun TuRBO or CMA-ES baselines.}
\label{tab:embedded-fhn-32d-continuous}
\small
\renewcommand{\arraystretch}{1.24}
\setlength{\tabcolsep}{3pt}
\begin{tabular}{@{}c c p{0.20\textwidth}c c c c@{}}
\toprule
$M/N$ & Configs & Best continuous method & \shortstack[c]{Best continuous\\evaluations} & \shortstack[c]{Success\\rate} & \shortstack[c]{BBHT reference\\evaluations} & \shortstack[c]{Continuous\\ / BBHT} \\
\midrule
0.001 & 100 & rank adaptive gaussian & \readableci{113.99}{110.92}{116.58} & 0.30 & \readableci{51.00}{50.02}{52.03} & \readableci{2.25}{2.19}{2.32} \\
0.003 & 100 & rank adaptive gaussian & \readableci{97.57}{93.68}{101.47} & 0.50 & \readableci{29.72}{29.15}{30.29} & \readableci{3.32}{3.17}{3.47} \\
0.01 & 100 & rank adaptive gaussian & \readableci{62.94}{58.73}{67.31} & 0.85 & \readableci{16.13}{15.84}{16.43} & \readableci{3.93}{3.68}{4.20} \\
0.03 & 100 & rank adaptive gaussian & \readableci{29.73}{27.49}{32.11} & 0.99 & \readableci{8.88}{8.71}{9.06} & \readableci{3.38}{3.13}{3.66} \\
0.1 & 100 & sobol online & \readableci{8.93}{8.07}{9.85} & 1.00 & \readableci{4.39}{4.29}{4.48} & \readableci{2.07}{1.85}{2.30} \\
\bottomrule
\end{tabular}
\end{table*}

\FloatBarrier

\section{Discussion}

The final evidence supports a narrow but strong conclusion. \qcphast{} is useful when the scientific task can be represented as finite-candidate rare-regime discovery, when the marked set is small enough that query efficiency matters, and when the predicate is accurate enough to support a marked-oracle abstraction. Grover/BBHT-style marked-set query search provides a query-model advantage that grows as the marked fraction decreases under exact access. Adaptive classical methods remain important; cross-entropy, subset-style search, GP active-search diagnostics, and smooth geometry can substantially reduce queries relative to random search and sometimes beat BBHT on easy regimes.

This balanced result is preferable to a universal-superiority claim. The spring-mass-damper benchmark and the synthetic ball geometry show that if the critical set is easy for a classical adaptive sampler, classical search can be competitive or superior. The FHN, Lorenz, Duffing, pendulum, Van der Pol, coupled FHN, and fragmented synthetic geometries show that the generic finite-pool policy gap is robust across a broader suite of nonlinear and boundary-like benchmarks. The direct controls show equally clearly that equation-aware structure can invalidate the need for that finite-pool layer.

\subsection{Contribution and supported regime}

\paragraph{What is inherited and what QC-PHAST contributes.}

The square-root marked-set query law is inherited from Grover and BBHT; reproducing it against random search is not the novelty. The contribution begins where that theorem stops. \qcphast{} makes the scientific reduction, access contract, comparator portfolio, threshold calibration, predicate error, finite-budget completeness, and oracle-cost headroom jointly auditable. It then tests the resulting claim across matched density sweeps and deliberately stronger access regimes. The scalable score-guided finite-pool replay ratio 2.71, the stronger scalar-score GP ratio 2.24, the search-only pilot GP stress ratio 1.36, the fully charged 1,750-configuration calibration ledger at 2.62, and the noisy-predicate reversals of 0.29 and 0.17 under their distinct access portfolios are parts of one regime map rather than independent headline claims.

This design changes the scientific question from ``does Grover beat random search?'' to ``after a simulator-derived first-hit task is defined, which information and cost assumptions preserve any reason to use a marked-set query layer?'' The answer is not encoded in the BBHT theorem. It depends on whether score feedback exposes geometry, whether the threshold can be specified without full-pool inspection, whether false positives overwhelm rare true marks, whether a structure-aware solver reaches the target directly, and whether coherent access fits inside the measured break-even budget. The benchmark and propositions quantify those dependencies.

\paragraph{QC-PHAST as evidence-gated algorithm selection.}

QC-PHAST should be understood as a transparent policy router over heterogeneous access models. Its decision state $c$ is populated by pilot observations and scientific metadata; its gates reject invalid interpretations before policies are ranked; and its objective chooses among admissible policies by p90 first-hit burden or measured total cost at fixed verification reliability. This is an ML contribution in the same broad sense as algorithm selection and active search: performance depends on matching an acquisition policy to instance characteristics and information access. It is not, however, a learned meta-selector. The present seven-system suite is too small and too structured to justify training a general router, and doing so would require held-out simulator families to avoid learning benchmark identity.

This distinction suggests a concrete future evaluation. A learned router would need a larger meta-dataset of simulator families, pre-registered routing features, train/validation/test separation by system family, regret relative to a virtual best admissible policy, and explicit feature-acquisition cost. Until that experiment exists, the deterministic gates in Algorithm~\ref{alg:qcphast} are more defensible: every route can be traced to a measured property such as smooth-structure success, $\widehat\eta_{\mathrm{fp}}/\widehat p$, or $\mu^\star$.

\paragraph{When the marked-set branch helps.}

The evidence suggests that the marked-set query branch is most compelling when five conditions hold. First, the scientific question can be stated as first discovery of at least one acceptable rare candidate. Second, the candidate pool is finite or can be made finite through a defensible design. Third, the score threshold creates a genuinely small marked fraction. Fourth, the marked set is not so geometrically simple that a classical adaptive sampler or equation-aware construction can find it almost immediately. Fifth, the marked predicate is exact, or its false-positive mass is small relative to $p$ and exact rejection remains affordable. Fragmented synthetic geometries fit this pattern better than smooth balls; FHN and Lorenz satisfy it only when their direct equation-level structure is unavailable to the search policy.

These conditions are practical rather than philosophical. A researcher can test them before claiming advantage. Generate a candidate pool, score it on a pilot subset or saved simulator run, estimate the marked fraction, run classical adaptive baselines, and only then interpret the BBHT query reference. If the classical baselines already find positives in a handful of queries, the problem may not need quantum search. If the marked fraction is tiny and adaptive baselines still need many verifications, the query-model comparison becomes more meaningful.

\subsection{Classical, threshold, and replay boundaries}

\paragraph{When classical adaptive search can win.}

The spring-mass-damper and smooth-ball results should not be hidden. They are useful negative controls. The spring score is smooth and directly tied to low-dimensional boundaries, while the ball geometry is exactly the kind of coherent target set that score-feedback methods can exploit. Cross-entropy, subset-style search, and GP active-search diagnostics can exploit that structure, and at less rare target fractions they can match or beat BBHT in practical finite settings. This does not contradict the query theory because these methods are not operating in an unstructured oracle model. They use score feedback, geometry, and surrogate assumptions.

The accurate statement is therefore narrower than a generic quantum-versus-classical comparison: under an exact finite marked-set query model, BBHT gives a square-root query reference, and in the \qcphast{} benchmark suite this reference is empirically favorable against the included finite-pool non-quantum baseline in most rare-regime configurations. The access and geometry ablations then identify the boundary cases: noisy predicates and smooth coherent targets can erase the advantage or drive the ratio below one. The narrower statement is more credible because it predicts its own failures.

\paragraph{When QC-PHAST should reject marked-set search.}

\qcphast{} should route directly to structure-aware non-quantum search when the score is an analytic coordinate distance, a monotone boundary, or a smooth low-dimensional surface that root finding, continuation, adaptive mesh refinement, or continuous BO can exploit. The direct controls make this concrete: Van der Pol, spring, pendulum, FHN, Duffing, and Lorenz have analytic constructions under their declared fixed-threshold conditions, while coupled FHN admits a root-following construction. The protocol should also reject a surrogate-only marked predicate when the estimated label error is comparable to the marked fraction. In that regime, the noisy-oracle phase diagram shows that false positives can dominate the amplified target set. Finally, the protocol must retain only a query-count interpretation when state preparation, reversible predicate synthesis, or qRAM-style loading exceed the oracle break-even multiplier.

These negative rules are part of the method. A practical user should first ask whether the transition has exploitable analytic structure, whether the predicate is verified, and whether the oracle-cost multiplier is physically plausible. If any answer is unfavorable, the correct conclusion is a classical workflow or a query-only diagnostic, not a quantum-advantage claim.

\paragraph{Threshold specification is part of the task.}

The quantile-controlled confidence sweep conditions on a chosen marked fraction. It answers: given a finite score landscape and a threshold that induces density $p$, how do the included policies scale with rarity? It does not answer how many simulator evaluations are required to discover or estimate that threshold. Full-pool quantile calibration would be circular in an online deployment because it reveals every score before search. The default fixed-threshold experiments and the pilot-threshold ablation bracket this distinction: the former represent scientifically specified criticality bands, while the latter charges finite pilot information to an uncertain empirical threshold.

Proposition~\ref{prop:pilot-coverage} shows that ultra-rare quantiles require pilot sizes on the same order as unstructured first discovery merely to observe the target tail once. Consequently, a practical application should prefer a threshold derived from physics, safety tolerances, or an independently validated calibration dataset. If no such threshold exists, the correct workflow may be sequential level-set estimation or rare-event probability estimation rather than QC-PHAST first-hit search. The marked fraction should then be reported as an outcome with uncertainty, not chosen after inspecting the evaluation pool.

\paragraph{Saved-score replay versus online discovery.}

Saved-score replay is valuable because every policy faces the same latent landscape and thousands of stochastic trials can be audited without simulator drift. Its cost is interpretive: generating the score bank is an offline expense that the first-hit query counter does not recover. The resulting ratios estimate policy behavior conditional on a prepared finite instance. They are appropriate for comparing acquisition rules and validating density trends, but they are not measurements of end-to-end simulator savings.

The online simulator challenge restores the propose--evaluate--update loop for two compact systems and records solver time, failures, and policy updates. Those experiments support the qualitative routing story, yet their sub-second evaluations remain far from production PDE or multiphysics workloads. A decisive end-to-end study would need an independently expensive simulator, lazy scoring only, matched continuous and finite-pool policies, repeated verified first-hit trials, and a compiled oracle resource model. The present paper identifies that study's acceptance criteria rather than claiming it has already been completed.

\paragraph{Why included non-quantum comparison matters.}

The final tables deliberately compare against the included non-quantum baseline per configuration. This is stricter than comparing against random search. In practical scientific computing, a researcher would not use random search if cross-entropy or Bayesian optimization reliably worked better. The manuscript therefore treats random search as a theoretical and intuitive reference, while the included non-quantum ratio is the fairness-critical number.

\subsection{Predicate and learned-surrogate boundaries}

\paragraph{Why oracle fidelity matters.}

The noisy-predicate ablation is the harshest result in the paper. At $M/N=0.001$, a 5\% observed-label flip rate is far larger than the true marked fraction. A search policy can then spend most of its budget chasing false positives. This matters more for quantum search than for ordinary classification because a wrong oracle changes the amplitude-amplification target itself. The practical implication is simple: \qcphast{} should never treat learned or noisy labels as an authoritative final oracle. Exact post-measurement verification is necessary but does not automatically restore the query margin when observed-mark purity is low. A learned model may rank candidates, compress a candidate pool, or propose a nonuniform state-preparation distribution in future work, but the final marked predicate must be tied to a verified simulator-derived score.

\paragraph{How to interpret learned triage.}

Learned triage is useful when a model can rank promising candidates and reduce verification burden. However, rare-event classification can be poorly calibrated, and high AUROC does not guarantee high precision at the rare threshold. The current learned diagnostics show this clearly. Some systems have strong learned classification performance, while others mainly benefit through early ranking. Therefore, the simulator-derived score remains the source of truth.

\subsection{Operational use and future work}

\paragraph{How to use QC-PHAST in a simulator workflow.}

In a practical simulator study, \qcphast{} should be used as a staged workflow rather than a single black-box algorithm. The first stage is scientific definition: decide what transition or rare regime matters and define a score that is physically interpretable. The second stage is a charged pre-flight structure test; if an analytic construction, root solver, continuation method, or continuous search finds a verified candidate, the marked-set branch is unnecessary. The third stage, used only when a finite-pool comparison remains scientifically justified, is candidate design and policy evaluation under explicit access contracts. The fourth stage is authoritative verification of every returned candidate with the original simulator or stability calculation.

This staged use is especially important for medical or biological applications. A canonical excitable-system benchmark is not clinical evidence. If QC-PHAST is later applied to a biomedical simulator, the score would need domain validation, patient-level data governance if patient data are used, leakage controls, endpoint definition, uncertainty analysis, and external validation. The current paper establishes a query-efficiency framework, not a clinical decision system.

\paragraph{Possible extensions.}

The most natural technical extension is structured amplitude amplification. The current BBHT layer uses a uniform finite-candidate oracle. Classical proposal distributions or learned rankings could, in principle, define nonuniform candidate preparation with higher initial marked probability, after which amplitude amplification could reduce query cost further. This paper does not implement that extension because state preparation and oracle synthesis would need their own assumptions. The idea is promising, but it should be developed as a separate theoretical and experimental contribution.

The present experiments address fixed-objective finite-pool discovery: the candidate set, score, and threshold remain fixed during each run. Dynamic environments and changing targets would require online candidate updates, time-dependent predicates, or trajectory-level objectives. Those extensions are compatible with the broader phase-space search motivation but are not implemented in this manuscript.

A second extension is targeted classical baseline expansion. The present work already includes analytic constructions, coupled-FHN root following, adaptive mesh, differential evolution, rank-Gaussian proposals, and continuous GP-LCB as routing controls; the default and stress runs also include Bayesian LCB. Future work could add full continuation packages, richer adaptive meshes, or system-specific bifurcation solvers for systems where smooth-score access is available. Those methods would answer a different question: not "how hard is marked-set search under a black-box predicate?" but "how much can system-specific smoothness reduce the need for search?" Both questions are valuable, but they should not be mixed without care.

The geometry ablation makes this extension concrete. For smooth balls and other coherent level sets, specialized classical methods are not optional; they are the correct competitors. For fragmented or thin boundary-like marked sets, finite-pool search remains a meaningful abstraction. This distinction should remain explicit because the classical-friendly cases are part of the evidence, not exceptions to hide.

\subsection{Oracle realism and end-to-end cost}

The largest gap between the query model and a physical quantum advantage is the marked-set oracle. BBHT assumes coherent access to a predicate that flips phase or marks amplitude when $s(x)\leq\tau$. For the analytic scores in this paper, that predicate is easy to evaluate classically after scores have been generated. For a quantum device, however, the predicate would need a reversible implementation of the relevant arithmetic, equilibrium calculation, stability test, surrogate verifier, or simulator approximation. For richer simulators, the cost may include state preparation, finite-precision arithmetic, reversible linear algebra, memory access, QRAM-like data loading, uncomputation, and error-correction overhead.

The reported ratios therefore have a precise interpretation: they are query-count ratios under a shared finite marked-predicate abstraction. They are not end-to-end runtime ratios. If each quantum oracle call costs far more than a classical verification, or if loading the candidate set and scores dominates the computation, the practical advantage can disappear. The break-even table quantifies this boundary rather than leaving it qualitative. A 2--3$\times$ query ratio allows only a 2--3$\times$ oracle-cost multiplier before parity is lost, and a single $O(N)$ state-preparation pass without amortization can cancel the advantage. Conversely, if a scientific workflow already exposes a cheap verified predicate or a reversible surrogate with acceptable error, the query-count layer becomes more relevant. This paper does not solve that engineering problem; it identifies when the search layer would be worth caring about before oracle construction costs are paid.

This cost separation is also why the benchmark uses canonical score functions. The goal is to isolate search behavior from oracle engineering. Treating oracle synthesis as future work is not meant to make it easy; it marks the boundary between a query-complexity benchmark and a hardware-implementation paper. A full hardware claim would need a resource estimate with gate counts, precision analysis, memory model, failure probability, and comparison against wall-clock classical optimization.

For a reversible numerical oracle, even a simple fixed-point explicit integrator would have cost that scales with the number of integration steps $L$, state dimension $m$, precision $b$, right-hand-side arithmetic cost $C_f(b)$, comparison cost $C_{\mathrm{cmp}}(b)$, and uncomputation. A schematic bound is
\begin{equation}
    C_{\mathrm{oracle}} = O\!\left(L\{C_f(b)+m b^2\}+C_{\mathrm{cmp}}(b)\right)
\end{equation}
Toffoli-like arithmetic operations for Euler-style stepping, with Runge--Kutta variants multiplying the right-hand-side evaluations and accumulation work. Clifford+T synthesis, magic-state factories, memory access, and error correction add further architecture-dependent overhead. This formula is not a compiled circuit estimate; it is included to show why a measured 2--3$\times$ query reduction leaves little room for an expensive coherent simulator oracle.

\section{Limitations}
\label{sec:limitations}

\paragraph{Query model.}
The BBHT results are query-count simulations under a finite marked-predicate model. End-to-end quantum runtime depends on oracle construction, state preparation, data loading, and hardware resource costs discussed above.

\paragraph{Noisy predicates.}
The access-model ablation shows that predicate noise can drive the rare-density ratio below one and reverse the comparison. The positive query-model claim therefore requires an exact marked predicate; post-measurement verification is retained in the noisy stress tests but does not establish noisy-oracle robustness.

\paragraph{Finite candidate sets.}
The theory and experiments apply to generated finite candidate pools. Continuous parameter spaces require candidate generation or discretization before the marked-set oracle is defined. The paper therefore does not claim global continuous optimization or exact bifurcation continuation. It claims query efficiency after a candidate set and a thresholded score have been specified.

\paragraph{Threshold calibration.}
The controlled density sweep uses full-pool empirical score quantiles to isolate the effect of $p$. Those thresholds are benchmark calibrators, not deployable online estimators. The pilot ablation shows that estimating an ultra-rare quantile from a small sample can sharply weaken the result. Applications should pre-specify a scientific threshold or account separately for calibration data and uncertainty.

\paragraph{Budget censoring.}
The adaptive scalable and GP diagnostics use finite query caps, with failures assigned the cap. Their means are conservative lower bounds on uncensored non-quantum first-hit burden, but they do not by themselves report post-cap success. The p90 and survival results must be interpreted with those caps visible.

\paragraph{Clinical scope.}
This study uses canonical dynamical-system benchmarks. It does not analyze patient cohorts, validate diagnosis, estimate treatment utility, or establish biomarkers. Applying QC-PHAST to clinical simulators or patient-level data would require separate data governance, endpoint definition, leakage controls, and validation.

\paragraph{Learned surrogates are triage tools.}
Learned models can prioritize candidates, but simulator verification remains necessary. The current learned-triage results are not uniformly strong enough to replace exact scoring.

\paragraph{Baseline scope.}
The final sweep includes scalable baselines across all 875 configurations. Bayesian LCB is included in default and stress runs, and the access-model ablation adds GP active-search diagnostics over the same 875-base-configuration design. The continuous-structure challenge adds analytic probes, adaptive mesh, differential evolution, rank-Gaussian proposals, and continuous GP-LCB, while the 32-dimensional embedded-FHN stress test adds off-grid random, Sobol, rank-adaptive Gaussian, local trust-region, and differential-evolution challengers over 500 configurations. This is still not an exhaustive classical optimization benchmark. Full system-specific continuation packages, specialized bifurcation solvers, and production BoTorch/TuRBO or CMA-ES pipelines remain outside the main finite-pool query model unless installed and run as named methods. For systems with reliable differentiable score access, those methods should be evaluated before \qcphast{} routes to its marked-set branch.

\paragraph{Benchmark scope.}
The benchmark suite is broad enough to test the framework across several canonical geometries, but it is not exhaustive. The online challenge adds stiff-ODE and compact flow-proxy simulators, but it still does not establish performance on production PDE-scale models, stochastic simulators, hybrid mechanistic-neural simulators, or externally validated biomedical models. The correct generalization is methodological: \qcphast{} applies when a candidate set, score, threshold, and verifier can be defined.

\paragraph{Selector scope.}
The routing procedure is evidence-gated and rule-based. It is not a trained algorithm-selection model, and the current benchmark does not estimate selector generalization to unseen simulator families. Claims of automated per-instance selection would require a larger meta-dataset and held-out family evaluation.

\section{Reproducibility}

The public code repository is \url{https://github.com/CVC-Lab/QC-PHAST-ML.git}. Artifact paths reported below are repository-relative paths in that reproducibility package, not absolute cluster or Overleaf paths. The arXiv folder contains only manuscript files, figures, tables, and references. The final confidence sweep saved:
\begin{itemize}
    \item \texttt{configs/qcphast\_final\_confidence\_sweep\_configs.json}
    \item \texttt{results/qcphast\_final\_confidence\_sweep.json}
    \item \texttt{logs/qcphast\_final\_confidence\_sweep-20260720-105837.log}
    \item \texttt{manifests/qcphast\_final\_confidence\_sweep\_manifest.json}
    \item generated tables, figures, and final reports
\end{itemize}
The final audit reports 875 configurations, 32 trials per configuration, 72 workers, no zero-positive final rows, and a clean final log scan. The full-capacity confirmation artifacts add 2,975 fixed-threshold configurations, 1,750 all-in pilot-calibration configurations, 1,575 continuous-routing configurations, and 875 noise base configurations. The additional ablation artifacts add:
\begin{itemize}
    \item \texttt{configs/qcphast\_access\_model\_ablation\_configs.json}
    \item \texttt{results/qcphast\_access\_model\_ablation.json}
    \item \texttt{logs/qcphast\_access\_model\_ablation-20260720-105911.log}
    \item \texttt{results/qcphast\_geometry\_ablation.json}
    \item \texttt{results/qcphast\_survival\_analysis.json}
    \item \texttt{results/qcphast\_oracle\_resource\_estimate.json}
    \item \texttt{results/qcphast\_oracle\_break\_even.json}
    \item \texttt{results/qcphast\_noise\_phase\_diagram.json}
    \item \texttt{results/qcphast\_classical\_structure\_challenge.json}
    \item \texttt{results/qcphast\_online\_simulator\_challenge.json}
    \item \texttt{results/qcphast\_qpu\_runtime\_sanity.json}
    \item \texttt{results/qcphast\_embedded\_fhn\_32d\_continuous\_stress.json}
    \item \texttt{results/qcphast\_arxiv\_june\_default\_learned\_oracles.json}
    \item \texttt{results/qcphast\_objective\_bridge\_analysis.json}
    \item \texttt{results/qcphast\_dynamical\_bridge\_extensions.json}
    \item \texttt{results/qcphast\_fixed\_threshold\_replay.json}
    \item \texttt{results/qcphast\_pilot\_threshold\_accounting.json}
    \item \texttt{results/qcphast\_fixed\_threshold\_fullcap\_20260721.json}
    \item \texttt{results/qcphast\_pilot\_threshold\_fullcap\_20260721.json}
    \item \texttt{results/qcphast\_structure\_fullcap\_20260721.json}
    \item \texttt{results/qcphast\_noise\_fullcap\_20260721.json}
    \item \texttt{results/qcphast\_hierarchical\_evidence.json}
    \item \texttt{results/qcphast\_bbht\_reference\_validation.json}
    \item \texttt{reports/example\_difficulty\_summary\_20260630.md}
    \item \texttt{reports/objective\_bridge\_analysis\_20260702.md}
    \item \texttt{reports/dynamical\_bridge\_extensions\_20260702.md}
    \item \texttt{reports/qcphast\_arxiv\_revision\_literature\_audit\_20260720.md}
\end{itemize}
Manuscript table values are copied from generated analysis tables rather than manually recomputed in the paper folder.

The final manifest records 64-character SHA-256 hashes for the final config and result files. These hashes are intended to make the reported tables auditable. Table~\ref{tab:artifact-hashes} prints the exact manifest values, and Table~\ref{tab:artifact-map} maps the main manuscript claims to generated analysis artifacts. The manuscript folder is intentionally limited to TeX, bibliography, figures, and table assets; raw results, logs, and scripts remain in the public reproducibility package.

The bibliography was also audited during manuscript preparation. DOI-bearing entries were checked against indexed metadata where available; non-DOI proceedings or books were checked against official proceedings, arXiv records, or library-style records. The reference audit is stored as \texttt{reports/reference\_audit\_20260630.md}, and the later bibliography-modernization log is stored as \texttt{reports/bibliography\_modernization\_20260702.md} in the public reproducibility package. The updated bibliography keeps only the most necessary pre-2000 sources and uses modern dynamical-systems, quantum-algorithm, Bayesian-optimization, rare-event, and quasi-Monte Carlo references for the surrounding context.

\begin{table*}[!htbp]
\centering
\caption{Final confidence-sweep reproducibility manifest. Values are read from the regenerated result and manifest inputs rather than recomputed in the manuscript directory. Paths are repository-relative within the reproducibility package.}
\label{tab:artifact-hashes}
\footnotesize
\renewcommand{\arraystretch}{1.18}
\setlength{\tabcolsep}{4pt}
\begin{tabular}{@{}p{0.18\textwidth}p{0.74\textwidth}@{}}
\toprule
Field & Value \\
\midrule
Configurations & 875 configurations, 32 trials/configuration, 72 workers \\
Scope & canonical dynamical systems only \\
Config file & \path{configs/qcphast_final_confidence_sweep_configs.json} \\
Config SHA-256 & \shortstack[l]{\texttt{2572221190d2846f5b84e98f3a872e2}\\\texttt{df42d69174e0836d97f7f75b2b099941c}} \\
Result file & \path{results/qcphast_final_confidence_sweep.json} \\
Result SHA-256 & \shortstack[l]{\texttt{722cd57e5b127a2af09bf30b187cef92}\\\texttt{a74e56e57e14af8db4c2711e2577083a}} \\
Run log & \path{logs/qcphast_final_confidence_sweep-20260720-105837.log} \\
\bottomrule
\end{tabular}
\end{table*}

\begin{table*}[!htbp]
\centering
\caption{Artifact map from manuscript claims to generated evidence. Paths are repository-relative in the reproducibility package.}
\label{tab:artifact-map}
\footnotesize
\renewcommand{\arraystretch}{1.16}
\setlength{\tabcolsep}{3pt}
\begin{tabular}{@{}p{0.22\textwidth}p{0.56\textwidth}p{0.14\textwidth}@{}}
\toprule
Claim or table & Source artifact & Manuscript location \\
\midrule
Offline density sweep & \path{results/qcphast_final_confidence_sweep.json}; \path{tables/final_by_fraction.tex} & Table~\ref{tab:final-by-fraction} \\
Paired hierarchical uncertainty and density exponents & \path{results/qcphast_hierarchical_evidence.json}; \path{tables/hierarchical_density.tex} & Table~\ref{tab:hierarchical-density} \\
Full-capacity fixed-threshold lazy-information replay and no-target handling & \path{results/qcphast_fixed_threshold_fullcap_20260721.json}; \path{configs/qcphast_fixed_threshold_fullcap_20260721_combine_configs.json}; \path{tables/fixed_threshold_replay.tex} & Table~\ref{tab:fixed-threshold-online} \\
Full-capacity all-in pilot-threshold calibration accounting & \path{results/qcphast_pilot_threshold_fullcap_20260721.json}; \path{configs/qcphast_pilot_threshold_fullcap_20260721_combine_configs.json}; \path{tables/pilot_threshold_all_in.tex} & Table~\ref{tab:pilot-all-in} \\
Full-capacity direct structure-aware transition certificates & \path{results/qcphast_fixed_threshold_fullcap_20260721.json}; \path{tables/transition_certificates.tex} & Table~\ref{tab:transition-certificates} \\
BBHT amplitude-subspace validation & \path{results/qcphast_bbht_reference_validation.json} & Appendix \\
Scalar-score access ablation & \path{results/qcphast_access_model_ablation.json} & Table~\ref{tab:access-ablation} \\
Full-capacity predicate-noise phase diagram & \path{results/qcphast_noise_fullcap_20260721.json}; \path{configs/qcphast_noise_fullcap_20260721_combine_configs.json}; \path{tables/noise_phase_diagram.tex} & Table~\ref{tab:noise-phase-diagram} \\
Full-capacity continuous and structure-aware routing challenge & \path{results/qcphast_structure_fullcap_20260721.json}; \path{configs/qcphast_structure_fullcap_20260721_combine_configs.json}; \path{tables/classical_structure_challenge.tex} & Table~\ref{tab:classical-structure-challenge} \\
\bottomrule
\end{tabular}
\end{table*}

\section{Conclusion}

\qcphast{} formulates rare-regime discovery in parameterized dynamical systems as evidence-gated finite-pool search selection with verified criticality predicates and explicit access models. The Grover/BBHT scaling law is inherited; the contribution is the auditable reduction and regime map around it. Across canonical excitable, nonlinear, smooth-control, and chaotic benchmarks, the final 875-configuration sweep gives a 2.71 point estimate for the included non-quantum / BBHT ratio at $M/N=0.001$ under exact finite-pool replay with a binary-oracle BBHT reference and scalable score-guided baselines; paired hierarchical resampling gives 2.71 [1.89, 3.68]. Four expanded confirmation sweeps span 7,175 base configurations and preserve the central boundary conditions: all-in pilot calibration with $B=256$ gives 2.62 [2.49, 2.77] under a restricted predicate-only portfolio, the predicate-only false-positive 5\% confirmation gives 0.17 [0.15, 0.20], and smooth analytic or equation-aware controls solve several systems directly. The stronger scientific-ML reading is therefore conditional: scalar-score GP access gives 2.24 [2.02, 2.47] with a 0.71 win rate, the pilot GP stress is explicitly search-only, noisy predicates can drive the ratio below one and reverse the comparison, and oracle/state-preparation costs can erase any hardware interpretation of a 2--3$\times$ query ratio. Direct FHN, Duffing, and Lorenz controls show why generic finite-pool gaps must not be read as claims about equation-aware continuous solvers. Theoretical purity and pilot-coverage calculations explain why predicate error and threshold calibration fail on the scale of $p$. The main lesson is therefore a routing rule: use classical structure when it is exposed, use scalar-score active search when it is informative, retain marked-set query search only for exact rare finite predicates, and make no runtime claim without a separate oracle resource estimate.

\FloatBarrier

\appendix

\section{Expanded Benchmark Notes}

\paragraph{Candidate designs.}
Default candidate pools use Sobol designs for most higher-dimensional systems and grids for the two-dimensional coupled FHN slice. The final confidence sweep samples from the saved source candidate pools at multiple size levels. Density sweeps use score quantiles to enforce target marked fractions.

\begin{table*}[!htbp]
\centering
\caption{Expanded benchmark definitions used by the default, stress, and final confidence-sweep evidence. Unlike the compact objective tables in the main benchmark narrative, this appendix table preserves the generated parameter bounds and candidate-library sizes needed for exact reproduction. Bounds are read from the generated configuration files; stress $N$ is shown only for systems included in the stress run.}
\label{tab:benchmark-suite}
\small
\renewcommand{\arraystretch}{1.22}
\setlength{\tabcolsep}{3pt}
\begin{tabular}{@{}p{0.14\textwidth}c p{0.24\textwidth}r r p{0.31\textwidth}@{}}
\toprule
System & $d$ & Parameter bounds & \shortstack[c]{Default\\$N$} & \shortstack[c]{Stress\\$N$} & Score and role \\
\midrule
FitzHugh-Nagumo & 4 & $\epsilon\in[0.02,0.25]$, $a\in[0.5,1.0]$, $b\in[0.5,1.2]$, $I_{\mathrm{ext}}\in[0,1.5]$ & 20,000 & 100,000 & Minimum distance of the leading real Jacobian eigenvalue to zero across real equilibria; reduced excitable-dynamics stability boundary. \\
Coupled FHN & 2 & $I_1\in[0,1.5]$, $k\in[0,0.5]$ with fixed $\epsilon_{1,2}=0.1$, $a_{1,2}=0.7$, $b_{1,2}=0.8$, $I_2=0.5$ & 10,000 & 40,000 & Leading-real-eigenvalue distance over equilibria returned by three fixed root starts; two-unit excitable interaction stability-proxy benchmark. \\
Van der Pol & 2 & $\mu\in[-1,1]$, $\omega\in[0.5,2.0]$ & 12,000 & -- & $s_{\mathrm{VDP}}=|\mu|$; clean relaxation-oscillator boundary around the zero-damping slice. \\
Duffing & 4 & $\alpha\in[-1,1]$, $\beta\in[0.5,2.0]$, $\gamma\in[-1,1]$, damping $\in[0.05,0.6]$ & 20,000 & 100,000 & Minimum leading-real-eigenvalue distance across real roots of $\beta x^3+\alpha x-\gamma=0$; nonlinear oscillator stability boundary. \\
Windy pendulum & 3 & torque $\in[-1.5,1.5]$, damping $\in[0,0.5]$, gravity $\in[0.8,1.2]$ & 16,000 & -- & $\min\{|\,|u|-g\,|, |d|\}$; saddle-node and conservative-boundary control benchmark. \\
Spring-mass-damper & 2 & stiffness $\in[0,2.0]$, damping $\in[0,0.5]$ & 12,000 & -- & $\min\{|k|, |c|\}$; smooth/easy control case where adaptive classical search can compete. \\
Lorenz Hopf boundary & 3 & $\sigma\in[8,12]$, $\beta\in[2,3]$, $\rho\in[10,40]$ & 16,000 & 100,000 & $|\rho-\rho_H|$ when $\sigma>\beta+1$; chaotic/stability-boundary benchmark. \\
\bottomrule
\end{tabular}
\end{table*}

\paragraph{Default thresholds versus density thresholds.}
The default and stress benchmarks use system-specific fixed thresholds supplied by their benchmark configurations. The final confidence sweep instead controls $M/N$ by a full-pool score quantile. This is why the final sweep is the right evidence for conditional density scaling, while default and stress tables are the right evidence for concrete benchmark instances. A full-pool quantile must not be described as an online discovery procedure: it is available only to the evaluator constructing the controlled experiment. The expanded fixed-threshold replay retains possible no-target pools, including five Lorenz no-target draws, while the all-in pilot-calibration control charges finite information to empirical calibration.

\paragraph{Small marked counts.}
At the rarest fraction and smallest candidate-pool sizes, some configurations have only a few marked candidates. The final validation found no zero-positive rows and a minimum marked count of two. These small counts are expected under rare-event search and are precisely where query efficiency matters.

\section{BBHT Query Simulation Details}

For a candidate set of size $N$ with $M>0$ marked candidates, the evaluator sets $\theta=\arcsin\sqrt{M/N}$ so that it can sample the correct measurement probability. Each simulated policy trial starts with BBHT scale $m=1$, samples an integer Grover-iteration count $j$ uniformly from the current range, adds $j$ marked-oracle calls and one verification check to the query counter, and samples success with probability $\sin^2((2j+1)\theta)$. If the attempt fails, $m$ is multiplied by $6/5$ and capped at $\sqrt{N}$. The trial stops at the first success or at the finite-set cap $N$.

Knowing $M$ inside the evaluator is not the same as giving $M$ to the simulated policy. The sampled BBHT schedule does not use $M$ to choose $j$; $M$ enters only the Bernoulli success probability that emulates measurement. The implementation does not simulate amplitudes over $N$ basis states, compile a reversible predicate, or return a concrete candidate index. Under an exact predicate, a sampled successful measurement is marked by construction. Under a statically corrupted predicate, the simulation additionally samples whether an observed marked measurement belongs to the true marked subset and charges another restart after failed verification. A $j=0$ attempt is charged one final candidate verification, and $M=0$ is retained as a no-target outcome rather than converted into a discovery. The reported BBHT values are therefore stochastic query-count references, not statevector or hardware experiments.

\begin{table}[!htbp]
\centering
\caption{Validation of the BBHT probability simulator. The exact two-dimensional marked/unmarked amplitude recurrence is compared with $\sin^2((2j+1)\theta)$ over the sampled finite-pool cases. A $j=0$ attempt is charged one final candidate verification; $M=0$ is retained as a no-target outcome rather than converted into a discovery.}
\label{tab:bbht-validation}
\footnotesize
\renewcommand{\arraystretch}{1.18}
\setlength{\tabcolsep}{5pt}
\begin{tabular}{lr}
\toprule
Diagnostic & Value \\
\midrule
Validation cases & 500 \\
Probability checks & 32000 \\
Maximum absolute probability error & 5.107e-15 \\
\bottomrule
\end{tabular}
\end{table}

The probability simulator agrees with the exact two-dimensional marked/unmarked amplitude recurrence to maximum absolute error $5.107\times10^{-15}$ across 500 finite-pool cases and 32,000 probability checks. This validates the probability layer used in the reported schedule simulation; it does not convert the study into a compiled quantum-circuit experiment.

\section{Expanded Limitation Notes}

\paragraph{Oracle construction.}
The query model assumes access to a binary predicate over the finite candidate set. Section~\ref{sec:limitations} gives the main cost-model caveats; this appendix records the practical consequence for reproducibility: the reported BBHT numbers should be compared as predicate-query counts, not as compiled circuit resource estimates.

\paragraph{Wall-clock time.}
The final sweep runs quickly on CPU because it evaluates query policies over saved simulator-score datasets. That wall-clock runtime should not be interpreted as the runtime of a future quantum implementation. The comparison metric is query count.

\paragraph{Scientific generalization.}
The benchmark suite is intentionally broad but not exhaustive. It covers canonical systems, not every scientific simulator. The correct generalization is methodological: \qcphast{} can be applied when a simulator, candidate set, score, and marked threshold can be defined.

\section{Code Availability}

The \qcphast{} research code has been uploaded to \url{https://github.com/CVC-Lab/QC-PHAST-ML.git}. The repository contains the experiment runners, shared utilities, reproducibility configurations, lightweight generated summary tables, and audit reports used to support this manuscript. Detailed execution commands are maintained in the repository README rather than duplicated in this paper.

\section{Supported Scope Boundaries}

The supported central claim is that \qcphast{} provides a reproducible query-efficiency framework for discovering rare regimes in parameterized dynamical systems. The scope boundaries are: no patient-level diagnosis, no treatment-utility claim, no clinical validation claim, no hardware quantum-speedup claim, and no universal-superiority claim over all classical or non-quantum methods.

\FloatBarrier

\bibliographystyle{plainnat}
\bibliography{references}

@article{grover1997,
  author = {Grover, Lov K.},
  title = {Quantum Mechanics Helps in Searching for a Needle in a Haystack},
  journal = {Physical Review Letters},
  volume = {79},
  number = {2},
  pages = {325--328},
  year = {1997},
  doi = {10.1103/PhysRevLett.79.325}
}

@article{boyer1998,
  author = {Boyer, Michel and Brassard, Gilles and H{\o}yer, Peter and Tapp, Alain},
  title = {Tight Bounds on Quantum Searching},
  journal = {Fortschritte der Physik},
  volume = {46},
  number = {4--5},
  pages = {493--505},
  year = {1998},
  doi = {10.1002/(SICI)1521-3978(199806)46:4/5<493::AID-PROP493>3.0.CO;2-P},
  eprint = {quant-ph/9605034},
  archivePrefix = {arXiv}
}

@incollection{brassard2002,
  author = {Brassard, Gilles and H{\o}yer, Peter and Mosca, Michele and Tapp, Alain},
  title = {Quantum Amplitude Amplification and Estimation},
  booktitle = {Quantum Computation and Information},
  series = {Contemporary Mathematics},
  volume = {305},
  pages = {53--74},
  publisher = {American Mathematical Society},
  year = {2002},
  doi = {10.1090/conm/305/05215},
  eprint = {quant-ph/0005055},
  archivePrefix = {arXiv}
}

@inproceedings{aaronson2020approxcount,
  author = {Aaronson, Scott and Rall, Patrick},
  title = {Quantum Approximate Counting, Simplified},
  booktitle = {Proceedings of the Symposium on Simplicity in Algorithms},
  year = {2020},
  doi = {10.1137/1.9781611976014.5},
  eprint = {1908.10846},
  archivePrefix = {arXiv}
}

@article{yoder2014fixedpoint,
  author = {Yoder, Theodore J. and Low, Guang Hao and Chuang, Isaac L.},
  title = {Fixed-Point Quantum Search with an Optimal Number of Queries},
  journal = {Physical Review Letters},
  volume = {113},
  number = {21},
  pages = {210501},
  year = {2014},
  doi = {10.1103/PhysRevLett.113.210501},
  eprint = {1409.3305},
  archivePrefix = {arXiv}
}

@article{suzuki2020,
  author = {Suzuki, Yohichi and Uno, Shumpei and Raymond, Rudy and Tanaka, Tomoki and Onodera, Tamiya and Yamamoto, Naoki},
  title = {Amplitude Estimation without Phase Estimation},
  journal = {Quantum Information Processing},
  volume = {19},
  number = {2},
  pages = {75},
  year = {2020},
  doi = {10.1007/s11128-019-2565-2}
}

@article{grinko2021,
  author = {Grinko, Dmitry and Gacon, Julien and Zoufal, Christa and Woerner, Stefan},
  title = {Iterative Quantum Amplitude Estimation},
  journal = {npj Quantum Information},
  volume = {7},
  number = {1},
  pages = {52},
  year = {2021},
  doi = {10.1038/s41534-021-00379-1}
}

@article{cerezo2021,
  author = {Cerezo, M. and Arrasmith, Andrew and Babbush, Ryan and Benjamin, Simon C. and Endo, Suguru and Fujii, Keisuke and McClean, Jarrod R. and Mitarai, Kosuke and Yuan, Xiao and Cincio, Lukasz and Coles, Patrick J.},
  title = {Variational Quantum Algorithms},
  journal = {Nature Reviews Physics},
  volume = {3},
  number = {9},
  pages = {625--644},
  year = {2021},
  doi = {10.1038/s42254-021-00348-9}
}

@article{bharti2022,
  author = {Bharti, Kishor and Cervera-Lierta, Alba and Kyaw, Thi Ha and Haug, Tobias and Alperin-Lea, Sumner and Anand, Abhinav and Degroote, Matthias and Heimonen, Hermanni and Kottmann, Jakob S. and Menke, Tim and Mok, Wai-Keong and Sim, Sukin and Kwek, Leong-Chuan and Aspuru-Guzik, Alan},
  title = {Noisy Intermediate-Scale Quantum Algorithms},
  journal = {Reviews of Modern Physics},
  volume = {94},
  number = {1},
  pages = {015004},
  year = {2022},
  doi = {10.1103/RevModPhys.94.015004}
}

@article{babbush2021beyondquadratic,
  author = {Babbush, Ryan and McClean, Jarrod R. and Newman, Michael and Gidney, Craig and Boixo, Sergio and Neven, Hartmut},
  title = {Focus beyond Quadratic Speedups for Error-Corrected Quantum Advantage},
  journal = {PRX Quantum},
  volume = {2},
  number = {1},
  pages = {010103},
  year = {2021},
  doi = {10.1103/PRXQuantum.2.010103},
  eprint = {2011.04149},
  archivePrefix = {arXiv}
}

@article{dimatteo2019qram,
  author = {Di Matteo, Olivia and Gheorghiu, Vlad and Mosca, Michele},
  title = {Fault-Tolerant Resource Estimation of Quantum Random-Access Memories},
  year = {2019},
  eprint = {1902.01329},
  archivePrefix = {arXiv}
}

@article{duan2024qram,
  author = {Duan, Bojia and Hsieh, Chang-Yu},
  title = {Compact and Classically Preprocessed Data-Loading Quantum Circuit as a Quantum Random Access Memory},
  journal = {Physical Review A},
  volume = {110},
  number = {1},
  pages = {012616},
  year = {2024},
  doi = {10.1103/PhysRevA.110.012616}
}

@article{cuccaro2004adder,
  author = {Cuccaro, Steven A. and Draper, Thomas G. and Kutin, Samuel A. and Moulton, David Petrie},
  title = {A New Quantum Ripple-Carry Addition Circuit},
  year = {2004},
  eprint = {quant-ph/0410184},
  archivePrefix = {arXiv}
}

@article{haener2018arithmetic,
  author = {H{\"a}ner, Thomas and Roetteler, Martin and Svore, Krysta M.},
  title = {Optimizing Quantum Circuits for Arithmetic},
  year = {2018},
  eprint = {1805.12445},
  archivePrefix = {arXiv}
}

@article{remaud2024adders,
  author = {Remaud, Maxime},
  title = {Optimizing T and CNOT Gates in Quantum Ripple-Carry Adders and Comparators},
  year = {2024},
  eprint = {2401.17921},
  archivePrefix = {arXiv}
}

@article{gidney2019magic,
  author = {Gidney, Craig and Fowler, Austin G.},
  title = {Efficient Magic State Factories with a Catalyzed {\textbar}CCZ{\textgreater} to 2{\textbar}T{\textgreater} Transformation},
  journal = {Quantum},
  volume = {3},
  pages = {135},
  year = {2019},
  doi = {10.22331/q-2019-04-30-135},
  eprint = {1812.01238},
  archivePrefix = {arXiv}
}

@inproceedings{regev2012faultyoracle,
  author = {Regev, Oded and Schiff, Liron},
  title = {Impossibility of a Quantum Speed-Up with a Faulty Oracle},
  booktitle = {Automata, Languages and Programming},
  series = {Lecture Notes in Computer Science},
  volume = {5125},
  pages = {773--781},
  publisher = {Springer},
  year = {2008},
  doi = {10.1007/978-3-540-70575-8_63},
  eprint = {1202.1027},
  archivePrefix = {arXiv}
}

@article{shenvi2003noisyoracle,
  author = {Shenvi, Neil and Brown, Kenneth R. and Whaley, K. Birgitta},
  title = {Effects of Noisy Oracle on Search Algorithm Complexity},
  year = {2003},
  eprint = {quant-ph/0304138},
  archivePrefix = {arXiv}
}

@article{lolck2024faultyoracle,
  author = {Lolck, David Rasmussen and Man\v{c}inska, Laura and Paraashar, Manaswi},
  title = {Quantum Advantage with Faulty Oracle},
  year = {2024},
  eprint = {2411.04931},
  archivePrefix = {arXiv}
}

@article{montanaro2015,
  author = {Montanaro, Ashley},
  title = {Quantum Speedup of Monte Carlo Methods},
  journal = {Proceedings of the Royal Society A},
  volume = {471},
  number = {2181},
  pages = {20150301},
  year = {2015},
  doi = {10.1098/rspa.2015.0301},
  eprint = {1504.06987},
  archivePrefix = {arXiv}
}

@article{montanaro2020branch,
  author = {Montanaro, Ashley},
  title = {Quantum Speedup of Branch-and-Bound Algorithms},
  journal = {Physical Review Research},
  volume = {2},
  number = {1},
  pages = {013056},
  year = {2020},
  doi = {10.1103/PhysRevResearch.2.013056},
  eprint = {1906.10375},
  archivePrefix = {arXiv}
}

@book{dalzell2025quantumalgorithms,
  author = {Dalzell, Alexander M. and McArdle, Sam and Berta, Mario and Bienias, Przemyslaw and Chen, Chi-Fang and Gily{\'e}n, Andr{\'a}s and Hann, Connor T. and Kastoryano, Michael J. and Khabiboulline, Emil T. and Kubica, Aleksander and Salton, Grant and Wang, Samson and Brand{\~a}o, Fernando G. S. L.},
  title = {Quantum Algorithms: A Survey of Applications and End-to-End Complexities},
  publisher = {Cambridge University Press},
  year = {2025},
  doi = {10.1017/9781009639651},
  eprint = {2310.03011},
  archivePrefix = {arXiv}
}

@article{liu2021,
  author = {Liu, Jin-Peng and Kolden, Herman Oie and Krovi, Hari K. and Loureiro, Nuno F. and Trivisa, Konstantina and Childs, Andrew M.},
  title = {Efficient Quantum Algorithm for Dissipative Nonlinear Differential Equations},
  journal = {Proceedings of the National Academy of Sciences},
  volume = {118},
  number = {35},
  pages = {e2026805118},
  year = {2021},
  doi = {10.1073/pnas.2026805118}
}

@article{auyeung2024quantumscientific,
  author = {Au-Yeung, Rhonda and Camino, B. and Rathore, O. and Kendon, Viv},
  title = {Quantum Algorithms for Scientific Computing},
  journal = {Reports on Progress in Physics},
  volume = {87},
  number = {11},
  pages = {116001},
  year = {2024},
  doi = {10.1088/1361-6633/ad85f0},
  eprint = {2312.14904},
  archivePrefix = {arXiv}
}

@article{wu2024nonlineardynamics,
  author = {Wu, Hsuan-Cheng and Wang, Jingyao and Li, Xiantao},
  title = {Quantum Algorithms for Nonlinear Dynamics: Revisiting Carleman Linearization with No Dissipative Conditions},
  year = {2024},
  eprint = {2405.12714},
  archivePrefix = {arXiv}
}

@article{bergstra2012,
  author = {Bergstra, James and Bengio, Yoshua},
  title = {Random Search for Hyper-Parameter Optimization},
  journal = {Journal of Machine Learning Research},
  volume = {13},
  pages = {281--305},
  year = {2012},
  url = {https://jmlr.org/papers/v13/bergstra12a.html}
}

@inproceedings{snoek2012,
  author = {Snoek, Jasper and Larochelle, Hugo and Adams, Ryan P.},
  title = {Practical Bayesian Optimization of Machine Learning Algorithms},
  booktitle = {Advances in Neural Information Processing Systems},
  volume = {25},
  year = {2012},
  eprint = {1206.2944},
  archivePrefix = {arXiv}
}

@inproceedings{balandat2020botorch,
  author = {Balandat, Maximilian and Karrer, Brian and Jiang, Daniel R. and Daulton, Samuel and Letham, Benjamin and Wilson, Andrew Gordon and Bakshy, Eytan},
  title = {BoTorch: A Framework for Efficient Monte-Carlo Bayesian Optimization},
  booktitle = {Advances in Neural Information Processing Systems},
  volume = {33},
  pages = {21524--21538},
  year = {2020},
  eprint = {1910.06403},
  archivePrefix = {arXiv}
}

@inproceedings{eriksson2019turbo,
  author = {Eriksson, David and Pearce, Michael and Gardner, Jacob R. and Turner, Ryan and Poloczek, Matthias},
  title = {Scalable Global Optimization via Local Bayesian Optimization},
  booktitle = {Advances in Neural Information Processing Systems},
  volume = {32},
  year = {2019},
  eprint = {1910.01739},
  archivePrefix = {arXiv}
}

@inproceedings{garnett2012active,
  author = {Garnett, Roman and Krishnamurthy, Yamuna and Xiong, Xuehan and Schneider, Jeff and Mann, Richard},
  title = {Bayesian Optimal Active Search and Surveying},
  booktitle = {Proceedings of the 29th International Conference on Machine Learning},
  year = {2012},
  url = {https://icml.cc/2012/papers/618.pdf}
}

@inproceedings{gotovos2013levelset,
  author = {Gotovos, Alkis and Casati, Nathalie and Hitz, Gregory and Krause, Andreas},
  title = {Active Learning for Level Set Estimation},
  booktitle = {Proceedings of the Twenty-Third International Joint Conference on Artificial Intelligence},
  year = {2013},
  url = {https://www.ijcai.org/Proceedings/13/Papers/202.pdf}
}

@inproceedings{shekhar2019multiscale,
  author = {Shekhar, Shubhanshu and Javidi, Tara},
  title = {Multiscale Gaussian Process Level Set Estimation},
  booktitle = {Proceedings of the Twenty-Second International Conference on Artificial Intelligence and Statistics},
  series = {Proceedings of Machine Learning Research},
  volume = {89},
  pages = {3283--3291},
  publisher = {PMLR},
  year = {2019},
  url = {https://proceedings.mlr.press/v89/shekhar19a.html}
}

@article{wang2023bo,
  author = {Wang, Xilu and Jin, Yaochu and Schmitt, Sebastian and Olhofer, Markus},
  title = {Recent Advances in Bayesian Optimization},
  journal = {ACM Computing Surveys},
  volume = {55},
  number = {13s},
  pages = {1--36},
  year = {2023},
  doi = {10.1145/3582078}
}

@inproceedings{xie2024costaware,
  author = {Xie, Qian and Astudillo, Raul and Frazier, Peter I. and Scully, Ziv and Terenin, Alexander},
  title = {Cost-Aware Bayesian Optimization via the Pandora's Box Gittins Index},
  booktitle = {Advances in Neural Information Processing Systems},
  year = {2024},
  eprint = {2406.20062},
  archivePrefix = {arXiv}
}

@article{xu2023greybox,
  author = {Xu, Wenjie and Jiang, Yuning and Svetozarevic, Bratislav and Jones, Colin N.},
  title = {Bayesian Optimization of Expensive Nested Grey-Box Functions},
  year = {2023},
  eprint = {2306.05150},
  archivePrefix = {arXiv}
}

@inproceedings{eriksson2021,
  author = {Eriksson, David and Jankowiak, Martin},
  title = {High-Dimensional Bayesian Optimization with Sparse Axis-Aligned Subspaces},
  booktitle = {Proceedings of the Thirty-Seventh Conference on Uncertainty in Artificial Intelligence},
  series = {Proceedings of Machine Learning Research},
  volume = {161},
  pages = {493--503},
  publisher = {PMLR},
  year = {2021},
  url = {https://proceedings.mlr.press/v161/eriksson21a.html}
}

@inproceedings{eriksson2021scbo,
  author = {Eriksson, David and Poloczek, Matthias},
  title = {Scalable Constrained Bayesian Optimization},
  booktitle = {Proceedings of the 24th International Conference on Artificial Intelligence and Statistics},
  series = {Proceedings of Machine Learning Research},
  volume = {130},
  pages = {730--738},
  publisher = {PMLR},
  year = {2021},
  url = {https://proceedings.mlr.press/v130/eriksson21a.html}
}

@inproceedings{ngo2025levelset,
  author = {Ngo, Giang and Nguyen, Dang and Gupta, Sunil},
  title = {Robust Transfer Learning for Active Level Set Estimation with Locally Adaptive Gaussian Process Prior},
  booktitle = {Proceedings of the 16th Asian Conference on Machine Learning},
  series = {Proceedings of Machine Learning Research},
  volume = {260},
  pages = {607--622},
  publisher = {PMLR},
  year = {2025},
  url = {https://proceedings.mlr.press/v260/ngo25a.html}
}

@inproceedings{pulatov2022algorithmselection,
  author = {Pulatov, Damir and Anastacio, Marie and Kotthoff, Lars and Hoos, Holger},
  title = {Opening the Black Box: Automated Software Analysis for Algorithm Selection},
  booktitle = {Proceedings of the First International Conference on Automated Machine Learning},
  series = {Proceedings of Machine Learning Research},
  volume = {188},
  pages = {6/1--18},
  publisher = {PMLR},
  year = {2022},
  url = {https://proceedings.mlr.press/v188/pulatov22a.html}
}

@inproceedings{kostovska2023psaas,
  author = {Kostovska, Ana and Cenikj, Gjorgjina and Vermetten, Diederick and Jankovic, Anja and Nikolikj, Ana and Skvorc, Urban and Korosec, Peter and Doerr, Carola and Eftimov, Tome},
  title = {{PS-AAS}: Portfolio Selection for Automated Algorithm Selection in Black-Box Optimization},
  booktitle = {Proceedings of the Second International Conference on Automated Machine Learning},
  series = {Proceedings of Machine Learning Research},
  volume = {224},
  pages = {11/1--17},
  publisher = {PMLR},
  year = {2023},
  url = {https://proceedings.mlr.press/v224/kostovska23a.html}
}

@book{rubinstein2004,
  author = {Rubinstein, Reuven Y. and Kroese, Dirk P.},
  title = {The Cross-Entropy Method: A Unified Approach to Combinatorial Optimization, Monte-Carlo Simulation, and Machine Learning},
  publisher = {Springer},
  year = {2004},
  doi = {10.1007/978-1-4757-4321-0}
}

@article{au2001,
  author = {Au, Siu-Kui and Beck, James L.},
  title = {Estimation of Small Failure Probabilities in High Dimensions by Subset Simulation},
  journal = {Probabilistic Engineering Mechanics},
  volume = {16},
  number = {4},
  pages = {263--277},
  year = {2001},
  doi = {10.1016/S0266-8920(01)00019-4}
}

@article{moustapha2021reliability,
  author = {Moustapha, Maliki and Marelli, Stefano and Sudret, Bruno},
  title = {Active Learning for Structural Reliability: Survey, General Framework and Benchmark},
  journal = {Structural Safety},
  pages = {102174},
  year = {2022},
  doi = {10.1016/j.strusafe.2021.102174},
  eprint = {2106.01713},
  archivePrefix = {arXiv}
}

@article{shyalika2024rareevent,
  author = {Shyalika, Chathurangi and Wickramarachchi, Ruwan and Sheth, Amit},
  title = {A Comprehensive Survey on Rare Event Prediction},
  journal = {ACM Computing Surveys},
  volume = {57},
  number = {3},
  pages = {1--39},
  year = {2024},
  doi = {10.1145/3699955},
  eprint = {2309.11356},
  archivePrefix = {arXiv}
}

@article{hansen2016cma,
  author = {Hansen, Nikolaus},
  title = {The CMA Evolution Strategy: A Tutorial},
  year = {2016},
  eprint = {1604.00772},
  archivePrefix = {arXiv}
}

@article{karniadakis2021,
  author = {Karniadakis, George Em and Kevrekidis, Ioannis G. and Lu, Lu and Perdikaris, Paris and Wang, Sifan and Yang, Liu},
  title = {Physics-Informed Machine Learning},
  journal = {Nature Reviews Physics},
  volume = {3},
  number = {6},
  pages = {422--440},
  year = {2021},
  doi = {10.1038/s42254-021-00314-5}
}

@article{brunton2020,
  author = {Brunton, Steven L. and Noack, Bernd R. and Koumoutsakos, Petros},
  title = {Machine Learning for Fluid Mechanics},
  journal = {Annual Review of Fluid Mechanics},
  volume = {52},
  number = {1},
  pages = {477--508},
  year = {2020},
  doi = {10.1146/annurev-fluid-010719-060214}
}

@article{willard2023,
  author = {Willard, Jared and Jia, Xiaowei and Xu, Shaoming and Steinbach, Michael and Kumar, Vipin},
  title = {Integrating Scientific Knowledge with Machine Learning for Engineering and Environmental Systems},
  journal = {ACM Computing Surveys},
  volume = {55},
  number = {4},
  pages = {1--37},
  year = {2023},
  doi = {10.1145/3514228}
}

@article{desai2021porthamiltonian,
  author = {Desai, Shaan and Mattheakis, Marios and Sondak, David and Protopapas, Pavlos and Roberts, Stephen},
  title = {Port-Hamiltonian Neural Networks for Learning Explicit Time-Dependent Dynamical Systems},
  journal = {Physical Review E},
  volume = {104},
  number = {3},
  pages = {034312},
  year = {2021},
  doi = {10.1103/PhysRevE.104.034312},
  eprint = {2107.08024},
  archivePrefix = {arXiv}
}

@inproceedings{takamoto2022pdebench,
  author = {Takamoto, Makoto and Praditia, Timothy and Leiteritz, Raphael and MacKinlay, Dan and Alesiani, Francesco and Pfl{\"u}ger, Dirk and Niepert, Mathias},
  title = {{PDEBench}: An Extensive Benchmark for Scientific Machine Learning},
  booktitle = {Advances in Neural Information Processing Systems},
  volume = {35},
  pages = {1596--1611},
  year = {2022},
  url = {https://proceedings.neurips.cc/paper_files/paper/2022/hash/0a9747136d411fb83f0cf81820d44afb-Abstract-Datasets_and_Benchmarks.html},
  eprint = {2210.07182},
  archivePrefix = {arXiv}
}

@article{ekanathan2024radau,
  author = {Ekanathan, Shreyas and Smith, Oscar and Rackauckas, Christopher},
  title = {A Fully Adaptive Radau Method for the Efficient Solution of Stiff Ordinary Differential Equations at Low Tolerances},
  year = {2024},
  eprint = {2412.14362},
  archivePrefix = {arXiv}
}

@article{choi2023qmcsoftware,
  author = {Choi, Sou-Cheng T. and Ding, Yuhan and Hickernell, Fred J. and Rathinavel, Jagadeeswaran and Sorokin, Aleksei G.},
  title = {Challenges in Developing Great Quasi-Monte Carlo Software},
  year = {2023},
  eprint = {2311.06162},
  archivePrefix = {arXiv}
}

@article{sorokin2023qmc,
  author = {Sorokin, Aleksei G. and Rathinavel, Jagadeeswaran},
  title = {On Bounding and Approximating Functions of Multiple Expectations Using Quasi-Monte Carlo},
  year = {2023},
  eprint = {2311.07555},
  archivePrefix = {arXiv}
}

@article{fitzhugh1961,
  author = {FitzHugh, Richard},
  title = {Impulses and Physiological States in Theoretical Models of Nerve Membrane},
  journal = {Biophysical Journal},
  volume = {1},
  number = {6},
  pages = {445--466},
  year = {1961},
  doi = {10.1016/S0006-3495(61)86902-6}
}

@article{nagumo1962,
  author = {Nagumo, Jin-Ichi and Arimoto, Suguru and Yoshizawa, Shuji},
  title = {An Active Pulse Transmission Line Simulating Nerve Axon},
  journal = {Proceedings of the IRE},
  volume = {50},
  number = {10},
  pages = {2061--2070},
  year = {1962},
  doi = {10.1109/JRPROC.1962.288235}
}

@article{izhikevich2006scholarpedia,
  author = {Izhikevich, Eugene M. and FitzHugh, Richard},
  title = {{FitzHugh-Nagumo} Model},
  journal = {Scholarpedia},
  volume = {1},
  number = {9},
  pages = {1349},
  year = {2006},
  doi = {10.4249/scholarpedia.1349},
  url = {http://www.scholarpedia.org/article/FitzHugh-Nagumo_model}
}

@book{izhikevich2007neuroscience,
  author = {Izhikevich, Eugene M.},
  title = {Dynamical Systems in Neuroscience: The Geometry of Excitability and Bursting},
  publisher = {MIT Press},
  address = {Cambridge, MA},
  year = {2007}
}

@article{cebrianlacasa2024sixdecades,
  author = {Cebri{\'a}n-Lacasa, Daniel and Parra-Rivas, Pedro and Ruiz-Reyn{\'e}s, Daniel and Gelens, Lendert},
  title = {Six Decades of the {FitzHugh-Nagumo} Model: A Guide Through Its Spatio-Temporal Dynamics and Influence Across Disciplines},
  journal = {Physics Reports},
  volume = {1096},
  pages = {1--39},
  year = {2024},
  doi = {10.1016/j.physrep.2024.09.014},
  eprint = {2404.11403},
  archivePrefix = {arXiv}
}

@article{vanderpol1926,
  author = {van der Pol, Balthasar},
  title = {On Relaxation-Oscillations},
  journal = {The London, Edinburgh, and Dublin Philosophical Magazine and Journal of Science},
  volume = {2},
  number = {11},
  pages = {978--992},
  year = {1926},
  doi = {10.1080/14786442608564127}
}

@article{lorenz1963,
  author = {Lorenz, Edward N.},
  title = {Deterministic Nonperiodic Flow},
  journal = {Journal of the Atmospheric Sciences},
  volume = {20},
  number = {2},
  pages = {130--141},
  year = {1963},
  doi = {10.1175/1520-0469(1963)020<0130:DNF>2.0.CO;2}
}

@book{strogatz2015,
  author = {Strogatz, Steven H.},
  title = {Nonlinear Dynamics and Chaos: With Applications to Physics, Biology, Chemistry, and Engineering},
  edition = {2},
  publisher = {Westview Press},
  year = {2015},
  isbn = {9780813349107}
}

@book{lee2013smoothmanifolds,
  author = {Lee, John M.},
  title = {Introduction to Smooth Manifolds},
  edition = {2},
  publisher = {Springer},
  year = {2013},
  doi = {10.1007/978-1-4419-9982-5}
}

@article{liessi2025matcont,
  author = {Liessi, Davide and Santi, Enrico and Vermiglio, Rossana and Thakur, Mayank and Meijer, Hil G. E. and Scarabel, Francesca},
  title = {New functionalities in MatCont: delay equations and Lyapunov exponents},
  year = {2025},
  eprint = {2504.12785},
  archivePrefix = {arXiv}
}

@article{govaerts2005matcont,
  author = {Govaerts, Willy and Kuznetsov, Yuri A. and Dhooge, Annick},
  title = {Numerical Continuation of Equilibria and Limit Cycles in {MATCONT}},
  journal = {SIAM Journal on Scientific Computing},
  volume = {27},
  number = {1},
  pages = {231--252},
  year = {2005},
  doi = {10.1137/030600746}
}

@article{echard2011akmcs,
  author = {Echard, Benjamin and Gayton, Nicolas and Lemaire, Maurice},
  title = {{AK-MCS}: An Active Learning Reliability Method Combining Kriging and Monte Carlo Simulation},
  journal = {Structural Safety},
  volume = {33},
  number = {2},
  pages = {145--154},
  year = {2011},
  doi = {10.1016/j.strusafe.2011.01.002}
}

@article{bect2012failure,
  author = {Bect, Julien and Ginsbourger, David and Li, Ling and Picheny, Victor and Vazquez, Emmanuel},
  title = {Sequential Design of Computer Experiments for the Estimation of a Probability of Failure},
  journal = {Statistics and Computing},
  volume = {22},
  number = {3},
  pages = {773--793},
  year = {2012},
  doi = {10.1007/s11222-011-9241-4}
}

\end{document}